\documentclass[twocolumn]{aastex631}

\usepackage{CJKutf8}
\usepackage{amsmath}
\usepackage{enumitem}
\usepackage{booktabs,pbox}
\usepackage{bm}
\usepackage{comment}
\usepackage{diagbox}
\usepackage[caption=false]{subfig}
\usepackage{chngpage}
\usepackage{booktabs}
\usepackage{type1cm}
\usepackage{txfonts}

\newcommand\PMO{Purple Mountain Observatory, Chinese Academy of Sciences, Nanjing 210023, China}
\newcommand\USTC{School of Astronomy and Space Sciences, University of Science and Technology of China, Hefei 230026, China}
\newcommand\NJU{School of Astronomy and Space Science, Nanjing University, Nanjing 210023, China}
\newcommand\NJULab{Key Laboratory of Modern Astronomy and Astrophysics (Nanjing University), Ministry of Education, China}
\newcommand\NAO{National Astronomical Observatories, Chinese Academy of Sciences, Beijing 100101, China}

\begin{document}

\title{The Soft X-ray Aspect of Gamma-ray Bursts in the Einstein Probe Era}

\author[0000-0001-7892-9790]{Hao-Xuan Gao}
\affiliation{\PMO}

\author[0000-0001-9648-7295]{Jin-Jun Geng}\thanks{E-mail: jjgeng@pmo.ac.cn}
\affiliation{\PMO}

\author[0009-0005-0170-192X]{Yi-Fang Liang}
\affiliation{\PMO}
\affiliation{\USTC}

\author[0000-0002-9615-1481]{Hui Sun}
\affiliation{\NAO}

\author[0000-0001-7943-4685]{Fan Xu}
\affiliation{\NJU}

\author[0000-0002-6299-1263]{Xue-Feng Wu}\thanks{E-mail: xfwu@pmo.ac.cn}
\affiliation{\PMO}
\affiliation{\USTC}

\author[0000-0001-7199-2906]{Yong-Feng Huang}%\thanks{E-mail: hyf@nju.edu.cn}
\affiliation{\NJU}
\affiliation{\NJULab}

\author[0000-0002-7835-8585]{Zi-Gao Dai}%\thanks{E-mail: daizg@ustc.edu.cn}
\affiliation{Department of Astronomy, School of Physical Sciences, University of Science and Technology of China, Hefei 230026, China}

%\author[0000-0002-9583-2947]{Bing Zhang}%\thanks{E-mail: zhang@physics.unlv.edu bing.zhang@unlv.edu }
%\affiliation{Nevada Center for Astrophysics, University of Nevada Las Vegas, Las Vegas, NV 89154, USA}
%\affiliation{Department of Physics and Astronomy, University of Nevada Las Vegas, Las Vegas, NV 89154, USA}

\author{Wei-Min Yuan}%\thanks{E-mail: wmy@nao.cas.cn}
\affiliation{\NAO}

\begin{abstract}

The Einstein Probe (EP) satellite, dedicated at time-domain high-energy astrophysics and multi-messenger astronomy, was recently launched and successfully put into operation.
The wide-field X-ray telescope (WXT, 0.5-4 keV) onboard has identified multiple gamma-ray burst (GRB) events, with an average duration of several hundred seconds.
This duration is several times longer than the average duration of long gamma-ray bursts (LGRBs) detected by the Neil Gehrels {\it Swift} Observatory, which typically stands at several tens of seconds.
Additionally, EP has detected some unknown X-ray transients whose connection to GRBs is uncertain, due to the absence of gamma-ray counterparts and efficient follow-up observation at multi-wavelengths.
Several main factors could account for the longer time, including the Doppler effect of off-axis viewing, the spectral lag effect of the synchrotron spectrum of cooling electrons, and some unknown prolonged intrinsic X-ray activities.
Our studies indicate that EP GRBs may primarily consist of off-axis viewed bursts, forming a unique population among the GRB zoo, yet the intrinsic origin for the specific bursts could not be excluded.
By analyzing the statistical properties of the historical LGRB samples, we explored observable properties of on-axis and off-axis LGRBs in the soft X-ray band.
The predicted characteristics of off-axis viewed GRBs, including the duration, energy fluence, low-energy spectral index, and the slopes of Amati and Yonetoku relations, could be tested with a larger sample of GRB events detected by EP in the future.

\end{abstract}

%\keywords{Gamma-ray bursts (629); High energy astrophysics (739); Non-thermal radiation sources (1119)}

\section{Introduction}
\label{sec:intro}

Gamma-ray bursts (GRBs), the most energetic stellar explosions in the Universe, have been observed and studied for nearly 60 years.
The prompt emission of GRBs is thought to come from the relativistic jet launched from the central compact remnant \citep{Blandford77,Eichler89,Piran04,Kumar15}.
It typically lasts for a few seconds (or less) and has a variable light curve consisting of several spikes.
However, the radiation mechanism responsible for the prompt emission remains an open question \citep{Zhang18}.
Both synchrotron radiation of non-thermal electrons \citep[e.g.,][]{Meszaros94,Tavani96,ZhangB11} and photospheric emission \citep[e.g.,][]{Rees05,Pe'er06,Ryde11} have been proposed to explain the GRB prompt emission.
Recently, \citet{Burgess20} suggest that synchrotron spectra from electrons in evolving (fast-to-slow) cooling regimes are capable of fitting 95\% of all time-resolved spectra of the brightest long GRBs observed by the gamma-ray burst monitor (GBM: 8 keV-40 MeV) on board the NASA $\it{Fermi}$ Gamma-Ray Observatory by comparing the theoretical results with observed data directly.

Before a relativistic outflow launched by the central engine can produce a successful GRB, it would inevitably propagate through the envelope or the ejected materials of the progenitor star~\citep{Bromberg11, Berger14, Nagakura14, Nakar17}. The structure of the jet may result from the jet formation mechanism itself \citep{Putten03, Vlahakis03, Aloy05}, or it may arise from the breakout process as the jet penetrates the stellar materials \citep{Levinson03, Zhang03, Lazzati05, Morsony10, Pescalli15, Geng16b}.
Structured jets, characterized by a narrow, highly relativistic inner core surrounded by less energetic, slower-moving wings at larger angles, have been extensively studied in the GRB community for over 20 years \citep[e.g.,][]{Meszaros98, Dai01, Lipunov01, Zhang02, Rossi02, Kumar03}. Two main types of structured jets, i.e., power-law like and Gaussian like jets, have have been discussed in the literature \citep{Granot17,Lazzati18,Troja19,Ryan20,Lamb17,Xiao17,Kathirgamaraju18,Meng18,Geng19,Li19,Gao22,Connor24}.

The duration of a burst is usually defined by the so-called ``$T_{90}$'', which is the time interval from 5 per cent to 95 per cent of the accumulated fluence.
It is found that the distribution of $T_{90}$ shows two Gaussian components with a separation line around 2 seconds in the logarithmic space \citep{Meegan92,Kouveliotou93}.
These two components are nowadays commonly classified into the short ($T_{90} < 2 $s) and long ($T_{90} > 2 $s) classes, which are widely believed to originate from relativistic outflows ejected during the merger of binary compact stars and the collapse of massive stars, respectively.
The number ratio of the two classes and the peak duration values of them are dependent on the energy bands and sensitivities of the instruments \citep[e.g.][]{Kouveliotou93,Sakamoto08b,Sakamoto11,Paciesas12,Zhang12d,Qin13}.

On the other hand, the cosmic rate and the luminosity function (LF) of GRBs allows us to test theories about their progenitors \citep[e.g.][]{Liang07,Pescalli15}.
These two functions have been derived for the population of long GRBs using various methods and samples of bursts \citep[e.g.][]{Daigne06,Guetta07,Firmani04,Salvaterra07,Salvaterra09b,Salvaterra12,Wanderman10,Yu15,Petrosian15,Lan21}.
Given the incomplete sampling of faint bursts and the low completeness in redshift measurements, \citet{Lan21} carefully selected a subsample of bright {\it Swift} bursts to revisit the GRB LF and redshift distribution, accounting for the probability of redshift measurement.
They also explored two general forms for the GRB LF: a broken power-law LF and a triple power-law LF.
Their results indicate that strong redshift evolution, either in luminosity (with an evolution index of $\delta= 1.92^{+0.25}_{-0.37}$) or in density ($\delta=1.26^{+0.33}_{-0.34}$), is necessary to adequately explain the observations, regardless of the assumed form of the GRB LF.

The spectrum of the prompt emission is non-thermal and often described by the so-called Band function empirically \citep{Band93}.
The Band function is characterized by three parameters, the low-energy and high-energy photon spectral indices and the peak energy ($E_\mathrm{p}$).
In past years, significant efforts have been dedicated to model the shape of prompt emission spectra \citep[e.g.,][]{Meszaros00,Pe'er06,Beloborodov10,Uhm14,Uhm16,Uhm18,Geng18b,Gao21}.
GRBs display correlations between the peak energy and other observational parameters.
The most famous one is the Amati relation \citep{Amati02}.
It connects the intrinsic hardness of $E_\mathrm{p,z}$, calculated as $E_{\mathrm{p,z}} = E_\mathrm{p}(1+z)$, and the isotropic equivalent total energy of $E_{\gamma,\mathrm{iso}}$ emitted in gamma-rays within the 1 to $10^{4}$ keV range.
The index of the Amati relation is about 0.5 \citep{Amati06,Nava12,Demianski17,Minaev20}.
Another important relation, the Yonetoku relation \citep{Yonetoku04}, shows the correlation between the intrinsic hardness of $E_\mathrm{p,z}$ and the peak luminosity of $L_{\mathrm{p,iso}}$.
The physical origin of these empirical relations remain under debate.
Some researchers have attempted to derive the on-axis and off-axis Amati relation indices by the analytical method or by means of numerical calculations \citep{Zhang02b,Granot02,Eichler04,Ramirez-Ruiz05,Dado12,Yamazaki04,Kocevski12,Mochkovitch15,Xu23}.
\citet{Xu23} provided a simple analytical derivation for both the Amati and Yonetoku relations within the standard fireball model, and this derivation was confirmed by numerical simulations. It was found that these relations strongly depend on the difference between the viewing angle and the jet opening angle when the jet is viewed off-axis.

Recently, the Einstein Probe (EP) \footnote{Einstein Probe is an international mission led by the Chinese Academy of Sciences (CAS) in collaboration with the European Space Agency (ESA), the Max-Planck-Institute for extraterrestrial Physics (MPE), Germany, and the Centre National d'$\acute{\text{E}}$tudes Spatiales (CNES), France.} satellite, a mission by the Chinese Academy of Sciences (CAS) dedicated at time-domain high-energy astrophysics and multi-messenger astronomy, was launched and successfully put into operation \citep{Yuan22,Yuan24}.
The mission's primary goals are to discover high-energy transients and monitor variable objects in the soft X-ray band, with sensitivity more than an order of magnitude greater than that of current orbiting instruments.
The EP is equipped with two scientific instruments: the wide-field X-ray telescope (WXT, 0.5-4 keV) and the follow-up X-ray telescope (FXT, 0.3-10 keV).
The WXT is designed to capture transients and monitor variable objects, while the FXT will perform deep follow-up observations of intriguing targets identified by the WXT and other facilities.
The successful operation of EP is expected to generate a wealth of observational data, including soft X-ray emissions during the main burst phase of gamma-ray bursts (GRBs), X-ray-rich GRBs that gamma-ray detectors may miss, and high-redshift GRBs.

EP has detected some GRB events, of which the average duration is approximately several hundred seconds \citep{Yin24,Zhang24,Zhou24a,Zhou24b}.
In contrast, previous studies on the duration of long bursts have shown that the average $T_{90}$ duration of LGRBs is approximately several tens of seconds, while the duration is known to be energy-dependent \citep{Qin13}.
Additionally, EP has detected some unknown X-ray transients that may be related to GRBs, but uncertain due to the absence of gamma-ray counterparts and efficient follow-up observation at multi-wavelengths.
These transients also have long durations, such as EP240414a and EP240416a \citep{Lian24b,Cheng24}.
It may indicate that a considerable sample of underlying LGRBs have relatively longer duration in the soft X-ray band, and point to the possibility of a distinct population among GRBs.

There are three possible origins of EP GRBs.
First, EP GRBs may consist of off-axis viewed LGRBs \citep{Troja17,Granot17,Haggard17,Ioka18}.
Due to the Doppler effect, off-axis LGRBs can exhibit longer durations compared to those of on-axis LGRBs.
The radiation energy of these bursts is concentrated in the soft X-ray energy band, while their hard X-ray emission may fall below the detection thresholds of other instruments.
This could explain the presence of unidentified X-ray transients with long durations detected by EP.
%Second, EP-detected bursts may simply represent typical LGRBs.
The second possibility comes from the so called ``spectral lag'' effect, i.e.,
the observed pulse gets increasingly broader in lower energy band~\citep{Cheng95,Norris96,Band97,Norris00,Wu00,Liang06}.
%A key characteristic of GRBs is the ``spectral lag'' observed between pulse light curves at different energy bands, where higher-energy light curves generally peak earlier than lower-energy light curves \citep{Cheng95,Norris96,Band97,Norris00,Wu00,Liang06}.
%This suggests that the duration of GRBs in the soft X-ray band may naturally be longer than in the hard X-ray band.
As a result, LGRBs observed by EP in the soft X-ray band may appear to have longer durations than their counterparts in the hard X-ray band.
An alternative possibility is that EP are detecting bursts with some prolonged X-ray activities, potentially induced the some energy release processes of the central engine, which remains highly uncertain.
In this work, we focus on the two former origins. The proportions of on-axis and off-axis bursts in EP-detected GRBs could be estimated through theoretical simulations, helping to distinguish between the two explanations.

%However, the proportions of on-axis and off-axis bursts in EP-detected GRBs can be estimated through theoretical simulations, providing valuable insights to distinguish between the two explanations mentioned above.}

In order to examine these two scenarios, we investigate the properties of LGRBs in the soft X-ray band from 0.5 to 4 keV that covers the energy band of EP-WXT. A brief overview of GRBs detected by EP is provided in Section \ref{sec:EP GRB}.
Several statistical characteristics of on-axis LGRBs are reviewed in Section \ref{sec:kc}.
The synchrotron radiation scenario and jet structure adopted in our calculations are briefly described in Section \ref{sec:model}.
In Section \ref{sec:MC}, the model parameters are constrained by matching the simulation results with the observed characteristics of on-axis GRBs, including the Amati and Yonetoku relations, the $T_{90}$ duration, and the redshift and luminosity distribution.
In Section \ref{sec:softXray}, we derive the properties of several observables for GRBs detected by EP and estimate the proportions of on-axis and off-axis bursts among the EP-detected GRBs.
Finally, we summarize our study in Section \ref{sec:conclusions}.
Throughout this paper a flat Lambda cold dark matter cosmological model with $H_0=70 \mathrm{~km}~\mathrm{~s}^{-1}~\mathrm{Mpc}^{-1}$, $\Omega_{\mathrm{m}}=0.3$, and $\Omega_{\Lambda}=0.7$ is adopted.

\section{A glance of EP GRBs}
\label{sec:EP GRB}

\begin{figure}
	\centering\includegraphics[scale=.35]{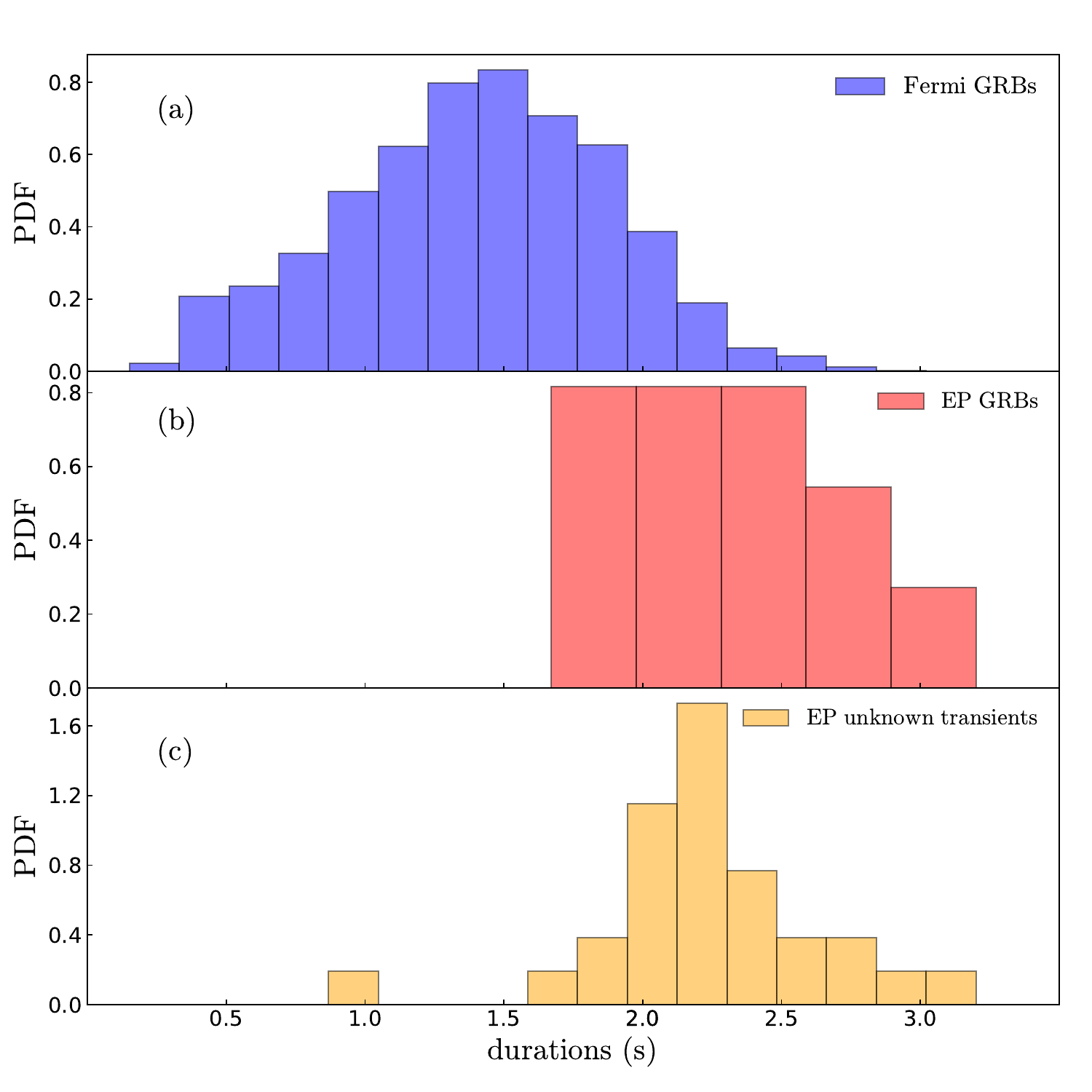}
    \caption{The duration distributions in the observer frame of Fermi GRBs, EP GRBs, and fast X-ray transients of unknown origin are compared.
Panel (a) illustrates the duration distribution of 1,964 Fermi LGRBs observed within the 50-300 keV energy band.
Panel (b) displays the duration distribution of 11 EP GRBs within the 0.5-4 keV energy band.
Panel (c) presents the duration distribution of 29 EP fast X-ray transients of unknown origin, also within the 0.5-4 keV energy band.}
    \label{fig:glance}
\end{figure}

By December 2024, WXT has detected around 11 X-ray transients with GRB origin as being consistent with the detections of the gamma-ray detectors.
The first GRB detected by WXT, EP240219a, has a $T_{90}$ duration of approximately 130 seconds in the 0.5-4 keV energy band \citep{Yin24}.
The second GRB observed by WXT, EP240315a, exhibited a $T_{90}$ duration of approximately 1,000 seconds within the same energy band \citep{Liu24}.
The durations of other 9 GRBs can be collected from the General Coordinates Network (GCN) \citep{Zhang24,Zhou24a,Zhou24b,Fu24,Li24b,Li24a,Tian24,LiuZY24,HuJW24c}.
Additionally, we gathered duration information from GCN for 29 fast X-ray transients detected by WXT, the origins of which remain unknown \citep{HuDF24,HuJW24a,HuJW24b,Cheng24,Pan24,Wu24,ZhangWJ24,WangY24,YangHN24a,YangHN24b,LianTY24,PengJQ24,ZhangYJ24,LiangYF24}.
We also collected durations of Fermi GRBs from \citet{Kienlin20}.
A total of 1964 long GRBs were selected based on the criterion that the $T_{90}$ duration of the burst exceeds 2 seconds.
The durations in the observer frame of Fermi GRBs, EP GRBs, and fast X-ray transients with unknown origins are presented in Figure \ref{fig:glance}.
The distributions of EP GRBs and unidentified fast X-ray transients are general distributions derived from the GCN data.
The average duration of GRBs detected by EP is approximately 280 s, significantly longer than the average duration of $\sim$30 s for GRBs observed by Fermi.
Notably, fast X-ray transients with unknown origins exhibit a duration distribution similar to that of GRBs detected by EP, with an average duration of approximately 270 s.

The isotropic energies and redshifts of EP GRBs can also be found from GCN.
Five GRBs with redshift measurements have been reported: EP240315a, EP240801a, EP241025a, EP241026a, and EP241030a \citep{Liu24,ZhengWK24,Svinkin24,Izzo24,LiRZ24}.
Their redshifts range from 1.411 to 4.859.
The isotropic energy in the 1-10,000 keV band for EP240315a is $6.4\times 10^{53}$ erg \citep{Liu24}.
The isotropic energies in the 10-10,000 keV band for EP241025a and EP241030a are $5.5\times 10^{53}$ erg and $2.8\times 10^{53}$ erg, respectively \citep{Svinkin24,Ridnaia24}.

\section{The characteristics of on-axis LGRBs}
\label{sec:kc}

%Here, we briefly describe the main characteristics of on-axis LGRBs, such as the luminosity and redshift distributions, the $T_{90}$ duration, as well as the Amati and Yonetoku relations.
Here, we provide a concise overview of the observational properties of LGRBs, which are assumed to be on-axis viewed bursts, including the luminosity and redshift distributions, $T_{90}$ durations, as well as Amati and Yonetoku relations.
A detailed description of replicating these characteristics is arranged in Section \ref{sec:MC}, while the plausible probability distributions of model parameters are determined.

\subsection{The luminosity and redshift distributions of LGRBs}
\label{sec:LF}

In the luminosity evolution model, the joint distribution function of redshift ($z$) and luminosity ($L$) for LGRBs detected by {\it Swift}/BAT writes as \citep{Lan21}
\begin{equation}
\begin{aligned}
f(L,z)=\frac{1}{N_{\text {exp }}}\frac{c}{H_0} \frac{D_{L}^{2}\left(z\right)}{\sqrt{\Omega_{\Lambda}+\Omega_{\mathrm{m}}\left(1+z\right)^{3}}}\Theta(P(L, z))\\
 \times  \frac{\psi(z)}{1+z} \phi(L, z),
\end{aligned}
\label{eq:fLz}
\end{equation}
where $c$ is the speed of light, $D_{L}(z)$ is the luminosity distance at $z$, $\Theta(P)$ represents the detection efficiency, $P(L,z)$ is the peak flux of the burst, $\psi(z)$ denotes the comoving event rate of GRBs in units of $\text{Mpc}^{-3}~\text{yr}^{-1}$,
$\phi(L, z)$ is the normalized GRB luminosity function, and detailed function forms of $\Theta(P)$, $P(L,z)$, $\psi(z)$, and $\phi(L, z)$ could be found in \cite{Lan21}.
The expected number of GRBs, $N_{\text {exp}}$, is then given by
\begin{equation}
\begin{aligned}
N_{\text {exp}}= & \frac{\Delta \Omega T}{4 \pi} \int_0^{z_{\max }} \int_{\max \left[L_{\min }, L_{\lim }(z)\right]}^{L_{\max }} \Theta(P(L, z)) \frac{\psi(z)}{1+z} \\
& \times \phi(L, z) \mathrm{d} L \mathrm{~d} V(z),
\end{aligned}
\end{equation}
where $\Delta \Omega$ is the {\it Swift}/BAT field of view, $T$ is the time of activity of {\it Swift} that covers the sample,
$L_{\min}$ and $L_{\max}$ are the minimum and maximum of the luminosity, $z_\mathrm{max}$ is the maximum
redshift, $\mathrm{d} V(z)$ is the comoving volume element,
$L_{\lim}(z)$ is luminosity threshold determined by the burst redshift and BAT's flux limit,
and the detailed values and expressions of them could be found in \cite{Lan21}.

\subsection{The detected $T_{90}$ distribution of LGRBs}
\label{sec:T90}

\citet{Horvath16} collected 997 GRBs from {\it Swift}/BAT (15-150 keV) and analyzed the log $T_{90}$ distribution of these bursts in the observer frame using various fitting functions.
The two log-normal functions are applicable and are adopted in this paper.
For the LGRBs component we are interested in, the function is $N(\mu,\sigma^{2})$, with $\mu=1.70$ and $\sigma=0.48$.

\subsection{The Amati relation and Yonetoku relation}
\label{sec:Amati}

\citet{Tsvetkova17} presented a catalog of GRBs with known redshifts detected by the Konus-Wind (KW) experiment between 1997 and 2016.
They analyzed the distribution of rest-frame GRB parameters and confirmed the Amati and Yonetoku relations for LGRBs.
Further, \citet{Tsvetkova21} reported the results of a systematic study of GRBs with reliable redshift estimates detected simultaneously by the KW experiment (in waiting mode) and the {\it Swift}/BAT telescope from January 2005 to the end of 2018.
In this work, 167 weak and relatively soft GRBs were added to the sample, extending the KW GRBs with known redshifts to a total of 338.
The data from this expanded sample were used to fit the parameters of the Amati and Yonetoku relations for LGRBs using a Bayesian method. The derived power-law indices of the Amati and Yonetoku relations for these GRBs are 0.481 and 0.428, respectively.

\section{The model of GRB jets}
\label{sec:model}

A GRB jet will experience internal energy dissipation, caused by internal shocks resulting from the collision between different outflow parts \citep{Rees94,Paczynski94} or magnetic reconnection \citep{Spruit01,ZhangYang11}.
Meanwhile, the electrons in the jet can be accelerated and form a non-thermal distribution in the energy space.
The distribution of accelerated electrons is assumed to be a power-law of
$Q (\gamma_{\rm e}^{\prime},t^{\prime}) = Q_0 (t^{\prime}) (\gamma_{\rm e}^{\prime} / \gamma_{\rm m}^{\prime})^{-p}$
for $\gamma_{\rm e}^{\prime} > \gamma_{\rm m}^{\prime}$, where $Q_0$ is related to the
injection rate by $N_{\rm inj}^{\prime} = \int_{\gamma_{\rm m}^{\prime}}^{\gamma_{\rm max}^{\prime}}
Q (\gamma_{\rm e}^{\prime},t^{\prime}) d \gamma_{\rm e}^{\prime}$
\footnote{$\gamma_{\rm max}^{\prime}$ is the maximum Lorentz factor of electrons and is given by the approximation $\gamma_{\rm max}^{\prime} \simeq 10^8 \left(\frac{B^{\prime}}{1~\mathrm{G}} \right)^{-0.5}$ \citep{Huang00}.} (the superscript prime $\prime$ denotes the quantities in the comoving frame hereafter).
These electrons will lose their energy owing to the synchrotron radiation and adiabatic cooling.
Therefore, their energy distribution shows a time evolution, which is described by the continuity equation \citep{Uhm14,Geng18b,Zhang19,Gao21,Gao24} of
\begin{equation}
\frac{\partial}{\partial t^{\prime}}\left(\frac{d N_{\mathrm{e}}}{d \gamma_{\mathrm{e}}^{\prime}}\right)+\frac{\partial}{\partial \gamma_{\mathrm{e}}^{\prime}}\left[\dot{\gamma}_{\mathrm{e}, \mathrm { tot }}^{\prime}\left(\frac{d N_{\mathrm{e}}}{d \gamma_{\mathrm{e}}^{\prime}}\right)\right]=Q\left(\gamma_{\mathrm{e}}^{\prime}, t^{\prime}\right).
\end{equation}
As the GRB jet rapidly expands with a bulk Lorentz factor of $\Gamma$, the toroidal-dominated magnetic field in the comoving frame decreases as $B^{\prime}=B_{0}^{\prime} (R / R_0)^{-q}$ \citep{Uhm14}, where $R_{0}$ is the radius where the jet begins to produce photons and $R$ is the radius of the expanding shell.
The non-thermal electrons will be scattered by the magnetic field and emit photons with the synchrotron radiation power of $P_{\mathrm{syn}}^{\prime}$ \citep{Rybicki79}.
The observed emission originates from jet elements at various latitudes, assuming a power-law structured jet described by
\begin{equation}
\begin{gathered}
\frac{dL}{d\Omega}= \begin{cases}L_{\mathrm{c}},&  \theta \leq \theta_{\mathrm{c}}, \\
L_{\mathrm{c}}\left(\theta / \theta_{\mathrm{c}}\right)^{-k_{\mathrm{L}}},&  \theta_{\mathrm{c}}<\theta \leq \theta_{\mathrm{m}},\end{cases} \\
\frac{d\Gamma}{d\Omega}= \begin{cases}\Gamma_{\mathrm{c}},&  \theta \leq \theta_{\mathrm{c}}, \\
\Gamma_{\mathrm{c}}\left(\theta / \theta_{\mathrm{c}}\right)^{-k_{\Gamma}}+1,& \theta_{\mathrm{c}}<\theta \leq \theta_{\mathrm{m}},\end{cases}
\end{gathered}
\end{equation}
where $L_{\mathrm{c}}$ and $\Gamma_{\mathrm{c}}$ are the luminosity and the Lorentz factor of the inner core, respectively.
$\theta_{\mathrm{m}}$ is the maximum of the half-opening angle.
The indices $k_\mathrm{L}$ and $k_{\Gamma}$ describe the angular distribution of luminosity density $L\left(\theta\right)$ and Lorentz factor $\Gamma\left(\theta\right)$ within the jet cone.
Note that to obtain the observed spectral flux, $F_{\nu_\mathrm{obs}}$, it is necessary to sum up the emission from electrons over the equal-arrival-time surface \citep{Geng16a}:
\begin{equation}
F_{\nu_{\text {obs }}}=\frac{1+z}{4 \pi D_L^2} \int_0^{\theta_\mathrm{m}} P_{\mathrm{syn}}^{\prime}\left(\nu^{\prime}\left(\nu_{\mathrm{obs}}\right)\right) \mathcal{D}^3 \frac{\sin \theta}{2} d \theta.
\end{equation}

\section{Monte Carlo simulations}
\label{sec:MC}

\begin{figure}
	\centering\includegraphics[scale=.6]{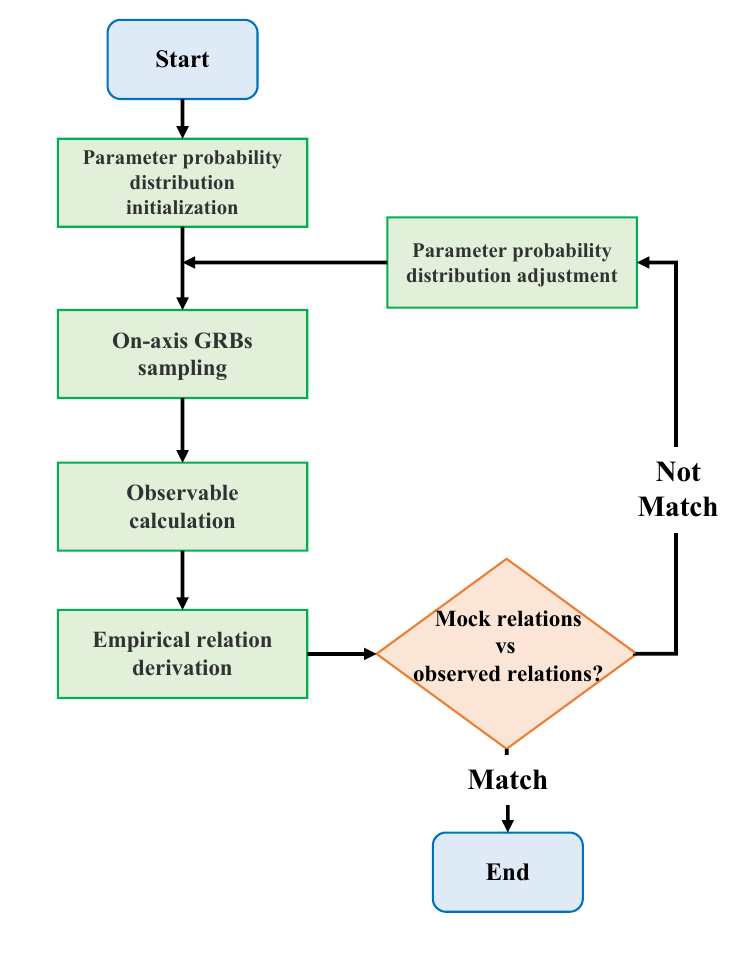}
    \caption{The flow chart for replicating statistical characteristics of on-axis LGRBs.}
    \label{fig:Flowchart}
\end{figure}

\begin{figure*}
\centerline{\includegraphics[width=1\textwidth,trim= 0 0 0 0 , clip]{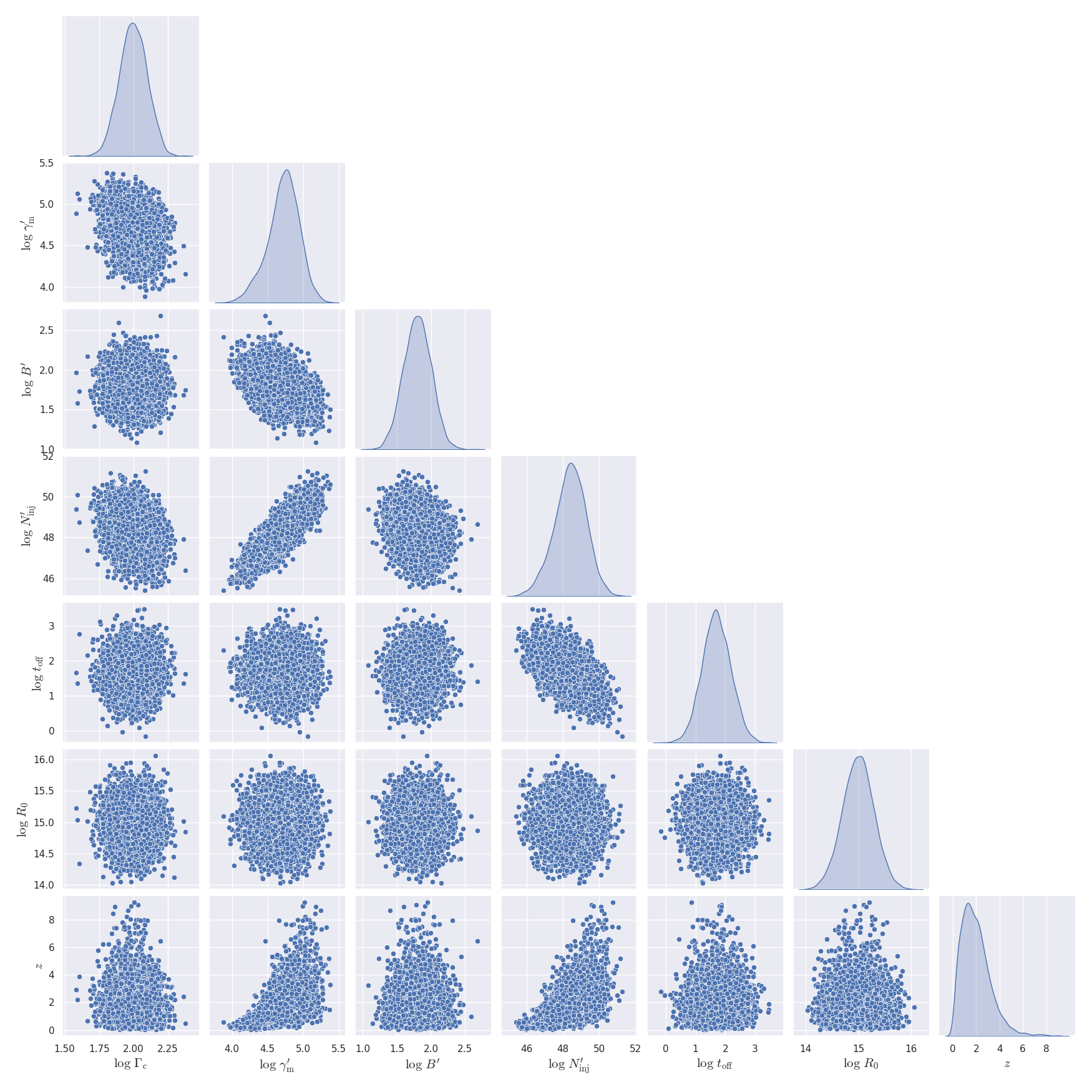}}
\caption{The optimal probability distribution of model input parameters, including \(\Gamma_\mathrm{c}\), \(\gamma_{\mathrm{m}}^{\prime}\), \(N_{\rm inj}^{\prime}\), \(B_{0}^{\prime}\), \(R_{0}\), \(t_{\mathrm{off}}\), and \(z\).}
\label{fig:Diagonalfig}
\end{figure*}

\begin{figure*}
\centerline{\includegraphics[width=1\textwidth,trim= 0 0 0 0 , clip]{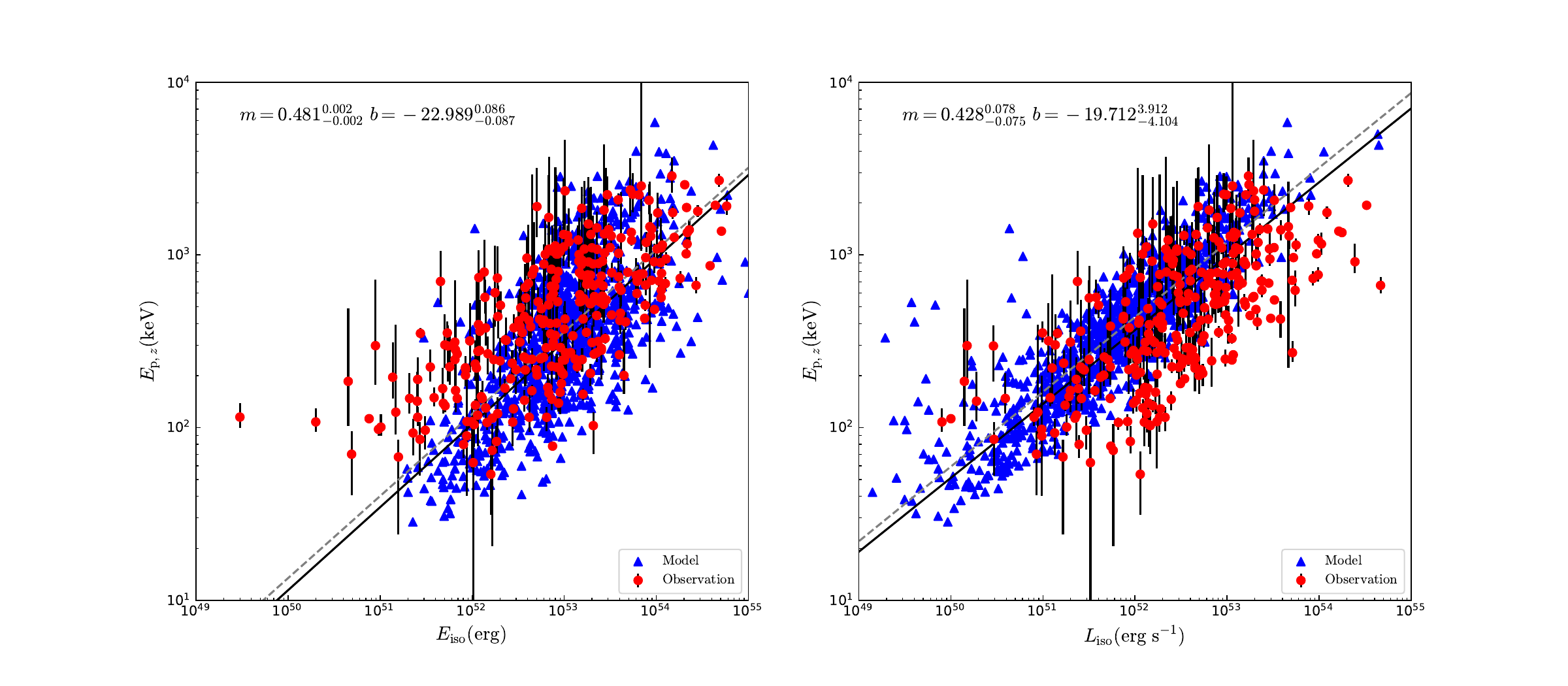}}
\caption{The distribution of mock and observed LGRBs in the $L_{\mathrm{p},\mathrm{iso}}$-$E_{\mathrm{p,z}}$ and $E_{\gamma,\mathrm{iso}}$-$E_{\mathrm{p,z}}$ planes.
The red circle dots represent observed LGRBs from \citet{Tsvetkova21}, and the blue triangular dots show our mock results for on-axis LGRBs.
The solid line and the dashed line correspond to the best-fit result for observed and mock samples, respectively.
}
\label{fig:Empirical-Relation}
\end{figure*}

\begin{figure}
	\centering\includegraphics[scale=.4]{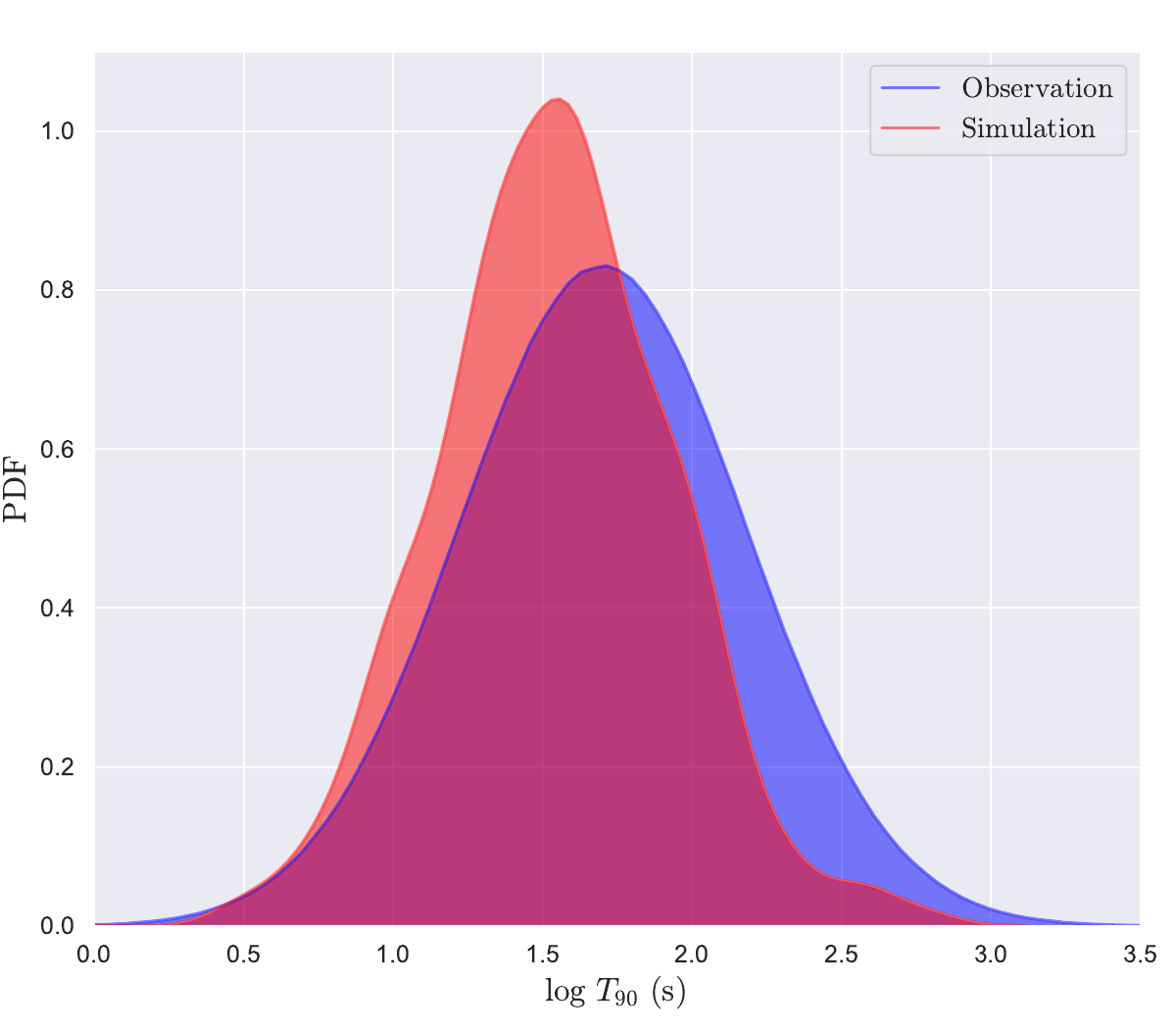}
    \caption{The probability distribution of log $T_{90}$ duration in the observer frame for both mock and observed LGRBs in the hard X-ray band.}
    The blue and red shadows represent observed LGRBs from \citet{Horvath16} and mock GRBs, respectively.
    \label{fig:T90}
\end{figure}

\begin{figure*}
\centerline{\includegraphics[width=1\textwidth,trim= 0 0 0 0 , clip]{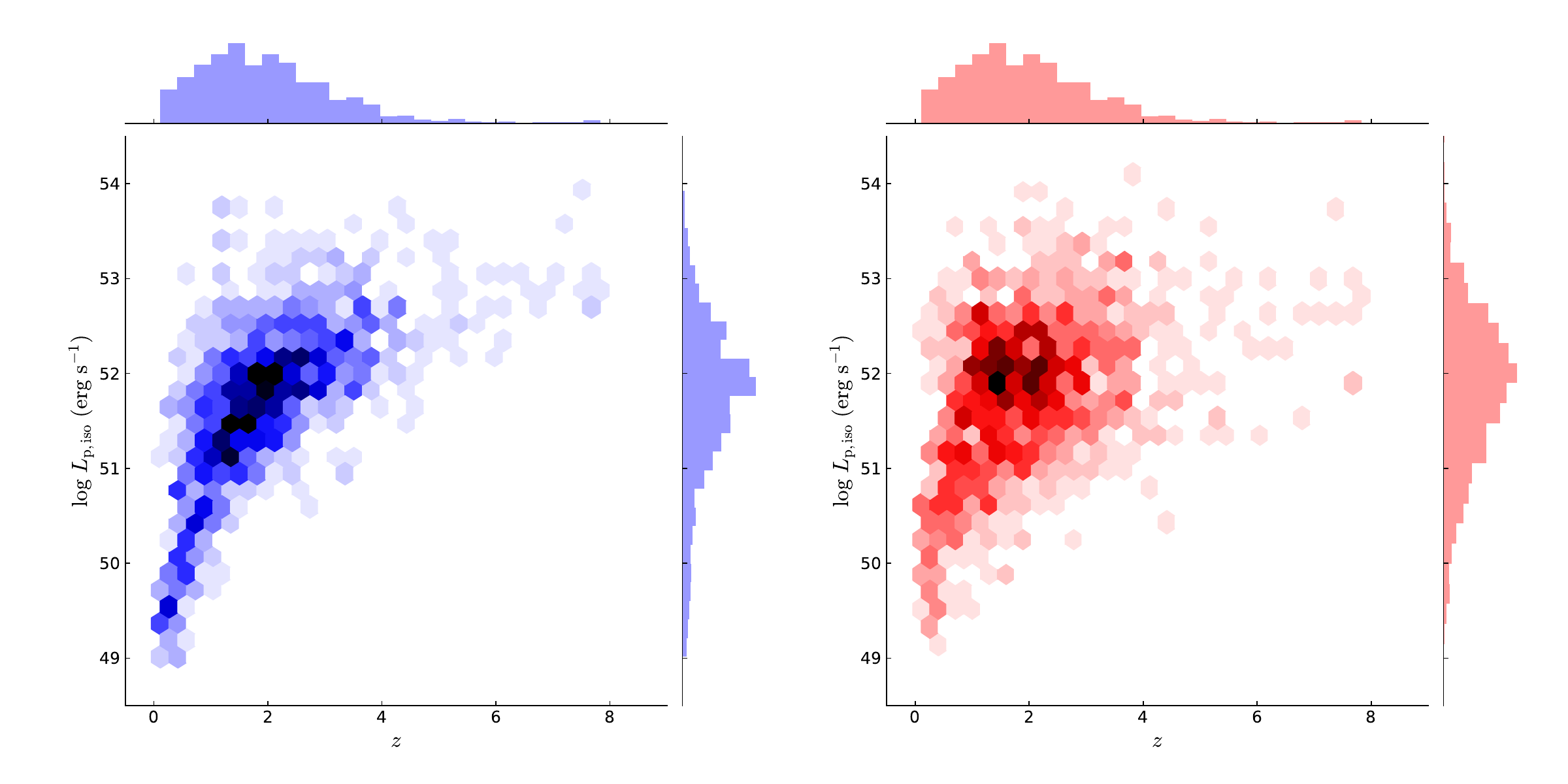}}
\caption{The left and right panels show the probability distribution of redshift and luminosity for observed GRBs and mock LGRBs, respectively.}
\label{fig:z-Liso2}
\end{figure*}

The synchrotron radiation scenario is expected to reproduce the empirical distributions and relations mentioned in Section \ref{sec:kc}.
To test this idea, Monte Carlo simulations were performed.
These simulations involve seven input parameters: \(\bm{\Gamma_\mathrm{c}}\), \(\gamma_{\mathrm{m}}^{\prime}\), \(N_{\rm inj}^{\prime}\), \(B_{0}^{\prime}\), \(R_{0}\), \(t_{\mathrm{off}}\), and \(z\).
Here, \(t_{\mathrm{off}}\) denotes the time in the observer frame at which electron injection ceases.
In this paper, we define a burst as an on-axis GRB if the viewing angle is less than the core angle of the GRB jet, and as an off-axis GRB if the viewing angle is greater than the core angle.
In our simulations, the core angle, \(\theta_\mathrm{c}\), is assumed to follow a normal Gaussian distribution of $N(2.5^{\circ},1^{\circ})$ \citep{Wang18}, the power-law index of accelerated electrons is set to \(p = 2.3\),
the parameter $q$ is fixed at $q=1$,
the $\theta_\mathrm{m}$ is set to $10\theta_\mathrm{c}$.
We assume that an on-axis observer's line of sight has an equal probability of pointing to any direction within the solid angle defined by the range of [0, $\theta_\mathrm{c}$], which can be expressed as
\begin{equation}
\frac{d P}{\sin\theta_\mathrm{v} d\theta_\mathrm{v} d\phi} = \frac{1}{\int_{0}^{2 \pi}\int_{0}^{\theta_{\mathrm{c}}}\sin\theta_\mathrm{v} d\theta_\mathrm{v} d\phi},
\label{eq:oa}
\end{equation}
where $\theta_\mathrm{v}$ and $\phi$ represent the viewing angle and azimuthal angle, respectively, measured relative to the jet central axis.
In this section, the parameters $k_\mathrm{L}$ and $k_{\Gamma}$ are fixed at 2 to replicate statistical characteristics of on-axis LGRBs.
The assumptions for these relevant parameters are summarized in Table \ref{Table:rp}.

\begin{table*}
\caption{Assumptions of relevant parameters in simulations for on-axis and off-axis bursts.}
\label{Table:rp}
\centering
\begin{tabular}{c c c c c c c c}
\hline\hline
Burst & $\theta_\mathrm{c}$ & $\theta_\mathrm{m}$ & $\theta_\mathrm{v}$ & $p$ & $q$    & $k_\mathrm{L}$  &$k_{\Gamma}$     \\

      &     &    &    &     &     &     & \\
\hline
On-axis  & $N(2.5^{\circ},1^{\circ})$  &$10\theta_\mathrm{c}$ & $\leq \theta_\mathrm{c}$   & 2.3     & 1  &2    &2\\
Off-axis   & $N(2.5^{\circ},1^{\circ})$  &$10\theta_\mathrm{c}$ & $>\theta_\mathrm{c}$  & 2.3     & 1  &\emph{U}(2, 4)    &\emph{U}(2, 4) \\
\hline
\end{tabular}
\tablecomments{The normal Gaussian distribution is denoted by $N(\mu, \sigma)$, while the uniform distribution is represented by $\emph{U}(a, b)$.
}
\end{table*}
%=\frac{1}{1-\cos{\theta_\mathrm{c}}}\sin{\theta_\mathrm{v}}

Given the substantial computational cost of each calculation and the potential for strong parameter degeneracies, we did not employ a rigorous Bayesian method for parameter constraint in this study.
Instead, we manually adjusted the probability distribution of physical parameters until a satisfactory visual agreement was achieved between the simulation outcomes and the observed empirical relations.
%Additionally, observational data indicate that most bursts exhibit a single emission episode, though approximately $(9-15)\%$ of GRBs display prompt emission signatures composed of two or more episodes with intervening quiescent periods \citep{Koshut95,Lazzati05,Burlon08,Bernardini13,Hu14}.
%\textbf{Additionally, approximately $(9-15)\%$ of GRBs display prompt emission signatures composed of two or more episodes with intervening quiescent periods \citep{Koshut95,Lazzati05,Burlon08,Bernardini13,Hu14,Lan18}. In other words, roughly $80\%$ of GRBs consist of a single emission episode.}
Additionally, observational data indicate that approximately $80\%$ of GRBs display prompt emission signatures composed of a single emission episode \citep{Koshut95,Lazzati05,Burlon08,Bernardini13,Hu14,Lan18}.
To streamline the simulation process, we modeled long-duration bursts as consisting of a single episode with one pulse, even though multiple pulses may be present within a single episode \citep{Lu12}.

The simulation process (also shown in Figure \ref{fig:Flowchart}) is as follows:

1. Generation of Input Parameters: Assuming that key parameters follow specific distributions, a set of seven input parameters are randomly generated.
Each set of parameters define a mock GRB.

2. Numerical Calculation: The model is then applied to numerically calculate the corresponding values of \(E_\mathrm{p}\), \(E_{\gamma,\mathrm{iso}}\), and \(L_{\mathrm{p,iso}}\) for the mock GRB.

3. Sample Creation: Steps 1 and 2 are repeated to create a large sample containing 1,000 GRBs.

4. Empirical Relations: The empirical relations or distributions for the mock GRBs are derived using the Bayesian method.

5. Probability Distribution Adjustment: The empirical relations of the mock GRBs are compared with the observed empirical relations. Based on these comparisons, the initial probability distributions of input parameters are adjusted to better match the observational relations.

These five steps are repeated iteratively until the empirical relations of the mock GRBs closely align with those determined by the observation.
Through this iterative process, plausible probability distributions of input parameters for the synchrotron emission scenario are determined.

\subsection{The generation of input parameters}
\label{sec:ParameterReduced}

%\textbf{We can roughly assume that the seven physical parameters in the model are subject to four main constraints. Specifically, the $T_{90}$ distribution of LGRBs constrains the injection time of shock-accelerated electrons. Equation (\ref{eq:fLz}) limits the redshift distribution of bursts and provides the correlation between redshift and luminosity, while Equation (\ref{eq:Ep}) links the burst luminosity to the peak energy. Therefore, if we assume that these parameters follow specific probability distributions (e.g., log-normal distribution), we can independently set the median values of only three.}

In this subsection, we constrain our model parameters using the observational properties of LGRBs outlined in Section 3.
The GRB spectrum, characterized by a Band function $N(E)$ with typical low-energy and high-energy spectral indices of $\alpha_\mathrm{s}=-1.5$ and $\beta_\mathrm{s}=-2.3$ \citep{Band93,Preece00,Kaneko06}, is also incorporated into the parameter constraint process.
We assume the seven physical parameters in the model are governed by four key observational constraints. First, the observed $T_{90}$ distribution of LGRBs constrains the injection timescale of shock-accelerated electrons. Second, Equation (\ref{eq:fLz}) governs the redshift distribution of bursts and encodes the redshift-luminosity correlation. Additionally, Equation (\ref{eq:Ep}) links the burst luminosity to the spectral peak energy. These constraints collectively imply that only three parameters can be independently assigned median values under our assumption of log-normal probability distributions for the parameters.

To streamline parameter constraints, we adopt the following strategy:

1. The redshift $z$ is sampled directly from the probability distribution in Equation (\ref{eq:fLz}).

2. The parameter $t_{\text{off}}$, which approximately represents the duration of the burst, is sampled from observed $T_{90}$ distribution of LGRBs (Subsection \ref{sec:T90}).

3. The bulk Lorentz factor ($\Gamma_\mathrm{c}$), magnetic field strength ($B_{0}^{\prime}$), and jet radius ($R_{0}$) are assumed to follow log-normal distributions, with their logarithmic values drawn from $N(\mu, \sigma^{2})$.

4. The remaining parameters, the injected rate of electrons $N_\text{inj}^{\prime}$ and the minimum electron Lorentz factor $\gamma_\mathrm{m}^{\prime}$, are derived analytically using the Yonetoku correlation rather than assuming log-normal priors.
The mock GRB distribution is then calibrated by iteratively adjusting the medians and variances of $\Gamma_\mathrm{c}$, $B_{0}^{\prime}$, and $R_0$ until consistency with observations is achieved. Below, we detail the derivation of $\gamma_\mathrm{m}^{\prime}$ and $N_\text{inj}^{\prime}$ from the other parameters.

The parameters $\gamma_\mathrm{m}^{\prime}$ and $N_\text{inj}^{\prime}$ are connected to the peak energy and the spectral morphology.
The peak energy $E_{\mathrm{p}}$ can always be estimated using the empirical Yonetoku correlation of
\begin{equation}
E_{\mathrm{p}}(1+z)/\mathrm{keV}=C(L_{\mathrm{p},\mathrm{iso}}/\mathrm{erg})^{0.428},
\label{eq:Ep}
\end{equation}
where the constant of $C$ equals $10^{-19.712}$, which is determined by fitting observation data of \citet{Tsvetkova21}. The redshift $z$ and isotropic peak luminosity $L_{\mathrm{p},\mathrm{iso}}$ are generated via Monte Carlo sampling based on Equation (\ref{eq:fLz}).
Then the minimum electron Lorentz factor $\gamma_\mathrm{m}^{\prime}$, governed by synchrotron radiation physics and Doppler boosting, is derived from \citep{Geng18b}
\begin{equation}
\gamma_{\mathrm{m}}^{\prime}=\sqrt{\frac{4 \pi m_{\mathrm{e}} c(1+z)E_{\text {p}}}{3 h q_{\mathrm{e}} B_{0}^{\prime} \Gamma_\mathrm{c}}},
\label{eq:gm}
\end{equation}
where $h$, $m_\mathrm{e}$, and $q_\mathrm{e}$ denote the Planck constant, electron rest mass, and electron charge.
%The synchrotron emission spectrum adopts a broken power-law form, empirically modeled by the Band function \citep{Band93}.
%We assume the low and high spectral indices $\bm{\alpha_\mathrm{s}}=-1.5$ and $\bm{\beta_\mathrm{s}}=-2.15$, consistent with synchrotron radiation in the fast-cooling regime for a particle distribution with power-law index $p=-2.3$.
The peak specific flux $F_{\nu_{\mathrm{obs}}}(\nu_\mathrm{p})$, determined by GRB spectral properties, is expressed as
\begin{equation}
F_{\nu_{\mathrm{obs}}}(\nu_\mathrm{p})=A h E_{\mathrm{p}} N(E_{\mathrm{p}}),
\label{eq:EN}
\end{equation}
where $\nu_\mathrm{p}=E_\mathrm{p}/h$ and $A=L_{\mathrm{p},\mathrm{iso}}/4 \pi D_L^2 /\int_{1 /(1+z)}^{10^4 /(1+z)} E N(E) d E$.

Then the electron injection rate can be derived from
\begin{equation}
N_{\mathrm{inj}}^{\prime}=F_{\nu_{\text {obs}}}(\nu_\mathrm{p})\frac{\sigma_{\mathrm{T}} c B_{0}^{\prime}\gamma_{\mathrm{m}}^{\prime}}{\sqrt{3}q_{\mathrm{e}}^{3}\Gamma_\mathrm{c}}\frac{2 D_L^2}{3(1+z)}.
\label{eq:Ninj2}
\end{equation}
where $\sigma_{\mathrm{T}}$ is the Thomson cross-section.

\subsection{The generation of observables}
\label{sec:go}

For each group of input parameters, the isotropic bolometric emission energy, the bolometric peak luminosity, the peak energy of the time-integrated spectrum, and the $T_{90}$ duration in the hard X-ray band are calculated.
The isotropic emission energy is defined as \citep{Bloom01}
\begin{equation}
E_{\gamma,\mathrm{iso}}=\frac{4 \pi D_{\mathrm{L}}^2 \int_{0}^{\Delta T}dt_\mathrm{obs} \int_{\nu_{1}/(1+z)}^{\nu_{2} /(1+z)} F_{\nu_{\mathrm{obs}}} d \nu_\mathrm{obs}}{1+z},
\end{equation}
where $\Delta T$ is the time during which the observed $\gamma$-ray flux surpasses the detection threshold ($\sim2\times10^{-8}\ \mathrm{erg\ cm^{-2}\ s^{-1}}$) of the {\it Swift}/BAT, with an exposure time set at 2 seconds.
The frequencies $\nu_{1}$ and $\nu_{2}$ correspond to the frequencies of the 10 keV and $10^{4}$ keV photons, respectively.
The bolometric peak luminosity $L_{\mathrm{p},\mathrm{iso}}$ is the peak value of the isotropic emission luminosity $L$ given by
\begin{equation}
L=4 \pi D_{\mathrm{L}}^2\int_{\nu_{1}/(1+z)}^{\nu_{2} /(1+z)} F_{\nu_{\mathrm{obs}}} d \nu_\mathrm{obs}.
\end{equation}
The peak energy, $E_{\mathrm{p}}$, of the time-integrated spectrum is obtained by fitting the time-integrated flux $\int_{0}^{\Delta T}F_{\nu_{\mathrm{obs}}}dt_\mathrm{obs}$ with the Band function.
To compute the $T_{90}$ duration, we calculate the photon count spectrum by using $C({\nu_\mathrm{obs}})=\frac{F_{\nu_\mathrm{obs}}}{h\nu_\mathrm{obs}}$,
where the unit of $C({\nu_\mathrm{obs}})$ is $\mathrm{cts} \cdot \mathrm{s}^{-1} \cdot \mathrm{Hz}^{-1}$.
$T_{90}$ is defined by the time interval between the epochs when 5\% and 9\% of the total fluence collected by the detector and can be expressed as
\begin{equation}
T_{90}=t_{95}-t_5.
\end{equation}

\subsection{The optimal probability distribution of input parameters}
\label{sec:op}

After numerous iterations, we identified the optimal probability distribution of input parameters, which are illustrated in Figure \ref{fig:Diagonalfig}.
The corresponding probability distributions of the mock GRBs are presented in Figures \ref{fig:Empirical-Relation}-\ref{fig:z-Liso2}.
As shown in Figure \ref{fig:Empirical-Relation}, the distribution of mock GRBs in the $L_{\mathrm{p},\mathrm{iso}}$-$E_{\mathrm{p,z}}$ and $E_{\gamma,\mathrm{iso}}$-$E_{\mathrm{p,z}}$ planes aligns well with those of the observed LGRBs.
The slopes of Amati and Yonetoku relations for mock GRBs are 0.475 and 0.434, which are close to the corresponding slope values of 0.481 and 0.428 of the observed LGRBs.
The probability distribution of the $T_{90}$ duration in the hard X-ray band for mock GRBs is exhibited in Figure \ref{fig:T90}.
The average $T_{90}$ duration for mock GRBs is about 40 s, which is close to the observed average value of $\sim$50 s for LGRBs.
A comparison of the probability distributions of redshift and luminosity between the mock LGRBs and the observed LGRBs is shown in Figure \ref{fig:z-Liso2}.
Although the median luminosity of the mock GRBs is slightly higher than that of the observed GRBs,  the distribution range of the mock GRBs in the $z$-$\mathrm{log} L_{\mathrm{p},\mathrm{iso}}$ plane closely matches that of the observed GRBs.
The median values and mean square errors of the optimal probability distribution for $\log \Gamma_\mathrm{c}$, $\log B^{\prime}$, $\log t_\mathrm{off}$, and $\log R_{0}$ are summarized in Table \ref{Table:op}.

\begin{comment}
\textbf{The parameter $\epsilon_{N}$ is set to be $\epsilon_{N}=10\times E_{\gamma,\mathrm{iso},\mathrm{es}}/E_{\gamma,\mathrm{iso}}$, and
$E_{\gamma,\mathrm{iso},\mathrm{es}}$ is estimated value by the equation of
\begin{equation}
E_{\mathrm{p}}=D\frac{{E_{\gamma,\mathrm{iso},\mathrm{es}}}^{0.481}}{1+z},
\label{eq:EpE}
\end{equation}
where the constant of $D$ equals $10^{-22.989}\ \mathrm{keV\cdot erg^{-1}}$, which is determined by fitting observation data of \citet{Tsvetkova21}.
The unit of $E_{\gamma,\mathrm{iso},\mathrm{es}}$ is $\mathrm{erg}$, and $E_{\mathrm{p}}$ is determined by Equation (\ref{eq:Ep}).
The corrected factor of $\epsilon_{\gamma}$ is adopted to be 1.1.}
\end{comment}

\begin{table*}
\caption{Parameters of the optimal probability distributions for physical variables of the jet.}
\label{Table:op}
\centering
\begin{tabular}{c c c c c c c}
\hline\hline
 & $\log\Gamma_\mathrm{c}$  & $\log (B^{\prime}/\text{G})$  & $\log (R_{0}/\text{cm})$ & $\log (t_\mathrm{off}/\text{s})$\\
\hline
$\mu$     &2.0    &1.8      &15    &1.7\\
$\sigma$  &0.1    &0.2      &0.3   &0.48\\
\hline
\end{tabular}
\end{table*}
%$\log N^{\prime}_\mathrm{inj}$ & $\log\gamma^{\prime}_\mathrm{m}$

\section{Observables in The Soft X-Ray Energy Band}
\label{sec:softXray}

With the plausible probability distribution of the physical parameters constrained from the on-axis LGRBs in Section \ref{sec:MC}, we conduct Monte Carlo simulations to investigate the observables in the soft X-ray band (0.5-4 keV) for EP LGRBs, including the $T_{90}$ duration, energy fluence, and low-energy spectral index.
EP LGRBs are defined as both on-axis and off-axis LGRBs whose radiation fluxes exceed the detection threshold of WXT ($\sim2\times10^{-9}\ \mathrm{erg\ cm^{-2}\ s^{-1}}$ for a 10-s exposure; \citealt{Huang24}).
Furthermore, we explore the dependence of observables on different jet structures, while the structure parameters $k_\mathrm{L}$ and $k_{\Gamma}$ are assumed to be in a range of [2, 4].
%In Section \ref{sec:MC}, we constraint the plausible probability distribution of the input parameters based on the statistical characteristics of currently observed on-axis LGRBs.
%In this section, Monte Carlo simulations were conducted to investigate several observables in the soft X-ray band (0.5-4 keV) for LGRBs, including the $T_{90}$ duration, energy fluence, and low-energy spectral index.
%The Amati and Yonetoku relations and the detection rate for EP LGRBs are also discussed,
%\textbf{while EP LGRBs are defined as both on-axis and off-axis LGRBs whose radiation fluxes exceed the detection threshold of WXT.}
%Furthermore, we explore the dependence of observables on different jet structures, while the structure parameters $k_\mathrm{L}$ and $k_{\Gamma}$ are assumed to be in a range of [2, 4].

\begin{comment}
The observer's line of sight is uniformly distributed over a solid angle of $4\pi$ steradians.
Assuming that the jet axis direction corresponds to $\theta_\mathrm{v}=0$ (where $\theta_\mathrm{v}$ is the viewing angle measured from the jet axis), the probability distribution of $\theta_\mathrm{v}$ follows
\begin{equation}
f(\theta_\mathrm{v})=\frac{1}{2}\sin{\theta_\mathrm{v}}, \quad \theta_\mathrm{v} \in [0, \pi].
\end{equation}
\end{comment}
In order to generate both on-axis and off-axis bursts, we assume that the observer's line of sight is uniformly distributed over a solid angle of $4\pi$ steradians, which can be expressed as
\begin{equation}
\frac{d P}{\sin\theta_\mathrm{v} d\theta_\mathrm{v} d\phi} = \frac{1}{4\pi}.
\end{equation}
However, when the viewing angle is too large, detecting the burst becomes challenging.
For a uniform jet with sharp edges, the ratio of off-axis to on-axis $E_{\gamma,\mathrm{iso}}$ is given by \citep{Granot17}
\begin{equation}
\frac{E_{\gamma,\mathrm{iso}}\left(\theta_{\mathrm{v}}\right)}{E_{\gamma,\mathrm{iso}}(0)} \approx  \frac{\left(\Gamma \theta_\mathrm{c}\right)^2}{\left(\Gamma \theta_{\mathrm{v}}\right)^6},\  \theta_{\text{v}}>2 \theta_\mathrm{c}.
\end{equation}
It is seen that a bright burst with $E_{\gamma,\mathrm{iso}}(0) \approx 10^{54}$ erg, its equivalent energy to an off-axis viewer would abruptly decrease, e.g., $E_{\gamma,\mathrm{iso}}(5\theta_\mathrm{c})$ becomes less than $10^{48}$ erg with typical values of $\Gamma$ ($\sim 100$) and $\theta_\mathrm{c}$ $\simeq$ several degrees, which is approaching the detection threshold of detectors.
Thus, in order to improve the simulation efficiency, we restrict our calculation within the range of $\theta_\mathrm{v}$ to [$0$, $10\theta_{\mathrm{c}}$].

Based on the assumptions outlined above, we conduct simulations of mock EP GRBs.
The probability distribution of observables for mock EP GRBs are discussed in Subsection \ref{sec:Dae}-\ref{sec:AY}.
Our simulations reveal that off-axis events constitute a significant fraction of the EP GRB population, and more details could be found in Subsection \ref{sec:count}.
This finding strongly suggests that off-axis LGRBs are likely the predominant source of the enigmatic long-duration bursts observed by EP.

%\textbf{Such extreme off-axis events ($\theta_\mathrm{v}\geq5\theta_{c}$) would appear as low-luminosity GRBs to the detector.}
%\textbf{To account for detectability limits, our simulations restrict $\theta_{v}$ to [$0$, $5\theta_\mathrm{c}$].}
%For $\theta_\mathrm{c}=2.5^{\circ}$ (maximum $\theta_\mathrm{v}=12.5^{\circ}$),
%\textbf{almost all of the bursts detected by the detector will be off-axis bursts if the detection threshold is ignored.
%However, simulations show that off-axis bursts comprise only $\sim$90\% of EP GRBs.
%This discrepancy arises because many off-axis bursts fall below EP's detection threshold of approximately $2\times10^{-9}\ \mathrm{erg\ cm^{-2}\ s^{-1}}$ for a 10-second exposure \citep{Huang24}.
%Our results suggest that off-axis LGRBs likely dominate the population of EP’s enigmatic long-duration bursts.}

%\textbf{The probability distribution of $T_{90}$ duration in the soft X-ray band for mock EP GRBs is shown in Figure \ref{fig:T90off}.}
\subsection{Duration and energy fluence}
\label{sec:Dae}
\begin{figure*}
\centering
\begin{adjustwidth}{-0.5cm}{-1.0cm}
    \subfloat{\includegraphics[width=10.cm,height=10.cm]{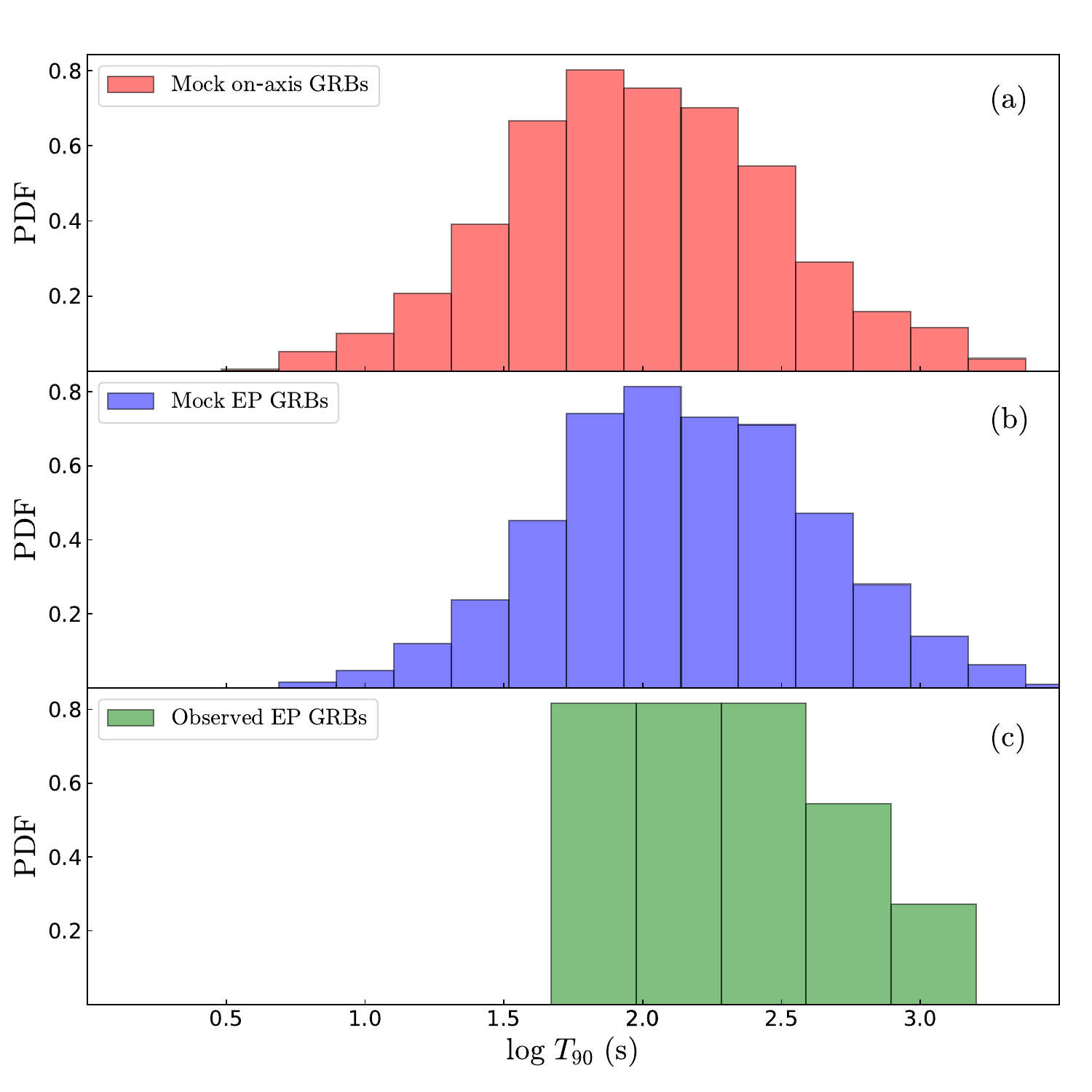}}%[width=9.1cm,height=10.6cm]{f6-1.pdf}}%[scale=0.35]
    \subfloat{\includegraphics[width=10.5cm,height=10.5cm]{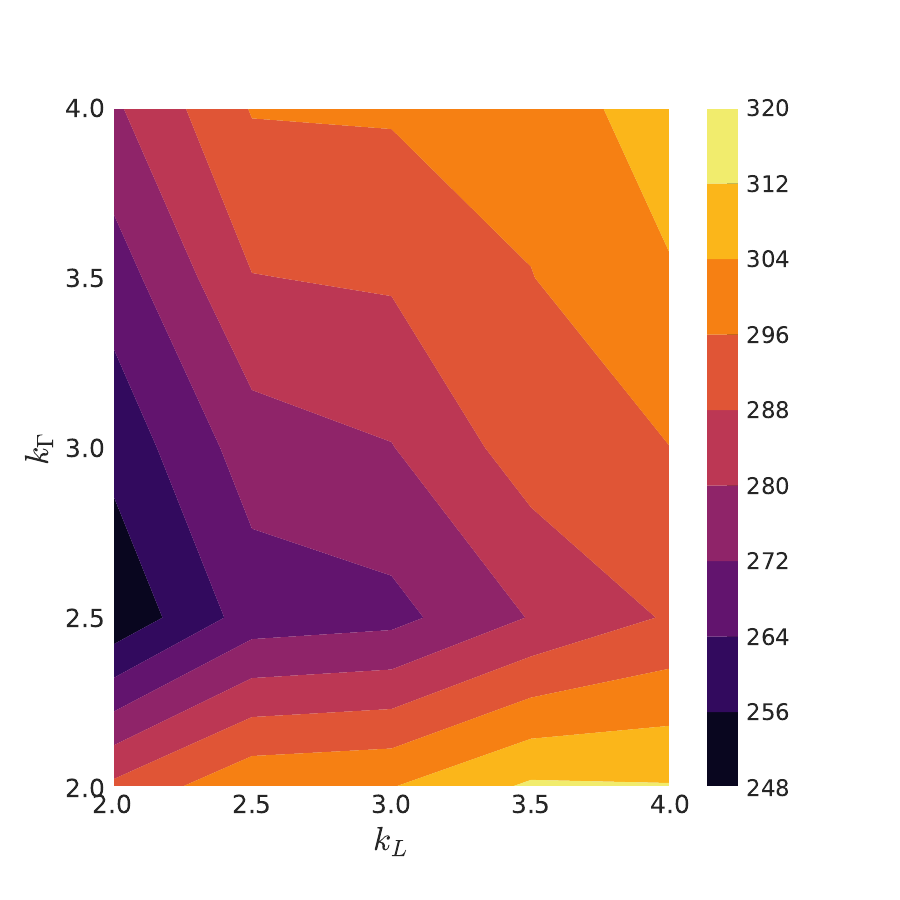}}
\end{adjustwidth}
\caption{
The $T_{90}$ duration of EP GRBs predicted by Monte Carlo simulations in the soft X-ray band.
Left panel: Panel (a) and (b) displays the probability distribution of log $T_{90}$ duration of mock on-axis LGRBs and mock EP LGRBs, with the indices $k_{L}$ and $k_{\Gamma}$ fixed at 2.
Panel (c) presents the log $T_{90}$ duration distribution of 11 EP GRBs within the same energy band.
Right panel: The dependence of the average $T_{90}$ duration for mock LGRBs of EP on the jet structures.}
\label{fig:T90off}
\end{figure*}

The probability distribution of $T_{90}$ duration in the soft X-ray band for mock EP GRBs is illustrated in Figure \ref{fig:T90off}.
The distributions of mock on-axis GRBs and observed EP GRBs are also incorporated for comparison.
For on-axis bursts, as shock-accelerated electrons cool down, the characteristic frequency of their synchrotron radiation will cross the hard X-ray band and then the lower energy bands.
The hard X-ray emission reaches peak flux before the soft X-ray emission and attenuates sooner.
This results in a longer average $T_{90}$ duration in the soft X-ray energy band compared to the hard X-ray band.
As a result, the average $T_{90}$ duration of on-axis GRBs falls within the ranges of $\sim$[180 s, 190 s], for different power indices associated with the jet structures, as illustrated in Table \ref{Table:ave}.
Its average $T_{90}$ duration is lower than that ($\sim$280 s) of GRBs observed by EP.

%\textbf{Assuming the uniform jet has a negligible half-opening angle, the observed burst duration $\Delta t_\mathrm{obs}$ relates to the comoving frame duration $\Delta t^{\prime}$ through relativistic time dilation and Doppler effects, which can be expressed by}
For off-axis bursts, the observed burst duration $\Delta t_\mathrm{obs}$ related to the comoving frame duration $\Delta t^{\prime}$ can be expressed by
\begin{equation}
\Delta t_\mathrm{obs}\approx(1+z)\Gamma(1-\beta \cos \theta)\Delta t^{\prime},
\label{eq:td}
\end{equation}
where $\theta$ is the angle between the observer's line of sight and the bulk velocity vectors of the uniform jet with a negligible half-opening angle.
However, in realistic scenarios where the jet has a finite half-opening angle $\theta_\mathrm{c}$, the angle $\theta$ in Equation (\ref{eq:td}) must be replaced by a weighted average angle.
This angle accounts for contributions from all jet elements within the opening angle and is defined by
\begin{equation}
\bar{\theta}=\frac{\int_{0}^{2 \pi}\int_{0}^{\theta_{\mathrm{c}}}\arccos\left(\sin\theta_\mathrm{v}\sin\theta\cos\varphi+\cos\theta_\mathrm{v}\cos\theta\right)\sin \theta d \theta d\varphi}{ \int_{0}^{\theta_{\mathrm{c}}} \sin \theta d \theta\int_{0}^{2 \pi}d\varphi},
\end{equation}
where $\varphi$ denotes the azimuthal angle of the jet element, measured relative to the jet central axis.
Assuming a typical angle of $\theta_\mathrm{c}=2.5^{\circ}$, we calculate the Doppler-weighted average angles for different observational scenarios.
For the on-axis case ($\theta_{\mathrm{v}}=0^{\circ}$), we find $\bar{\theta}_{\mathrm{on}}\approx1.6^{\circ}$, while the off-axis case ($\theta_{\mathrm{v}}=5^{\circ}$) yields $\bar{\theta}_{\mathrm{off}}\approx5.2^{\circ}$.
Using Equation (\ref{eq:td}) with a typical Lorentz factor of $\Gamma \approx 100$, the ratio of observed durations becomes
\begin{equation}
\frac{t_\mathrm{obs}(\bar{\theta}_{\mathrm{off}})}{t_\mathrm{obs}(\bar{\theta}_{\mathrm{on}})}=\frac{1-\beta \cos \bar{\theta}_{\mathrm{off}}}{1-\beta \cos \bar{\theta}_{\mathrm{on}}}\approx8.
\end{equation}
As illustrated in Figure \ref{fig:T90off},
the average $T_{90}$ duration for mock EP bursts is found to lie within the ranges of $\sim$[250 s, 310 s], which has a good match with that of observed EP GRBs.
Note that the hard X-ray photons observed by the on-axis observer will appear as soft X-ray photons to the off-axis observer, due to the Doppler effect.
The duration ratio of off-axis bursts to $\it{Swift}$ bursts falls within the range of $\sim$[6, 8], which is close to the estimate provided by the analytical equation.

\begin{figure*}
\centering
\begin{adjustwidth}{-0.5cm}{-1.0cm}
    \subfloat{\includegraphics[width=10.cm,height=10.cm]{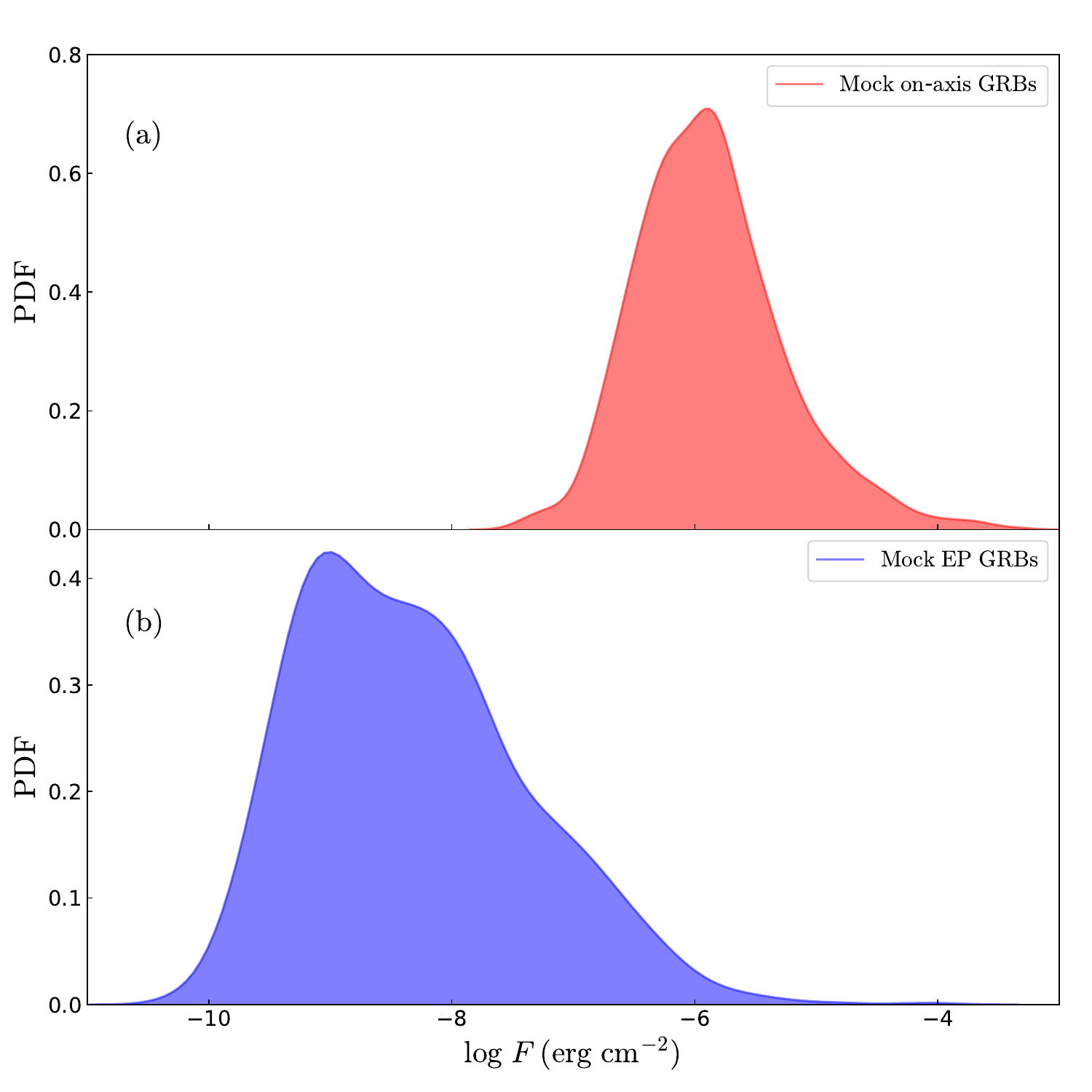}}%[width=9.1cm,height=10.6cm]{f6-1.pdf}}%[scale=0.35]
    \subfloat{\includegraphics[width=10.5cm,height=10.5cm]{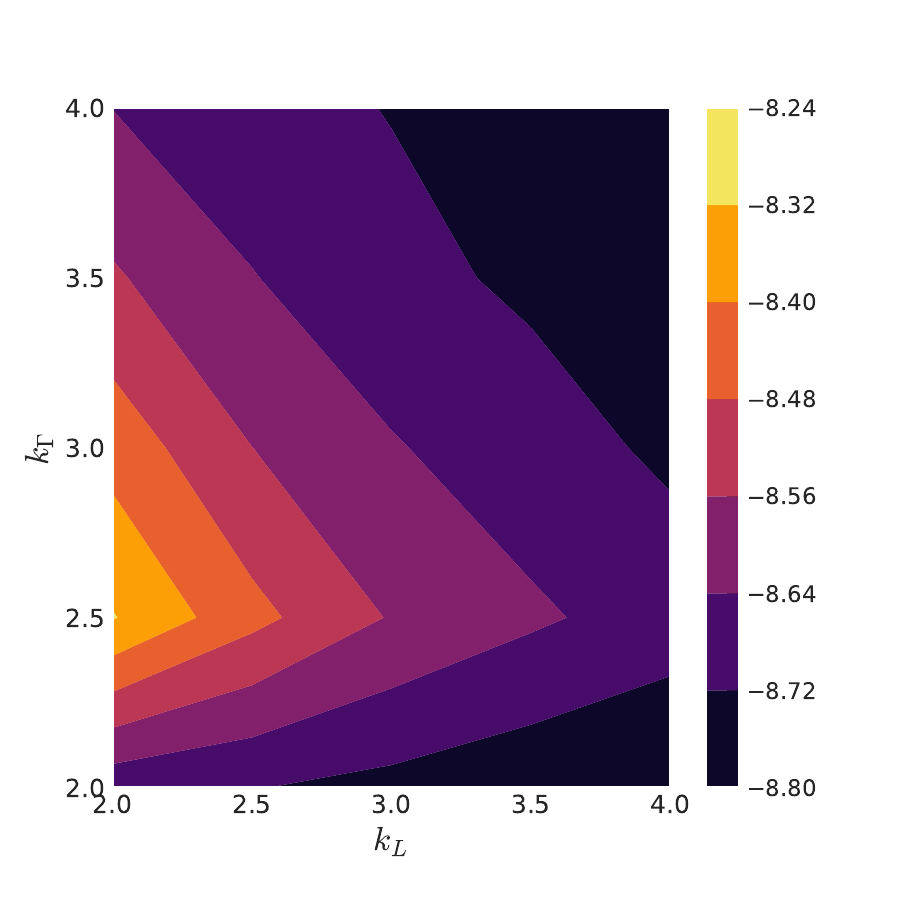}}
\end{adjustwidth}
\caption{
The energy fluence ($F$) of EP GRBs predicted by Monte Carlo simulations in the soft X-ray band.
Left panel: Panels (a) and (b) display the probability distribution of $\log F$ of mock on-axis LGRBs and mock EP LGRBs, with the indices $k_{L}$ and $k_{\Gamma}$ fixed at 2.
Right panel: The dependence of the average energy fluence for mock LGRBs of EP on the jet structures.}
\label{fig:Fluenceoff}
\end{figure*}

The possibility distribution function of log $F$ for on-axis bursts and EP bursts is presented in Figure \ref{fig:Fluenceoff}.
Owing to the beaming effect, only a small portion of photons can be detected by an off-axis observer,
resulting in a lower median energy fluence ($\sim 10^{-9}$ $\mathrm{erg\cdot cm^{-2}}$) for mock EP LGRBs compared to mock on-axis LGRBs.
Changes in the structure parameters of GRB jets have slight influence on the average energy fluence of EP LGRBs.

\subsection{Low-energy spectral index}
\label{sec:alpha}

\begin{figure*}
\centering
\begin{adjustwidth}{-0.5cm}{-1.0cm}
    \subfloat{\includegraphics[width=10.cm,height=10.cm]{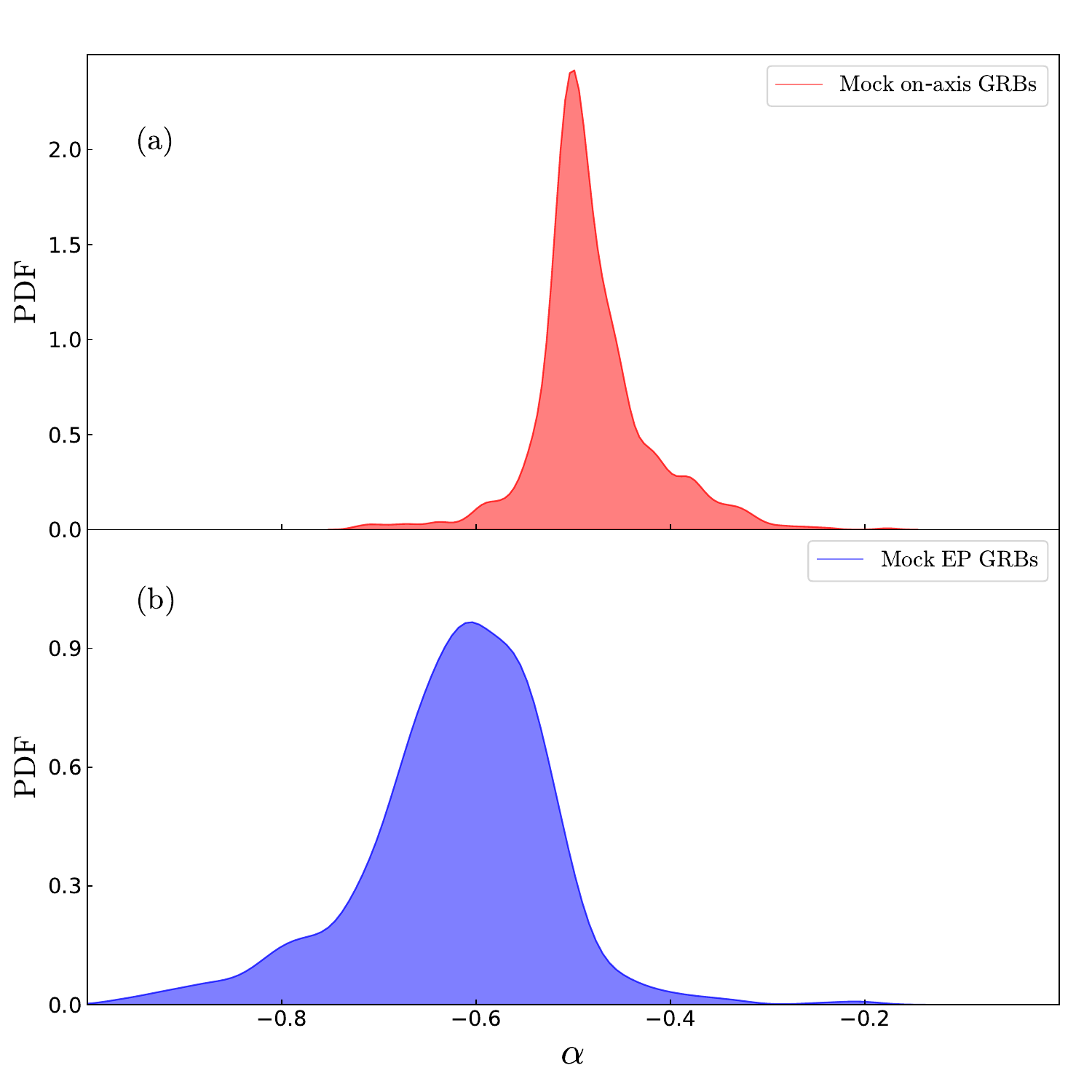}}%[width=9.1cm,height=10.6cm]{f6-1.pdf}}%[scale=0.35]
    \subfloat{\includegraphics[width=10.5cm,height=10.5cm]{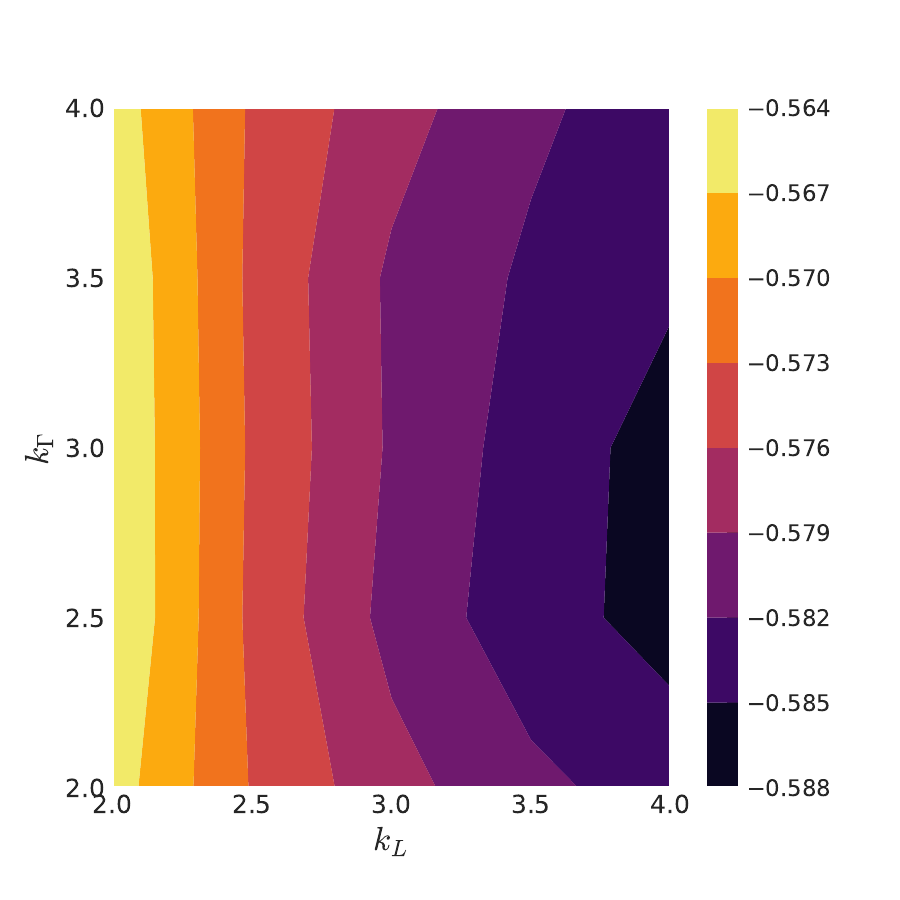}}
\end{adjustwidth}
\caption{
The low-energy spectral index $\alpha$ ($F_\nu \propto \nu^{\alpha}$) predicted by Monte Carlo simulations for EP GRBs.
Left panel: Panels (a) and (b) display the probability distribution of $\alpha$ of mock on-axis LGRBs and mock EP LGRBs, with the indices $k_{L}$ and $k_{\Gamma}$ fixed at 2.
Note that the mock probability distribution for LGRBs is scaled by a factor of 0.2.
Right panel: The dependence of the average $\alpha$ for mock LGRBs of EP on the jet structures.}
\label{fig:alphaoff}
\end{figure*}

The probability distribution of the low-energy spectral index $\alpha$ of mock EP GRBs is presented in Figure \ref{fig:alphaoff}.
The distributions for mock on-axis bursts are also shown as a comparison.
%The low-energy spectral index $\alpha$ ($F_\nu \propto \nu^{\alpha}$) is also calculated.
%As shown in Figure \ref{fig:alphaoff},
For on-axis bursts,
the $\alpha$ values are primarily distributed around $-0.5$, which is expected in the fast-cooling regime of the synchrotron scenario.
In our model,  the surrounding magnetic field of the jet decreases, and the fast-cooling electrons can have a harder energy spectrum, resulting in $\alpha$ values greater than $-0.5$ \citep{Uhm14,Geng18a}.

%the observed soft X-ray emission (of $h\nu_{\mathrm{obs}}$) corresponds to the higher comoving energy radiation (of $(1+z)h\nu_{\mathrm{obs}}/\mathcal{D}$), which has a softer energy spectral index.
%A larger a weighted average angle leads to a smaller Doppler factor of
%\begin{equation}
%\mathcal{D}\approx1/[\Gamma(1-\beta\cos\overline{\theta})].
%\end{equation}
%The photon energy in the comoving frame, which corresponds to the observed soft X-ray emission, increases and approaches that of the hard X-ray.
%Consequently, the low-energy spectral index of the observed spectrum decreases.
%As illustrated in Figure \ref{fig:alphaoff}, the average value of $\alpha$ for mock EP LGRBs is approximately -0.6, which is lower than that for mock on-axis LGRBs, and is almost independent of the structure parameters of GRB jets.
For mock EP LGRBs, the average value of $\alpha$ is approximately -0.6, which exhibits a systematic softening compared to mock on-axis LGRBs.
This discrepancy can be explained through Doppler effects.
For an off-axis observer, the increased weighted average angle compared to the on-axis configuration leads to a lower Doppler factor of $\mathcal{D}\approx1/[\Gamma(1-\beta\cos\overline{\theta})]$.
Then the observed soft X-ray emission (of $h\nu_{\mathrm{obs}}$) corresponds to the higher comoving energy radiation (of $(1+z)h\nu_{\mathrm{obs}}/\mathcal{D}$), which has a softer energy spectral index.
Consequently, off-axis observers will measure a systematically softer $\alpha$ compared to on-axis counterparts.

Notably, the distribution of the low-energy spectral index of mock GRBs may differ from that of actually observed events.
The statistics of $\alpha$ are in contention with a synchrotron origin for GRBs \citep{Preece98}.
This contention was eased by detailed treatment of the cooling processes of electrons \citep{Derishev01,Daigne11,Uhm14,ZhaoXH14,Geng18b}, such as taking into account the synchrotron self-Compton (SSC) cooling and the decreasing of the jet magnetic field.
However, the SSC cooling process is ignored in this paper in order to reduce the consumption of computing resources.
The careful treatment of the cooling processes should be performed in future investigations.

\begin{comment}
\begin{figure}
	\centering\includegraphics[scale=.35]{f6-1.pdf}
    \caption{\textbf{The $T_{90}$ duration of EP GRBs predicted by Monte Carlo simulations in the soft X-ray band.
Panel (a) and (b) displays the probability distribution of log $T_{90}$ duration of mock on-axis LGRBs and mock EP LGRBs, with the indices $k_{L}$ and $k_{\Gamma}$ fixed at 2.
Panel (c) presents the log $T_{90}$ duration distribution of 11 EP GRBs within the same energy band.}}
    \label{fig:T90off}
\end{figure}
\end{comment}

\begin{table*}
\caption{The averages of observables of EP GRBs for different jet's structures.}
\label{Table:ave}
\centering
\begin{tabular}{|c|c|c|c|c|c|c|c|c|c|c|}
\hline {Bursts}  & \multicolumn{5}{|c|}{On-axis GRBs} \\
\hline $(k_{L}, k_{\Gamma})$ & $T_{90}/\mathrm{s}$ & $\log (F/\mathrm{erg\ cm^{-2}})$ & $\alpha$ &$S_\mathrm{A}$ & $S_\mathrm{Y}$ \\%($\mathrm{erg\ cm^{-2}}$)
\hline $(2, 2)$ & 188	&-5.88	&-0.48	&0.475	&0.433
 \\
\hline $(2, 3)$ & 185	&-5.89	&-0.49	&0.479	&0.434\\
\hline $(2, 4)$ & 183	&-5.89 &-0.48	&0.479 &0.434\\
\hline $(3, 2)$ & 187	&-5.88	&-0.48	&0.478 &0.434\\
\hline $(3, 3)$ & 185	&-5.89	&-0.48	&0.479	&0.434 \\
\hline $(3, 4)$ & 184 &-5.90	&-0.48	&0.478	&0.434\\
\hline $(4, 2)$ & 186 &-5.89	&-0.48 &0.478	 &0.434\\
\hline $(4, 3)$ & 184	&-5.89 &-0.48	&0.478	&0.434\\
\hline $(4, 4)$ & 184	&-5.90	&-0.48	&0.478	&0.434\\
\hline
\end{tabular}
\tablecomments{The slopes of Amati and Yonetoku relations is denoted by $S_\mathrm{A}$ and $S_\mathrm{Y}$.
}
\end{table*}

\subsection{The Amati and Yonetoku relations for off-axis LGRBs}
\label{sec:AY}

\begin{figure*}
\centering
\begin{adjustwidth}{-1.2cm}{-1.0cm}
\centering
    \subfloat{\includegraphics[width=10.5cm,height=10.6cm]{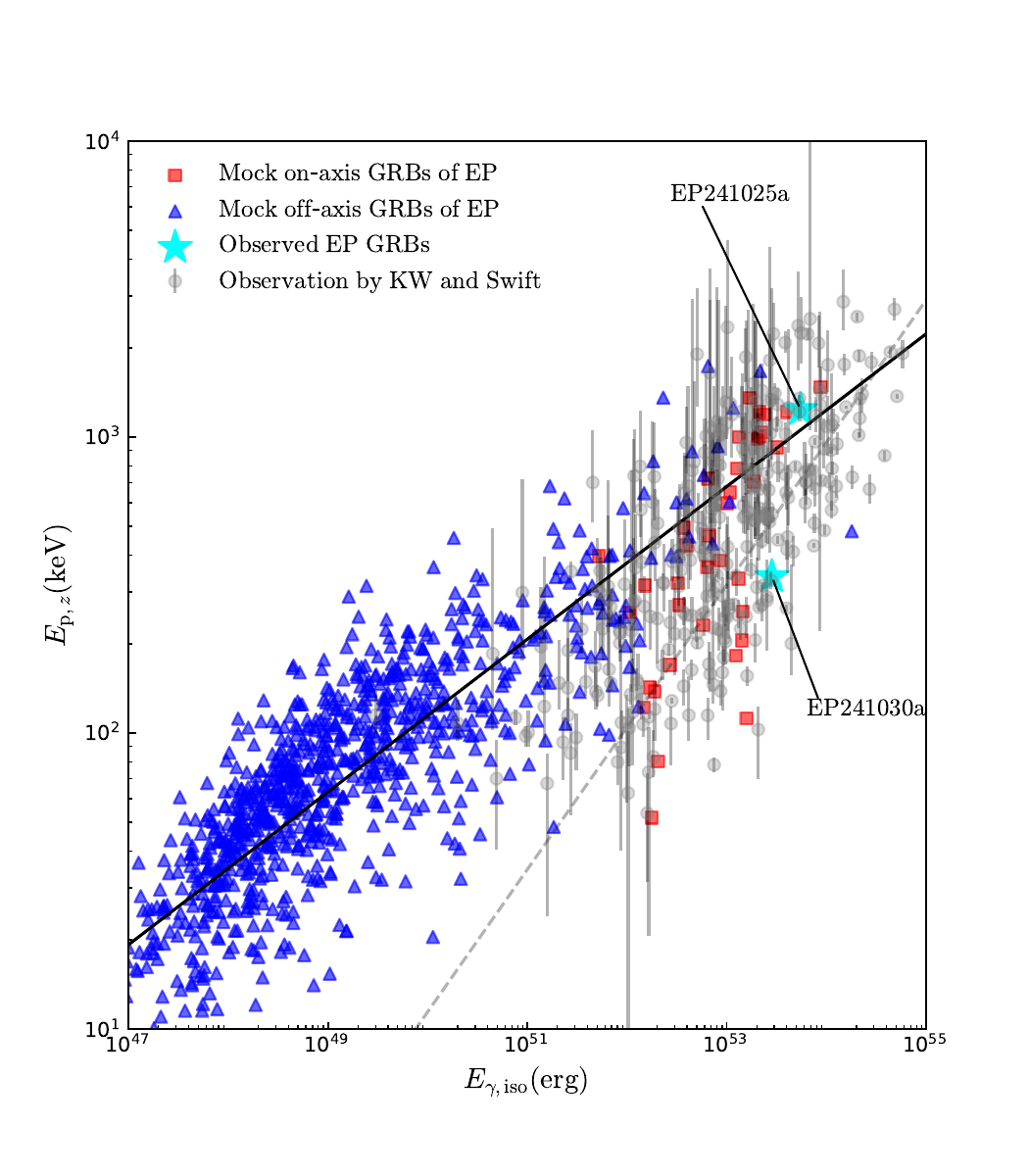}}
    \subfloat{\includegraphics[width=10.5cm,height=10.6cm]{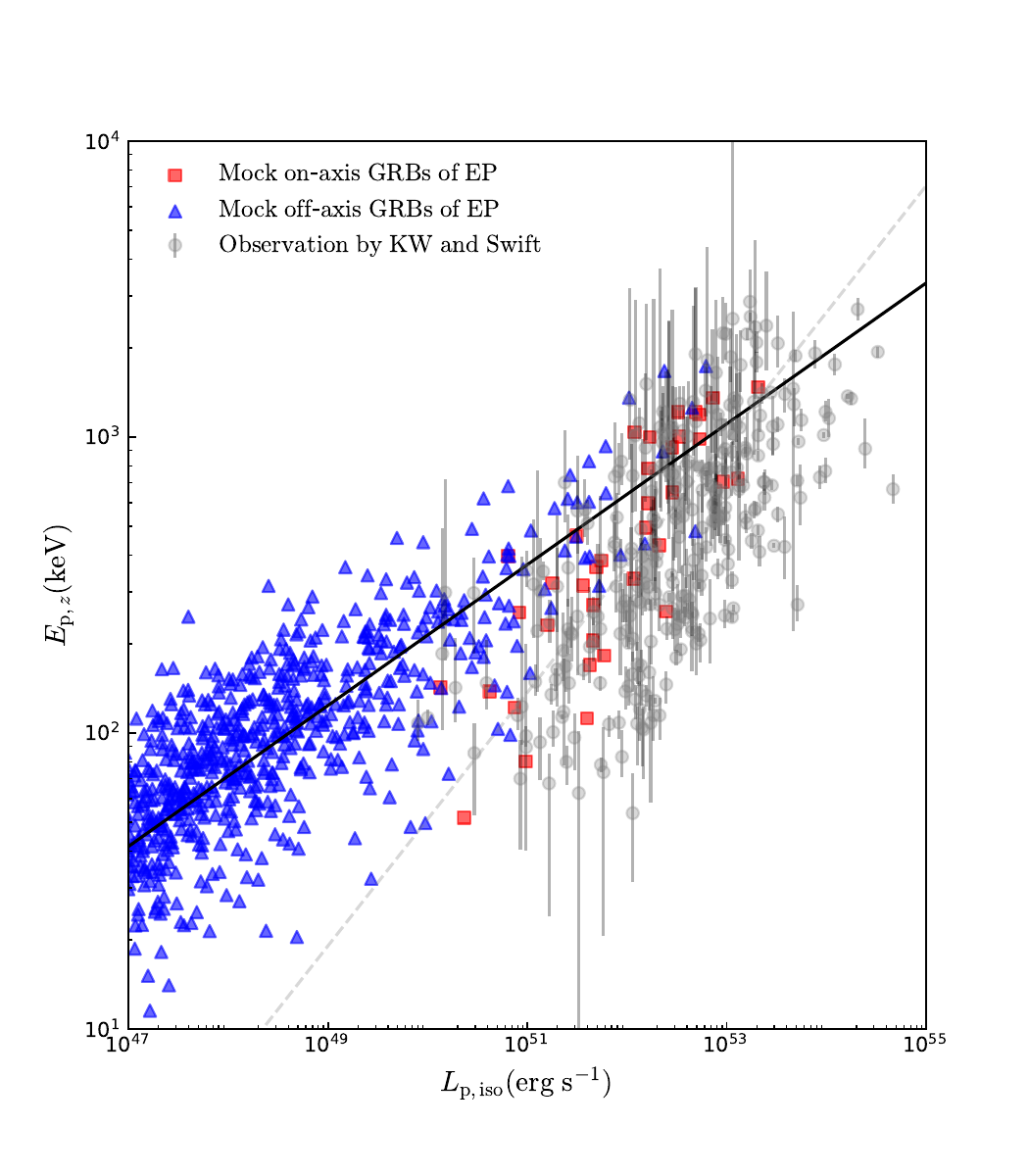}} \\
    \subfloat{\includegraphics[width=10.5cm,height=10.6cm]{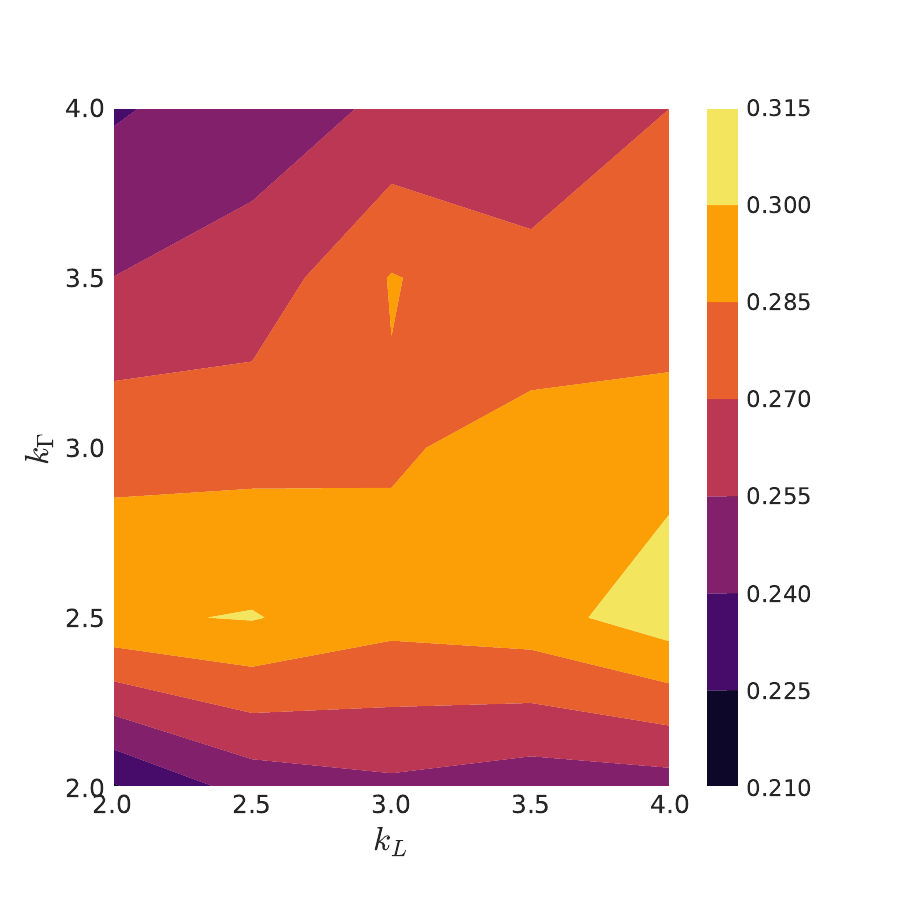}}
    \subfloat{\includegraphics[width=10.5cm,height=10.6cm]{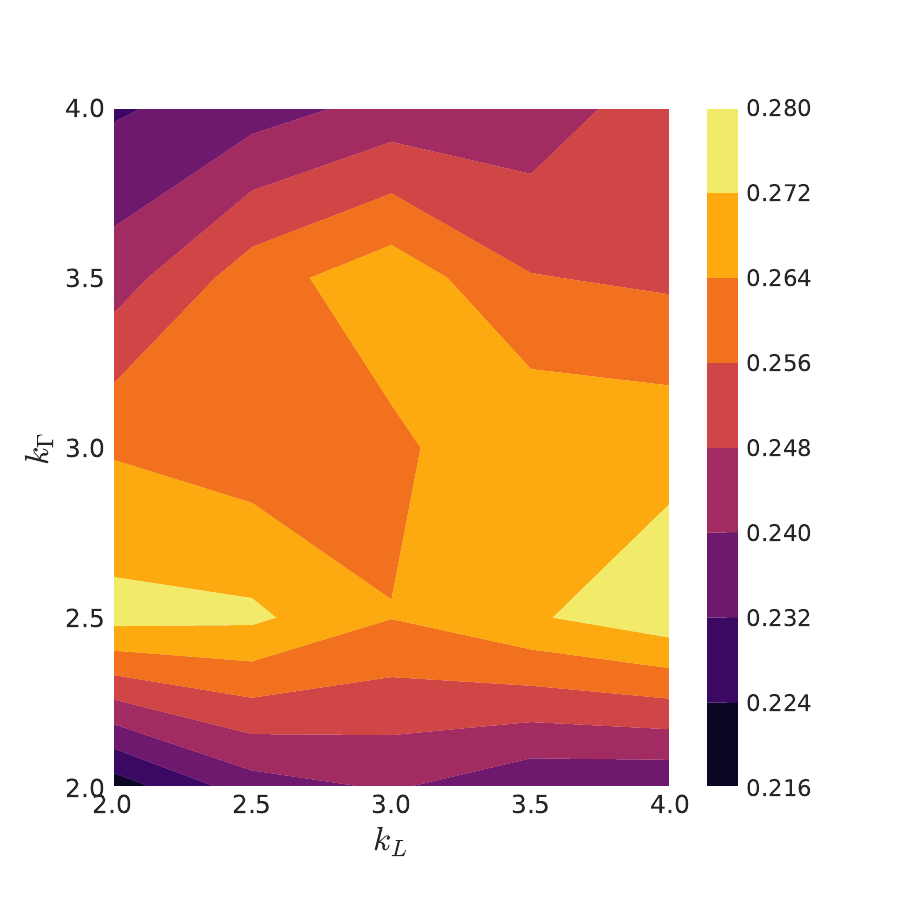}}
\end{adjustwidth}
\caption{Slopes of the Amati and Yonetoku relations predicted by Monte Carlo simulations for EP GRBs.
Upper left and right panels: The distribution of mock LGRBs of EP in the $L_{\mathrm{p},\mathrm{iso}}$-$E_{\mathrm{p,z}}$ and $E_{\gamma,\mathrm{iso}}$-$E_{\mathrm{p,z}}$ planes.
The grey circle dots represent observed LGRBs from \citet{Tsvetkova21}, and the red and blue triangular dots show our mock results for EP LGRBs.
The solid line and the dashed line correspond to the best-fit result for mock and observed samples, with the indices $k_\mathrm{L}$ and $k_{\Gamma}$ fixed at 2.
Lower left and right panels: The dependence of the slopes of the Amati (left) and Yonetoku (right) relations for EP LGRBs on the jet structures.}
\label{fig:ERoff}
\end{figure*}

The mock distribution of EP LGRBs in the $L_{\mathrm{p},\mathrm{iso}}$-$E_{\mathrm{p,z}}$ and $E_{\gamma,\mathrm{iso}}$-$E_{\mathrm{p,z}}$ planes is shown in Figure \ref{fig:ERoff}.
This figure indicates that most EP LGRBs should be low-luminosity GRBs, with $E_{\gamma,\mathrm{iso}}$ values below $10^{48}$ erg.
Only a small fraction of them fall within the $10^{49}$-$10^{55}$ erg range due to the beaming effect.
\citet{Xu23} provide a simple analytical derivation of the Amati and Yonetoku relations for both on-axis and off-axis GRBs.
Their findings show that these relations depend on the value of $\theta_\mathrm{v}-\theta_\mathrm{c}$ when the jet is viewed off-axis.
They also suggest that the slopes of both the Amati and Yonetoku relations for off-axis GRBs are lower than those for on-axis GRBs,
and a similar result can be observed in the upper left and upper right panels in Figure \ref{fig:ERoff}.
For EP bursts, the best-fit Amati relation is $E_\mathrm{p} \propto E_{\gamma,\mathrm{iso}}^{0.298}$, and the best-fit Yonetoku relation is $E_\mathrm{p} \propto L_{\mathrm{p},\mathrm{iso}}^{0.275}$, with the indices $k_\mathrm{L}$ and $k_{\Gamma}$ fixed at 2.
For various power indices associated with the jet structure, the optimal slopes of the Amati and Yonetoku relations are found to lie within the ranges of $\sim$[0.22, 0.31] and $\sim$[0.21, 0.28], respectively.
On the other hand, two EP bursts, including EP241025a and EP241030a, have been plotted in the upper left panel of Figure \ref{fig:ERoff}.
More bursts are required to determine the slope of Amati relation for observed EP GRBs.

\begin{figure*}
\centering
\begin{adjustwidth}{-1.2cm}{-1.0cm}
    \subfloat{\includegraphics[scale=.65]{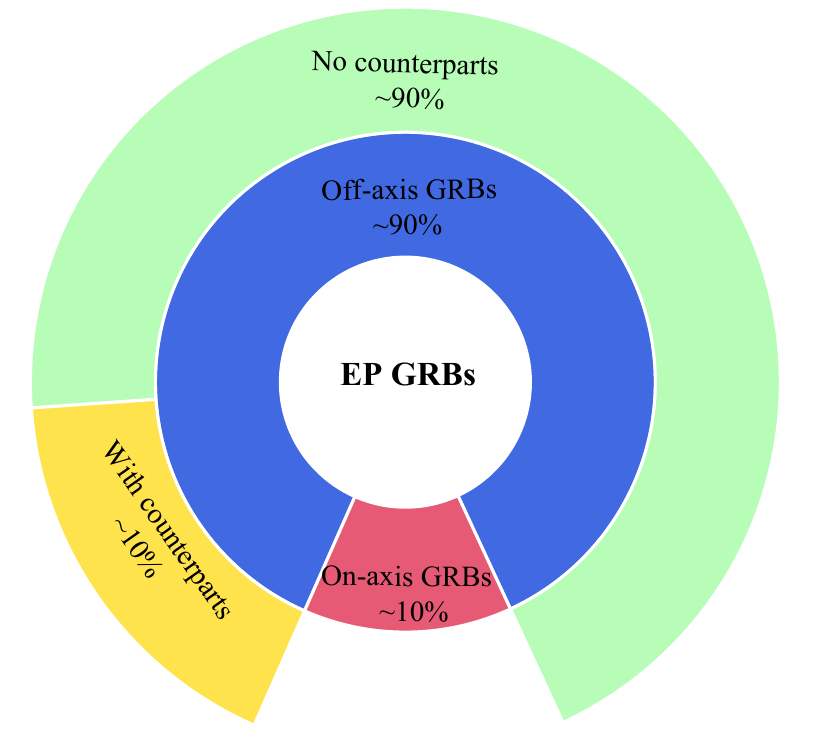}}
    \subfloat{\includegraphics[scale=.75]{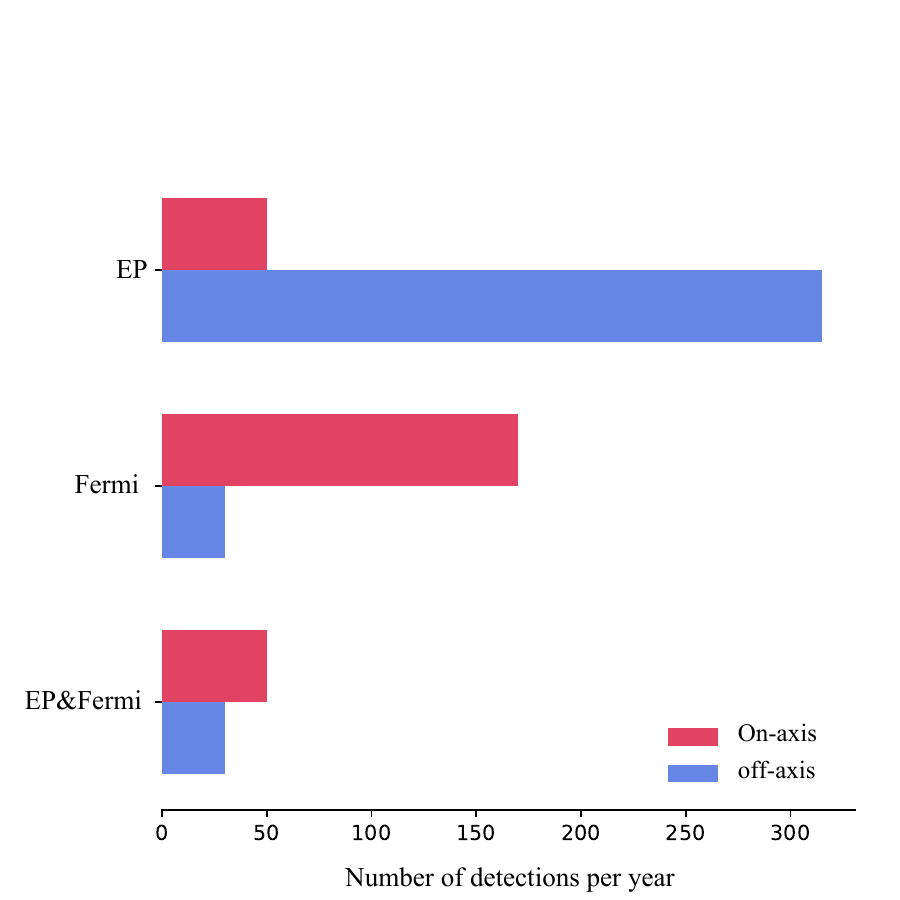}}
\end{adjustwidth}
\caption{Left panel: The estimated distribution of different components in the LGRB sample observed by EP. The inner ring shows the proportion of the off-axis LGRBs (red) and the on-axis LGRBs (blue) among the EP GRBs.
The outer ring shows the proportion of bursts with (yellow) and without (green) gamma-ray counterparts among the off-axis GRBs.
Right panel: The estimated detection rates of the on-axis (red) or off-axis GRBs (blue) for EP, $\it{Fermi}$, and the joint of EP and $\it{Fermi}$.
}
\label{fig:proportion}
\end{figure*}

\subsection{The detection rate of EP GRBs}
\label{sec:count}

We now estimate detection rate of LGRBs for EP, especailly for off-axis LGRBs.
The average number of LGRBs detected by {\it Swift} each year is approximately 70 \citep{Lan21}.
Assuming that LGRBs observed by {\it Swift} are primarily on-axis bursts, approximately 700 on-axis LGRBs and nearly 7,000 off-axis LGRBs are expected to occur across the celestial sphere each year.
If the detection sensitivity limit is not taken into account, the detector with a field of view of 3,600 $\mathrm{deg}^{2}$, similar to that of EP, will catch nearly 500 off-axis bursts and 50 on-axis bursts a year.
In practice, however, only about 60\% of off-axis bursts exceed the EP detection threshold, while the majority of on-axis bursts do.
Consequently, EP is expected to detect $\sim370$ LGRBs per year, with $\sim90\%$ of these being off-axis events, as illustrated in Figure \ref{fig:proportion}.

We also examine the number of off-axis LGRBs detected by EP that have gamma-ray counterparts.
Monte Carlo simulations indicate a $\sim10\%$ probability that an off-axis burst observed by EP would also be detectable by $\it{Fermi}$ with a $5\sigma$ detection threshold of $\sim0.5$ $\mathrm{ph\ cm^{-2}\ s^{-1}}$, and approximately 30 off-axis bursts per year are expected to be jointly detected by both EP and $\it{Fermi}$.

\begin{comment}
In the estimation, we assume an ideal operational mode for the sky survey conducted by both EP and $\it{Fermi}$.
However, in reality, the sky survey is occasionally interrupted by other observation programs, causing some GRBs to be missed by EP and $\it{Fermi}$.
Additionally, the two satellites follow different orbits and do not always observe the same regions of the sky. As a result, a burst detected by EP may not be observed by $\it{Fermi}$.
On the other hand, in the absence of the afterglow observation, it is hard to identify a burst as an off-axis GRB \citep{Alexander18,Kasliwal17,Kathirgamaraju18,Li19}.
The actual off-axis LGRB detection rate will be lower than our simulation result.

For our estimation, we use the $5\sigma$ detection threshold for $\it{Fermi}$. However, some bursts with fluxes below this threshold may still be confirmed as GRBs. Additionally, an X-ray transient missed by $\it{Fermi}$ could potentially be detected by KW and be identified as a GRB. Therefore, the probability that an off-axis burst observed by EP is confirmed as a GRB may be higher than our estimated value.
\end{comment}

%Several factors may contribute to the discrepancy between the simulated and observed detection rates.
Several factors may account for deviations between the simulated and observed detection rates.
First, an ideal operational mode are assumed for the sky survey conducted by both EP and $\it{Fermi}$. However, in reality, the sky survey is occasionally interrupted by observations for targets of opportunity, causing some GRBs to be missed by EP or $\it{Fermi}$. These additional scientific objectives may occupy a significant portion of EP's observation time.
Second, the two satellites follow different orbits and do not always observe the same regions of the sky. Consequently, a considerable fraction of bursts detected by EP may not be observed by $\it{Fermi}$.
Third, in the absence of the gamma-ray observation, it is hard to identify an X-ray transient as a GRB.
Many off-axis bursts detected by EP could be misclassified as fast X-ray transients with unknown origins.
Additionally, some bursts detected by \textit{Swift} may be off-axis, but our estimation assumes that all bursts are on-axis. This factor could lead to an overestimation of both the off-axis and on-axis bursts across the entire sky.

Given these factors, the predicted detection rate in our simulations is expected to be higher than the actual observed rate.
As mentioned in Section \ref{sec:EP GRB}, 11 fast X-ray transients detected by EP in 2024 have been confirmed as GRBs through the detection of their gamma-ray counterparts.
Our simulation results indicate that EP is expected to detect approximately 80 LGRBs per year with confirmed gamma-ray counterparts.
This predicted value exceeds the current observational count, which is consistent with our initial expectations.

%As mentioned in Section \ref{sec:alpha}, nearly 90\% of bursts detected by EP are off-axis events.
Currently, a fast X-ray transient detected by EP can be classified as a GRB if its gamma-ray counterpart has been observed by other satellites, such as $\it{Fermi}$ and KW \citep{Yin24,Liu24}.
This classification approach may lead to many off-axis bursts not being identified as GRBs.
Consequently, the proportion of off-axis bursts among EP-detected GRBs may be lower than 90\%.
As more telescopes and detectors are deployed for follow-up observations of EP-detected transients, an increasing number of these transients will be classified as GRBs through afterglow observations.

As discussed in Subsection \ref{sec:Dae}, the average $T_{90}$ duration in the soft-Xray band for off-axis LGRBs ranges from approximately 250 to 310 seconds, depending on the jet structure parameters, and is comparable to the average duration of GRBs detected by EP.
Moreover, off-axis LGRBs constitute roughly 90\% of the total LGRBs observed by EP.
These findings suggest that the GRBs detected by EP may be primarily off-axis LGRBs.
This conclusion can be further tested with a larger sample of GRB events detected by EP in the future, by comparing the predicted average value of various observables, such as duration, energy fluence, low-energy spectral index, and the slopes of the Amati and Yonetoku relations.

\section{Conclusions and Discussion}
\label{sec:conclusions}

EP has identified some GRB events, with a average duration of $\mathcal{O}(100)$ seconds, several times longer than the average duration of LGRBs observed by {\it Swift}.
It is urgent to understand the physical origin of the intriguing EP GRBs.
Based on statistical characteristics of LGRBs accumulated in the past, we constrain the parameters of GRB jets within the synchrotron emission scenario and investigate the distributions of observables in the soft X-ray band for LGRBs detected by EP, including the duration, energy fluence, and low-energy spectral index.
The Amati and Yonetoku relations for EP LGRBs are also studied.
Additionally, the dependence of observables on different jet structures is explored.
Our simulation results suggest that the new population with particularly long durations observed by EP may primarily consist of off-axis LGRBs.

Assuming a viewing angle of within $10\theta_\mathrm{c}$, we examine the observables in the soft X-ray band during the main burst for both on-axis and off-axis LGRBs.
For on-axis LGRBs, the average $T_{90}$ duration is approximately 180 s.
A larger viewing angle results in a greater average angle between the motion direction of jet elements and the observer's line of sight.
Due to the Doppler effect, off-axis LGRBs exhibit a longer $T_{90}$ duration.
As off-axis LGRBs account for $\sim$90\% of burst observed by EP,
the average duration of EP bursts falls within the ranges of $\sim$[250, 310] s for various power-law indices associated with the jet structure.
The duration ratio of LGRBs observed by EP in the soft X-ray band to on axis LGRBs observed in the gamma-ray and hard X-ray band is $\sim$ 7.

Monte Carlo simulations reveal that the average energy fluence of on-axis LGRBs is about $10^{-6}$ $\mathrm{erg\cdot cm^{-2}}$.
Due to the beaming effect, only a small portion of photons are detected by an off-axis observer, resulting in a lower average energy fluence of approximately $10^{-9}$ $\mathrm{erg\cdot cm^{-2}}$ for EP LGRBs.
We find that the structure parameters of GRB jets have little influence on the average energy fluence of EP LGRBs.

The values of low-energy spectral index, $\alpha$, for the on-axis LGRBs are primarily distributed around -0.5, which is expected in the fast-cooling regime of the synchrotron scenario.
In our scenario, the surrounding magnetic field of the jet decreases, and the fast-cooling electrons can have a harder energy spectrum, with $\alpha$ that can be greater than -0.5.
For the LGRBs detected by EP, the average value of $\alpha$ is approximately -0.6, and $\alpha$ exhibits a broad distribution range from -0.9 to 0, and is almost independent of the structure parameters of GRB jets.

The Amati and Yonetoku relations for EP LGRBs are also explored.
It is seen that the EP LGRBs form a unique population in the $E_{\gamma,\rm{iso}} - E_{\rm p,z}$ and $L_{\rm{p,iso}} - E_{\rm p,z}$ diagrams.
Due to the beaming effect, only a small fraction of EP LGRBs fall within the $10^{49}-10^{55}$ erg range.
The slopes of both the Amati and Yonetoku relations for these bursts are lower than those for on-axis LGRBs.
For different structure parameters of GRB jets, the optimal slopes of Amati and Yonetoku relations are found to lie within the ranges of $\sim$[0.22, 0.31] and $\sim$[0.21, 0.28], respectively.
As more GRB data from EP is collected, this hypothesis can be tested by directly comparing these simulated features, including the distribution of $T_{90}$ duration and energy fluence, the slopes of the Amati and Yonetoku relations, with the observational results.

Based on the average number of LGRBs detected by {\it Swift} each year, the optimistic detection rate of LGRBs by EP/WXT is approximately 370 per year, among which the off-axis LGRBs account for approximately 90\%.
We also estimate the number of off-axis LGRBs detected by EP that have gamma-ray counterparts, and approximately 30 off-axis bursts per year are expected to be jointly detected by both EP and $\it{Fermi}$.
The idealized assumption of sky survey operations in simulations may cause deviations between the simulated and observed detection rates.

In the main text, a power-law structured jet is assumed, and it is found that various jet parameters have only a minor influence on the distribution of observables for off-axis bursts. It is expected that Monte Carlo simulations based on a Gaussian jet will yield similar results to those obtained in the power-law structured jet case.

In the literature, it has been argued that there are many orphan afterglows in the sky, which could be produced by relativistic outflows whose motion is not pointed toward us \citep{Rhoads97}.
These outflows are essentially off-axis GRB jets.
They do not show up as GRBs simply because our line of sight is not on the axis.
However, their afterglow may still be observable when the outflow is significantly decelerated at late stages so that softer photons are emitted into a much wider solid angle.
Such afterglows are called orphan afterglows because they are not associated with any GRBs.
Observing these orphan afterglows may potentially help measure the beaming angle of normal GRBs \citep{Meszaros99b}, but no firm detection of these fascinating phenomena has been established yet despite extensive observational efforts \citep{Ho22}.
We argue that the X-ray transients detected by EP may be the early counterpart of orphan afterglows.
It is thus solicited that these X-ray transients be followed and monitored by large telescopes at multi-wavelengths, which may hopefully lead to the discovery of orphan afterglows.
Note that orphan afterglows may also be produced by ``failed GRBs'', i.e. isotropic baryon-contaminated fireballs with the Lorentz factor much less than 100 \citep{Huang02}.
These two kinds of orphan afterglows, i.e., off-axis ones and failed-GRB-induced events, could be discriminated by examining their decaying behaviors at late stages \citep{Huang02}.

Extragalactic fast X-ray transients (FXTs), short-duration ($\sim$ks) X-ray flashes of unknown origin, may also be linked to orphan afterglows. \citet{Wichern24} have explored the possible connection between FXTs and the afterglows of off-axis, merger-induced GRBs. Their analysis suggests that a slightly off-axis viewing angle of
$\theta_{\mathrm{obs}} \approx(2.2-3) \theta_{\mathrm{c}}$ and a structured jet are necessary to explain the shallow temporal indices ($|\alpha| \leqslant 0.3$) observed in FXT light curves, which can not be accounted for a uniform jet at any viewing angle.
Future observations of FXTs by EP will help clarify the potential connection between GRBs and FXTs and may eventually identify the progenitors of some FXTs.

\begin{comment}
\textbf{However, the long duration observed in the soft X-ray band may be attributed to the ``spectral lag'' effect commonly seen in typical GRBs.
This potential origin of the EP-detected GRBs is not excluded in our simulations.
Comprehensive multi-band follow-up observations will be crucial for determining the origin of EP GRBs and distinguishing whether they are viewed off-axis or on-axis.}
\end{comment}

The long duration in the soft X-ray band may also hint at the intrinsic characteristics of some LGRBs themselves. Except for ultra-relativistic jets launched from the GRB central engine sufficiently discussed in the literature, there is growing evidence that the mildly relativistic jets~\citep{Sun24} or weak jets~\citep{SunH24} could play significant roles. Radiation from these jets may be concentrated in the soft X-ray band and obscured by emissions from strong jets, making it difficult to identify them in the prompt and afterglow phase.
The fraction of these intrinsic long GRBs in EP GRBs remains highly uncertain and invokes further studies.

The successful launch of EP in early 2024 opened up a new window into the transient X-ray sky.
EP is expected to yield a substantial amount of observational data in the near future, including soft X-ray emissions during the main burst of GRBs, X-ray-rich GRBs that gamma-ray detectors may miss, and high-redshift GRBs.
This will help us to unveil the physical origins of GRBs and shed light on the mysteries of the early universe.

\begin{acknowledgments}

We thank the anonymous referee for constructive suggestions.
We also appreciate Bing Zhang for his valuable input and Chen-Ran Hu for the helpful discussions.
This work is based on data obtained with Einstein Probe, a space mission supported by Strategic Priority Program on Space Science of Chinese Academy of Sciences, in collaboration with ESA, MPE and CNES (Grant No. XDA15310000).
This study is partially supported by the National Natural Science Foundation of China (grant Nos. 12273113, 12321003, 12233002, 12393812, 12393813),
the National SKA Program of China (grant Nos. 2022SKA0130100, 2020SKA0120300, 2020SKA0120302),
the International Partnership Program of Chinese Academy of Sciences for Grand Challenges (114332KYSB20210018), the National Key R\&D Program of China (2021YFA0718500),
the Strategic Priority Research Program of the Chinese Academy of Sciences (grant No. XDB0550400).
Jin-Jun Geng acknowledges support from the Youth Innovation Promotion Association (2023331).
Hao-Xuan Gao acknowledges support from Jiangsu Funding Program for Excellent Postdoctoral Talent.
Yong-Feng Huang also acknowledges the support from the Xinjiang Tianchi Program.

\end{acknowledgments}

\bibliographystyle{aasjournal}
\bibliography{reference}

\begin{thebibliography}{}
\expandafter\ifx\csname natexlab\endcsname\relax\def\natexlab#1{#1}\fi
\providecommand{\url}[1]{\href{#1}{#1}}
\providecommand{\dodoi}[1]{doi:~\href{http://doi.org/#1}{\nolinkurl{#1}}}
\providecommand{\doeprint}[1]{\href{http://ascl.net/#1}{\nolinkurl{http://ascl.net/#1}}}
\providecommand{\doarXiv}[1]{\href{https://arxiv.org/abs/#1}{\nolinkurl{https://arxiv.org/abs/#1}}}

\bibitem[{{Aloy} {et~al.}(2005){Aloy}, {Janka}, \& {M{\"u}ller}}]{Aloy05}
{Aloy}, M.~A., {Janka}, H.~T., \& {M{\"u}ller}, E. 2005, \aap, 436, 273,
  \dodoi{10.1051/0004-6361:20041865}

\bibitem[{{Amati}(2006)}]{Amati06}
{Amati}, L. 2006, \mnras, 372, 233, \dodoi{10.1111/j.1365-2966.2006.10840.x}

\bibitem[{{Amati} {et~al.}(2002){Amati}, {Frontera}, {Tavani}, {in't Zand},
  {Antonelli}, {Costa}, {Feroci}, {Guidorzi}, {Heise}, {Masetti}, {Montanari},
  {Nicastro}, {Palazzi}, {Pian}, {Piro}, \& {Soffitta}}]{Amati02}
{Amati}, L., {Frontera}, F., {Tavani}, M., {et~al.} 2002, \aap, 390, 81,
  \dodoi{10.1051/0004-6361:20020722}

\bibitem[{{Band} {et~al.}(1993){Band}, {Matteson}, {Ford}, {Schaefer},
  {Palmer}, {Teegarden}, {Cline}, {Briggs}, {Paciesas}, {Pendleton}, {Fishman},
  {Kouveliotou}, {Meegan}, {Wilson}, \& {Lestrade}}]{Band93}
{Band}, D., {Matteson}, J., {Ford}, L., {et~al.} 1993, \apj, 413, 281,
  \dodoi{10.1086/172995}

\bibitem[{Band(1997)}]{Band97}
Band, D.~L. 1997, The Astrophysical Journal, 486, 928, \dodoi{10.1086/304566}

\bibitem[{{Beloborodov}(2010)}]{Beloborodov10}
{Beloborodov}, A.~M. 2010, \mnras, 407, 1033,
  \dodoi{10.1111/j.1365-2966.2010.16770.x}

\bibitem[{{Berger}(2014)}]{Berger14}
{Berger}, E. 2014, \araa, 52, 43, \dodoi{10.1146/annurev-astro-081913-035926}

\bibitem[{{Bernardini} {et~al.}(2013){Bernardini}, {Campana}, {Ghisellini},
  {D'Avanzo}, {Burlon}, {Covino}, {Ghirlanda}, {Melandri}, {Salvaterra},
  {Vergani}, {D'Elia}, {Fugazza}, {Sbarufatti}, \&
  {Tagliaferri}}]{Bernardini13}
{Bernardini}, M.~G., {Campana}, S., {Ghisellini}, G., {et~al.} 2013, \apj, 775,
  67, \dodoi{10.1088/0004-637X/775/1/67}

\bibitem[{{Blandford} \& {Znajek}(1977)}]{Blandford77}
{Blandford}, R.~D., \& {Znajek}, R.~L. 1977, \mnras, 179, 433,
  \dodoi{10.1093/mnras/179.3.433}

\bibitem[{Bloom {et~al.}(2001)Bloom, Frail, \& Sari}]{Bloom01}
Bloom, J.~S., Frail, D.~A., \& Sari, R. 2001, The Astronomical Journal, 121,
  2879, \dodoi{10.1086/321093}

\bibitem[{{Bromberg} {et~al.}(2011){Bromberg}, {Nakar}, {Piran}, \&
  {Sari}}]{Bromberg11}
{Bromberg}, O., {Nakar}, E., {Piran}, T., \& {Sari}, R. 2011, \apj, 740, 100,
  \dodoi{10.1088/0004-637X/740/2/100}

\bibitem[{{Burgess} {et~al.}(2020){Burgess}, {B{\'e}gu{\'e}}, {Greiner},
  {Giannios}, {Bacelj}, \& {Berlato}}]{Burgess20}
{Burgess}, J.~M., {B{\'e}gu{\'e}}, D., {Greiner}, J., {et~al.} 2020, Nature
  Astronomy, 4, 174, \dodoi{10.1038/s41550-019-0911-z}

\bibitem[{{Burlon} {et~al.}(2008){Burlon}, {Ghirlanda}, {Ghisellini},
  {Lazzati}, {Nava}, {Nardini}, \& {Celotti}}]{Burlon08}
{Burlon}, D., {Ghirlanda}, G., {Ghisellini}, G., {et~al.} 2008, \apjl, 685,
  L19, \dodoi{10.1086/592350}

\bibitem[{{Cheng} {et~al.}(2024){Cheng}, {Wang}, {Yuan}, {Liu}, {Ling},
  {Zhang}, {Jin}, {Chen}, {Cui}, {Fan}, {Hu}, {Hu}, {Huang}, {Li}, {Lian},
  {Liu}, {Liu}, {Lv}, {Mao}, {Pan}, {Pan}, {Sun}, {Wang}, {Wu}, {Xu}, {Xu},
  {Yang}, {Zhang}, {Zhang}, {Zhang}, {Zhang}, {Zhao}, {Kuulkers},
  {Santovincenzo}, {O'Brien}, {Nandra}, {Rau}, {Cordier}, \& {Einstein Probe
  Team}}]{Cheng24}
{Cheng}, H.~Q., {Wang}, W.~X., {Yuan}, W., {et~al.} 2024, GRB Coordinates
  Network, 36138, 1

\bibitem[{{Cheng} {et~al.}(1995){Cheng}, {Ma}, {Cheng}, {Lu}, \&
  {Zhou}}]{Cheng95}
{Cheng}, L.~X., {Ma}, Y.~Q., {Cheng}, K.~S., {Lu}, T., \& {Zhou}, Y.~Y. 1995,
  \aap, 300, 746

\bibitem[{{Dado} \& {Dar}(2012)}]{Dado12}
{Dado}, S., \& {Dar}, A. 2012, \apj, 749, 100,
  \dodoi{10.1088/0004-637X/749/2/100}

\bibitem[{{Dai} \& {Lu}(2001)}]{Dai01}
{Dai}, Z.~G., \& {Lu}, T. 2001, The Astrophysical Journal,, 551, 249,
  \dodoi{10.1086/320056}

\bibitem[{{Daigne} {et~al.}(2011){Daigne}, {Bo{\v{s}}njak}, \&
  {Dubus}}]{Daigne11}
{Daigne}, F., {Bo{\v{s}}njak}, {\v{Z}}., \& {Dubus}, G. 2011, \aap, 526, A110,
  \dodoi{10.1051/0004-6361/201015457}

\bibitem[{{Daigne} {et~al.}(2006){Daigne}, {Rossi}, \&
  {Mochkovitch}}]{Daigne06}
{Daigne}, F., {Rossi}, E.~M., \& {Mochkovitch}, R. 2006, \mnras, 372, 1034,
  \dodoi{10.1111/j.1365-2966.2006.10837.x}

\bibitem[{{Demianski} {et~al.}(2017){Demianski}, {Piedipalumbo}, {Sawant}, \&
  {Amati}}]{Demianski17}
{Demianski}, M., {Piedipalumbo}, E., {Sawant}, D., \& {Amati}, L. 2017, \aap,
  598, A112, \dodoi{10.1051/0004-6361/201628909}

\bibitem[{{Derishev} {et~al.}(2001){Derishev}, {Kocharovsky}, \&
  {Kocharovsky}}]{Derishev01}
{Derishev}, E.~V., {Kocharovsky}, V.~V., \& {Kocharovsky}, V.~V. 2001, \aap,
  372, 1071, \dodoi{10.1051/0004-6361:20010586}

\bibitem[{{Eichler} \& {Levinson}(2004)}]{Eichler04}
{Eichler}, D., \& {Levinson}, A. 2004, \apjl, 614, L13, \dodoi{10.1086/425310}

\bibitem[{{Eichler} {et~al.}(1989){Eichler}, {Livio}, {Piran}, \&
  {Schramm}}]{Eichler89}
{Eichler}, D., {Livio}, M., {Piran}, T., \& {Schramm}, D.~N. 1989, \nat, 340,
  126, \dodoi{10.1038/340126a0}

\bibitem[{{Firmani} {et~al.}(2004){Firmani}, {Avila-Reese}, {Ghisellini}, \&
  {Tutukov}}]{Firmani04}
{Firmani}, C., {Avila-Reese}, V., {Ghisellini}, G., \& {Tutukov}, A.~V. 2004,
  \apj, 611, 1033, \dodoi{10.1086/422186}

\bibitem[{{Fu} {et~al.}(2024){Fu}, {Jiang}, {Hu}, {Ling}, {Zhang}, {Liu},
  {Jin}, {Zhang}, {Cheng}, {Chen}, {Cui}, {Fan}, {Hu}, {Huang}, {Li}, {Liu},
  {Liu}, {Lv}, {Lian}, {Mao}, {Pan}, {Pan}, {Sun}, {Wang}, {Wang}, {Wu}, {Xu},
  {Xu}, {Yang}, {Yuan}, {Zhang}, {Zhang}, {Zhang}, {Zhao}, {Chen}, {Jia},
  {Cui}, {Han}, {Li}, {Song}, {Zhao}, {Zhang}, {Zhang}, {Kuulkers},
  {Santovincenzo}, {O'Brien}, {Nandra}, {Rau}, {Cordier}, \& {Einstein Probe
  Team}}]{Fu24}
{Fu}, Y.~C., {Jiang}, S.~Q., {Hu}, J.~W., {et~al.} 2024, GRB Coordinates
  Network, 37088, 1

\bibitem[{{Gao} {et~al.}(2021){Gao}, {Geng}, \& {Huang}}]{Gao21}
{Gao}, H.-X., {Geng}, J.-J., \& {Huang}, Y.-F. 2021, Astronomy \& Astrophysics,
  656, A134, \dodoi{10.1051/0004-6361/202141647}

\bibitem[{Gao {et~al.}(2024)Gao, Geng, Sun, Li, Huang, \& Wu}]{Gao24}
Gao, H.-X., Geng, J.-J., Sun, T.-R., {et~al.} 2024, The Astrophysical Journal,
  971, 81, \dodoi{10.3847/1538-4357/ad5443}

\bibitem[{{Gao} {et~al.}(2022){Gao}, {Geng}, {Hu}, {Hu}, {Lan}, {Chang},
  {Zhang}, {Zhang}, {Huang}, \& {Wu}}]{Gao22}
{Gao}, H.-X., {Geng}, J.-J., {Hu}, L., {et~al.} 2022, \mnras, 516, 453,
  \dodoi{10.1093/mnras/stac2215}

\bibitem[{{Geng} {et~al.}(2018{\natexlab{a}}){Geng}, {Huang}, {Wu}, {Song}, \&
  {Zong}}]{Geng18a}
{Geng}, J.-J., {Huang}, Y.-F., {Wu}, X.-F., {Song}, L.-M., \& {Zong}, H.-S.
  2018{\natexlab{a}}, \apj, 862, 115, \dodoi{10.3847/1538-4357/aacd05}

\bibitem[{{Geng} {et~al.}(2018{\natexlab{b}}){Geng}, {Huang}, {Wu}, {Zhang}, \&
  {Zong}}]{Geng18b}
{Geng}, J.-J., {Huang}, Y.-F., {Wu}, X.-F., {Zhang}, B., \& {Zong}, H.-S.
  2018{\natexlab{b}}, \apjs, 234, 3, \dodoi{10.3847/1538-4365/aa9e84}

\bibitem[{{Geng} {et~al.}(2016{\natexlab{a}}){Geng}, {Wu}, {Huang}, {Li}, \&
  {Dai}}]{Geng16a}
{Geng}, J.~J., {Wu}, X.~F., {Huang}, Y.~F., {Li}, L., \& {Dai}, Z.~G.
  2016{\natexlab{a}}, \apj, 825, 107, \dodoi{10.3847/0004-637X/825/2/107}

\bibitem[{{Geng} {et~al.}(2019){Geng}, {Zhang}, {K{\"o}lligan}, {Kuiper}, \&
  {Huang}}]{Geng19}
{Geng}, J.-J., {Zhang}, B., {K{\"o}lligan}, A., {Kuiper}, R., \& {Huang}, Y.-F.
  2019, \apjl, 877, L40, \dodoi{10.3847/2041-8213/ab224b}

\bibitem[{{Geng} {et~al.}(2016{\natexlab{b}}){Geng}, {Zhang}, \&
  {Kuiper}}]{Geng16b}
{Geng}, J.-J., {Zhang}, B., \& {Kuiper}, R. 2016{\natexlab{b}}, \apj, 833, 116,
  \dodoi{10.3847/1538-4357/833/1/116}

\bibitem[{{Granot} {et~al.}(2017){Granot}, {Guetta}, \& {Gill}}]{Granot17}
{Granot}, J., {Guetta}, D., \& {Gill}, R. 2017, \apjl, 850, L24,
  \dodoi{10.3847/2041-8213/aa991d}

\bibitem[{{Granot} {et~al.}(2002){Granot}, {Panaitescu}, {Kumar}, \&
  {Woosley}}]{Granot02}
{Granot}, J., {Panaitescu}, A., {Kumar}, P., \& {Woosley}, S.~E. 2002, \apjl,
  570, L61, \dodoi{10.1086/340991}

\bibitem[{{Guetta} \& {Della Valle}(2007)}]{Guetta07}
{Guetta}, D., \& {Della Valle}, M. 2007, \apjl, 657, L73,
  \dodoi{10.1086/511417}

\bibitem[{{Haggard} {et~al.}(2017){Haggard}, {Nynka}, {Ruan}, {Kalogera},
  {Cenko}, {Evans}, \& {Kennea}}]{Haggard17}
{Haggard}, D., {Nynka}, M., {Ruan}, J.~J., {et~al.} 2017, \apjl, 848, L25,
  \dodoi{10.3847/2041-8213/aa8ede}

\bibitem[{Ho {et~al.}(2022)Ho, Perley, Yao, Svinkin, de~Ugarte~Postigo, Perley,
  Kann, Burns, Andreoni, Bellm, {et~al.}}]{Ho22}
Ho, A.~Y., Perley, D.~A., Yao, Y., {et~al.} 2022, The Astrophysical Journal,
  938, 85, \dodoi{10.3847/1538-4357/ac8bd0}

\bibitem[{{Horv{\'a}th} \& {T{\'o}th}(2016)}]{Horvath16}
{Horv{\'a}th}, I., \& {T{\'o}th}, B.~G. 2016, \apss, 361, 155,
  \dodoi{10.1007/s10509-016-2748-6}

\bibitem[{{Hu} {et~al.}(2024{\natexlab{a}}){Hu}, {Zheng}, {Wang}, {Liu},
  {Cheng}, {Jin}, \& {Einstein Probe Team}}]{HuDF24}
{Hu}, D.~F., {Zheng}, T.~C., {Wang}, B.~T., {et~al.} 2024{\natexlab{a}}, GRB
  Coordinates Network, 38335, 1

\bibitem[{{Hu} {et~al.}(2024{\natexlab{b}}){Hu}, {Huang}, {Liu}, {Dai},
  {Zhang}, {Zhang}, {Liu}, \& {Einstein Probe Team}}]{HuJW24c}
{Hu}, J.~W., {Huang}, M.~Q., {Liu}, Z.~Y., {et~al.} 2024{\natexlab{b}}, GRB
  Coordinates Network, 38239, 1

\bibitem[{{Hu} {et~al.}(2024{\natexlab{c}}){Hu}, {Wang}, {He}, {Yang}, {Cheng},
  {Yuan}, \& {Einstein Probe Team}}]{HuJW24b}
{Hu}, J.~W., {Wang}, Y., {He}, H., {et~al.} 2024{\natexlab{c}}, GRB Coordinates
  Network, 37834, 1

\bibitem[{{Hu} {et~al.}(2024{\natexlab{d}}){Hu}, {Zhao}, {Liu}, {Ling},
  {Zhang}, {Jin}, {Cheng}, {Chen}, {Cui}, {Fan}, {Hu}, {Huang}, {Li}, {Lian},
  {Liu}, {Liu}, {Lv}, {Mao}, {Pan}, {Pan}, {Sun}, {Wang}, {Wang}, {Wu}, {Xu},
  {Xu}, {Yang}, {Yuan}, {Zhang}, {Zhang}, {Zhang}, {Zhang}, {Kuulkers},
  {Santovincenzo}, {O'Brien}, {Nandra}, {Rau}, {Cordier}, \& {Einstein Probe
  Team}}]{HuJW24a}
{Hu}, J.~W., {Zhao}, D.~H., {Liu}, Y., {et~al.} 2024{\natexlab{d}}, GRB
  Coordinates Network, 36053, 1

\bibitem[{{Hu} {et~al.}(2014){Hu}, {Liang}, {Xi}, {Peng}, {Lu}, {L{\"u}}, \&
  {Zhang}}]{Hu14}
{Hu}, Y.-D., {Liang}, E.-W., {Xi}, S.-Q., {et~al.} 2014, \apj, 789, 145,
  \dodoi{10.1088/0004-637X/789/2/145}

\bibitem[{{Huang} {et~al.}(2024){Huang}, {Zhang}, {Tao}, {Li}, {Zhao}, {Yin},
  {Wen}, {Xiao}, {Zhang}, {Zhang}, {Xiong}, {Bu}, {Cang}, {Cao}, {Chen},
  {Ding}, {Gao}, {Gao}, {Hou}, {Jia}, {Jin}, {Li}, {Li}, {Li}, {Li}, {Liu},
  {Ma}, {Pan}, {Qi}, {Rao}, {Sun}, {Tang}, {Tang}, {Wang}, {Xu}, {Yang},
  {Yang}, {Yang}, {Zhang}, {Zhang}, {Zhou}, {Zhao}, {Zhao}, {Zhao}, \&
  {Zhao}}]{Huang24}
{Huang}, Y., {Zhang}, J., {Tao}, L., {et~al.} 2024, Experimental Astronomy, 57,
  3, \dodoi{10.1007/s10686-024-09924-0}

\bibitem[{{Huang} {et~al.}(2002){Huang}, {Dai}, \& {Lu}}]{Huang02}
{Huang}, Y.~F., {Dai}, Z.~G., \& {Lu}, T. 2002, \mnras, 332, 735,
  \dodoi{10.1046/j.1365-8711.2002.05334.x}

\bibitem[{{Huang} {et~al.}(2000){Huang}, {Gou}, {Dai}, \& {Lu}}]{Huang00}
{Huang}, Y.~F., {Gou}, L.~J., {Dai}, Z.~G., \& {Lu}, T. 2000, \apj, 543, 90,
  \dodoi{10.1086/317076}

\bibitem[{{Ioka} \& {Nakamura}(2018)}]{Ioka18}
{Ioka}, K., \& {Nakamura}, T. 2018, Progress of Theoretical and Experimental
  Physics, 2018, 043E02, \dodoi{10.1093/ptep/pty036}

\bibitem[{{Izzo} {et~al.}(2024){Izzo}, {Kuhn}, {Rossi}, {Malesani}, {D'Elia},
  \& {Cusano}}]{Izzo24}
{Izzo}, L., {Kuhn}, O., {Rossi}, A., {et~al.} 2024, GRB Coordinates Network,
  37925, 1

\bibitem[{Kaneko {et~al.}(2006)Kaneko, Preece, Briggs, Paciesas, Meegan, \&
  Band}]{Kaneko06}
Kaneko, Y., Preece, R.~D., Briggs, M.~S., {et~al.} 2006, \apjs, 166, 298,
  \dodoi{10.1086/505911}

\bibitem[{{Kathirgamaraju} {et~al.}(2018){Kathirgamaraju}, {Barniol Duran}, \&
  {Giannios}}]{Kathirgamaraju18}
{Kathirgamaraju}, A., {Barniol Duran}, R., \& {Giannios}, D. 2018, \mnras, 473,
  L121, \dodoi{10.1093/mnrasl/slx175}

\bibitem[{{Kocevski}(2012)}]{Kocevski12}
{Kocevski}, D. 2012, \apj, 747, 146, \dodoi{10.1088/0004-637X/747/2/146}

\bibitem[{{Koshut} {et~al.}(1995){Koshut}, {Kouveliotou}, {Paciesas}, {van
  Paradijs}, {Pendleton}, {Briggs}, {Fishman}, \& {Meegan}}]{Koshut95}
{Koshut}, T.~M., {Kouveliotou}, C., {Paciesas}, W.~S., {et~al.} 1995, \apj,
  452, 145, \dodoi{10.1086/176286}

\bibitem[{{Kouveliotou} {et~al.}(1993){Kouveliotou}, {Meegan}, {Fishman},
  {Bhat}, {Briggs}, {Koshut}, {Paciesas}, \& {Pendleton}}]{Kouveliotou93}
{Kouveliotou}, C., {Meegan}, C.~A., {Fishman}, G.~J., {et~al.} 1993, \apjl,
  413, L101, \dodoi{10.1086/186969}

\bibitem[{{Kumar} \& {Granot}(2003)}]{Kumar03}
{Kumar}, P., \& {Granot}, J. 2003, \apj, 591, 1075, \dodoi{10.1086/375186}

\bibitem[{{Kumar} \& {Zhang}(2015)}]{Kumar15}
{Kumar}, P., \& {Zhang}, B. 2015, \physrep, 561, 1,
  \dodoi{10.1016/j.physrep.2014.09.008}

\bibitem[{{Lamb} \& {Kobayashi}(2017)}]{Lamb17}
{Lamb}, G.~P., \& {Kobayashi}, S. 2017, \mnras, 472, 4953,
  \dodoi{10.1093/mnras/stx2345}

\bibitem[{{Lan} {et~al.}(2021){Lan}, {Wei}, {Zeng}, {Li}, \& {Wu}}]{Lan21}
{Lan}, G.-X., {Wei}, J.-J., {Zeng}, H.-D., {Li}, Y., \& {Wu}, X.-F. 2021,
  \mnras, 508, 52, \dodoi{10.1093/mnras/stab2508}

\bibitem[{{Lan} {et~al.}(2018){Lan}, {L{\"u}}, {Zhong}, {Zhang}, {Rice},
  {Cheng}, {Du}, {Li}, {Lin}, {Lu}, \& {Liang}}]{Lan18}
{Lan}, L., {L{\"u}}, H.-J., {Zhong}, S.-Q., {et~al.} 2018, \apj, 862, 155,
  \dodoi{10.3847/1538-4357/aacda6}

\bibitem[{{Lazzati}(2005)}]{Lazzati05}
{Lazzati}, D. 2005, \mnras, 357, 722, \dodoi{10.1111/j.1365-2966.2005.08687.x}

\bibitem[{{Lazzati} {et~al.}(2018){Lazzati}, {Perna}, {Morsony},
  {Lopez-Camara}, {Cantiello}, {Ciolfi}, {Giacomazzo}, \&
  {Workman}}]{Lazzati18}
{Lazzati}, D., {Perna}, R., {Morsony}, B.~J., {et~al.} 2018, \prl, 120, 241103,
  \dodoi{10.1103/PhysRevLett.120.241103}

\bibitem[{{Levinson} \& {Eichler}(2003)}]{Levinson03}
{Levinson}, A., \& {Eichler}, D. 2003, \apjl, 594, L19, \dodoi{10.1086/378487}

\bibitem[{{Li} {et~al.}(2024{\natexlab{a}}){Li}, {Liu}, {Huang}, {Peng}, {Jin},
  \& {Eintein Probe Team}}]{Li24b}
{Li}, D.~Y., {Liu}, Z.~Y., {Huang}, M.~Q., {et~al.} 2024{\natexlab{a}}, GRB
  Coordinates Network, 37864, 1

\bibitem[{{Li} {et~al.}(2024{\natexlab{b}}){Li}, {Xu}, {Wang}, {Zhang}, {Hu},
  {Jin}, {Ling}, {Zhang}, {Liu}, {Zhang}, {Cheng}, {Chen}, {Cui}, {Fan}, {Hu},
  {Hu}, {Huang}, {Liu}, {Liu}, {Lv}, {Lian}, {Mao}, {Pan}, {Pan}, {Sun},
  {Wang}, {Wang}, {Wu}, {Xu}, {Yang}, {Yuan}, {Zhang}, {Zhang}, {Zhang},
  {Zhao}, {Chen}, {Jia}, {Cui}, {Han}, {Li}, {Song}, {Zhao}, {Zhang}, {Zhang},
  {Kuulkers}, {Santovincenzo}, {O'Brien}, {Nandra}, {Rau}, {Cordier}, \&
  {Einstein Probe Team}}]{Li24a}
{Li}, D.~Y., {Xu}, X.~P., {Wang}, B.~T., {et~al.} 2024{\natexlab{b}}, GRB
  Coordinates Network, 37492, 1

\bibitem[{{Li} {et~al.}(2019){Li}, {Geng}, {Huang}, \& {Li}}]{Li19}
{Li}, L.-B., {Geng}, J.-J., {Huang}, Y.-F., \& {Li}, B. 2019, \apj, 880, 39,
  \dodoi{10.3847/1538-4357/ab275d}

\bibitem[{{Li} {et~al.}(2024{\natexlab{c}}){Li}, {Lin}, {Li}, {Feng}, {Wang},
  {Song}, {Mao}, \& {Bai}}]{LiRZ24}
{Li}, R.~Z., {Lin}, H., {Li}, S.~S., {et~al.} 2024{\natexlab{c}}, GRB
  Coordinates Network, 38027, 1

\bibitem[{{Lian} {et~al.}(2024{\natexlab{a}}){Lian}, {Pan}, {Ling}, {Liu},
  {Zhang}, {Jin}, {\%Cheng}, {Chen}, {Cui}, {Fan}, {Hu}, {Hu}, {\%Huang}, {Li},
  {Liu}, {Liu}, {Lv}, {Mao}, {\%Pan}, {Sun}, {Wang}, {Wang}, {Wu}, {Xu},
  {\%Xu}, {Yang}, {Yuan}, {Zhang}, {Zhang}, {Zhang}, {Zhang}, {Zhao}, {Yang},
  {Dai}, {Liang}, {\%Chen}, {Jia}, {Zhang}, {Kuulkers}, {Santovincenzo},
  {\%O'Brien}, {Nandra}, {Rau}, {Cordier}, \& {Einstein Probe Team}}]{Lian24b}
{Lian}, T.~Y., {Pan}, X., {Ling}, Z.~X., {et~al.} 2024{\natexlab{a}}, GRB
  Coordinates Network, 36091, 1

\bibitem[{{Lian} {et~al.}(2024{\natexlab{b}}){Lian}, {Pan}, {Ling}, {Liu},
  {Ling}, {Zhang}, {Jin}, {Cheng}, {Chen}, {Cui}, {Fan}, {Hu}, {Hu}, {Huang},
  {Li}, {Liu}, {Liu}, {Lv}, {Mao}, {Pan}, {Sun}, {Wang}, {Wang}, {Wu}, {Xu},
  {Xu}, {Yang}, {Yuan}, {Zhang}, {Zhang}, {Zhang}, {Zhang}, {Zhao}, {Yang},
  {Dai}, {Fang}, {Chen}, {Jia}, {Zhang}, {Kuulkers}, {Santovincenzo},
  {O'Brien}, {Nandra}, {Rau}, {Cordier}, \& {Einstein Probe Team}}]{LianTY24}
---. 2024{\natexlab{b}}, GRB Coordinates Network, 36086, 1

\bibitem[{{Liang} {et~al.}(2007){Liang}, {Zhang}, {Virgili}, \&
  {Dai}}]{Liang07}
{Liang}, E., {Zhang}, B., {Virgili}, F., \& {Dai}, Z.~G. 2007, \apj, 662, 1111,
  \dodoi{10.1086/517959}

\bibitem[{{Liang} {et~al.}(2006){Liang}, {Zhang}, {O'Brien}, {Willingale},
  {Angelini}, {Burrows}, {Campana}, {Chincarini}, {Falcone}, {Gehrels}, {Goad},
  {Grupe}, {Kobayashi}, {M{\'e}sz{\'a}ros}, {Nousek}, {Osborne}, {Page}, \&
  {Tagliaferri}}]{Liang06}
{Liang}, E.~W., {Zhang}, B., {O'Brien}, P.~T., {et~al.} 2006, \apj, 646, 351,
  \dodoi{10.1086/504684}

\bibitem[{{Liang} {et~al.}(2024){Liang}, {Liu}, {Mao}, {Jin}, {Liu}, {Ling},
  {Zhang}, {Cheng}, {Chen}, {Cui}, {Fan}, {Hu}, {W.}, {Huang}, {Li}, {Liu},
  {Lv}, {Lian}, {Pan}, {Pan}, {Sun}, {Wang}, {Wang}, {Xu}, {Xu}, {Yang},
  {Yuan}, {Zhang}, {Zhang}, {Zhang}, {Zhang}, {Zhao}, {Chen}, {Jia}, {Cui},
  {Han}, {Li}, {Song}, {Zhao}, {Zhang}, {Zhang}, {Kuulkers}, {Santovincenzo},
  {O'Brien}, {Nandra}, {Rau}, {Cordier}, \& {Einstein Probe Team}}]{LiangYF24}
{Liang}, Y.~F., {Liu}, H.~Y., {Mao}, X., {et~al.} 2024, GRB Coordinates
  Network, 37214, 1

\bibitem[{{Lipunov} {et~al.}(2001){Lipunov}, {Postnov}, \&
  {Prokhorov}}]{Lipunov01}
{Lipunov}, V.~M., {Postnov}, K.~A., \& {Prokhorov}, M.~E. 2001, Astronomy
  Reports, 45, 236, \dodoi{10.1134/1.1353364}

\bibitem[{{Liu} {et~al.}(2024{\natexlab{a}}){Liu}, {Sun}, {Xu}, {Svinkin},
  {Delaunay}, {Tanvir}, {Gao}, {Zhang}, {Chen}, {Wu}, {Zhang}, {Yuan}, {An},
  {Bruni}, {Frederiks}, {Ghirlanda}, {Hu}, {Li}, {Li}, {Li}, {Malesani},
  {Piro}, {Raman}, {Ricci}, {Troja}, {Vergani}, {Wu}, {Yang}, {Zhang}, {Zhu},
  {de Ugarte Postigo}, {Demin}, {Dobie}, {Fan}, {Fu}, {Fynbo}, {Geng},
  {Gianfagna}, {Hu}, {Huang}, {Jiang}, {Jonker}, {Julakanti}, {Kennea},
  {Kokomov}, {Kuulkers}, {Lei}, {Leung}, {Levan}, {Li}, {Li}, {Littlefair},
  {Liu}, {Lysenko}, {Ma}, {Martin-Carrillo}, {O'Brien}, {Parsotan},
  {Quirola-Vasquez}, {Ridnaia}, {Ronchini}, {Rossi}, {Mata-Sanchez},
  {Schneider}, {Shen}, {Thakur}, {Tohuvavohu}, {Torres}, {Tsvetkova}, {Ulanov},
  {Wei}, {Xiao}, {Yin}, {Bai}, {Burwitz}, {Cai}, {Chen}, {Chen}, {Chen},
  {Chen}, {Chen}, {Chen}, {Cheng}, {Cui}, {Cui}, {Dai}, {Dai}, {Eder}, {Fan},
  {Feldman}, {Feng}, {Feng}, {Friedrich}, {Gao}, {Guan}, {Han}, {Han}, {Hou},
  {Hu}, {Hu}, {Huang}, {Huo}, {Hutchinson}, {Ji}, {Jia}, {Jia}, {Jiang}, {Jin},
  {Jin}, {Jin}, {Keereman}, {Lerman}, {Li}, {Li}, {Li}, {Li}, {Li}, {Lian},
  {Liang}, {Ling}, {Liu}, {Liu}, {Liu}, {Liu}, {Liu}, {Lu}, {LU}, {Luo}, {Ma},
  {Ma}, {Mao}, {Mao}, {McHugh}, {Meidinger}, {Nandra}, {Osborne}, {Pan}, {Pan},
  {Ravasio}, {Rau}, {Rea}, {Rehman}, {Sanders}, {Santovincenzo}, {Song}, {Su},
  {Sun}, {Sun}, {Sun}, {Tan}, {Tang}, {Tao}, {Tong}, {Wang}, {Wang}, {Wang},
  {Wang}, {Wang}, {Wang}, {Wang}, {Wang}, {Wei}, {Willingale}, {Xiong}, {Xu},
  {Xu}, {Xu}, {Xu}, {Xu}, {Xue}, {Xue}, {Yan}, {Yang}, {Yang}, {Yang}, {Yang},
  {Yu}, {Zhang}, {Zhang}, {Zhang}, {Zhang}, {Zhang}, {Zhang}, {Zhang}, {Zhang},
  {Zhang}, {Zhao}, {Zhao}, {Zhao}, {Zhao}, {Zhou}, {Zhou}, {Zhu}, {Zhu}, \&
  {Zuo}}]{Liu24}
{Liu}, Y., {Sun}, H., {Xu}, D., {et~al.} 2024{\natexlab{a}}, arXiv e-prints,
  arXiv:2404.16425, \dodoi{10.48550/arXiv.2404.16425}

\bibitem[{{Liu} {et~al.}(2024{\natexlab{b}}){Liu}, {Huang}, {Dai}, {Xu},
  {Zhang}, {Yuan}, \& {Einstein Probe Team}}]{LiuZY24}
{Liu}, Z.~Y., {Huang}, M.~Q., {Dai}, C.~Y., {et~al.} 2024{\natexlab{b}}, GRB
  Coordinates Network, 38214, 1

\bibitem[{{Lu} {et~al.}(2012){Lu}, {Wei}, {Liang}, {Zhang}, {L{\"u}}, {L{\"u}},
  {Lei}, \& {Zhang}}]{Lu12}
{Lu}, R.-J., {Wei}, J.-J., {Liang}, E.-W., {et~al.} 2012, \apj, 756, 112,
  \dodoi{10.1088/0004-637X/756/2/112}

\bibitem[{{Meegan} {et~al.}(1992){Meegan}, {Fishman}, {Wilson}, {Paciesas},
  {Pendleton}, {Horack}, {Brock}, \& {Kouveliotou}}]{Meegan92}
{Meegan}, C.~A., {Fishman}, G.~J., {Wilson}, R.~B., {et~al.} 1992, \nat, 355,
  143, \dodoi{10.1038/355143a0}

\bibitem[{{Meng} {et~al.}(2018){Meng}, {Geng}, {Zhang}, {Wei}, {Xiao}, {Liu},
  {Gao}, {Wu}, {Liang}, {Huang}, {Dai}, \& {Zhang}}]{Meng18}
{Meng}, Y.-Z., {Geng}, J.-J., {Zhang}, B.-B., {et~al.} 2018, \apj, 860, 72,
  \dodoi{10.3847/1538-4357/aac2d9}

\bibitem[{M{\'e}sz{\'a}ros {et~al.}(1999)M{\'e}sz{\'a}ros, Rees, \&
  Wijers}]{Meszaros99b}
M{\'e}sz{\'a}ros, P., Rees, M., \& Wijers, R. 1999, New Astronomy, 4, 303

\bibitem[{{M{\'e}sz{\'a}ros} \& {Rees}(2000)}]{Meszaros00}
{M{\'e}sz{\'a}ros}, P., \& {Rees}, M.~J. 2000, \apj, 530, 292,
  \dodoi{10.1086/308371}

\bibitem[{{M{\'e}sz{\'a}ros} {et~al.}(1994){M{\'e}sz{\'a}ros}, {Rees}, \&
  {Papathanassiou}}]{Meszaros94}
{M{\'e}sz{\'a}ros}, P., {Rees}, M.~J., \& {Papathanassiou}, H. 1994, \apj, 432,
  181, \dodoi{10.1086/174559}

\bibitem[{{M{\'e}sz{\'a}ros} {et~al.}(1998){M{\'e}sz{\'a}ros}, {Rees}, \&
  {Wijers}}]{Meszaros98}
{M{\'e}sz{\'a}ros}, P., {Rees}, M.~J., \& {Wijers}, R.~A.~M.~J. 1998, \apj,
  499, 301, \dodoi{10.1086/305635}

\bibitem[{{Minaev} \& {Pozanenko}(2020)}]{Minaev20}
{Minaev}, P.~Y., \& {Pozanenko}, A.~S. 2020, \mnras, 492, 1919,
  \dodoi{10.1093/mnras/stz3611}

\bibitem[{{Mochkovitch} \& {Nava}(2015)}]{Mochkovitch15}
{Mochkovitch}, R., \& {Nava}, L. 2015, \aap, 577, A31,
  \dodoi{10.1051/0004-6361/201424490}

\bibitem[{{Morsony} {et~al.}(2010){Morsony}, {Lazzati}, \&
  {Begelman}}]{Morsony10}
{Morsony}, B.~J., {Lazzati}, D., \& {Begelman}, M.~C. 2010, \apj, 723, 267,
  \dodoi{10.1088/0004-637X/723/1/267}

\bibitem[{{Nagakura} {et~al.}(2014){Nagakura}, {Hotokezaka}, {Sekiguchi},
  {Shibata}, \& {Ioka}}]{Nagakura14}
{Nagakura}, H., {Hotokezaka}, K., {Sekiguchi}, Y., {Shibata}, M., \& {Ioka}, K.
  2014, \apjl, 784, L28, \dodoi{10.1088/2041-8205/784/2/L28}

\bibitem[{Nakar \& Piran(2016)}]{Nakar17}
Nakar, E., \& Piran, T. 2016, The Astrophysical Journal, 834, 28,
  \dodoi{10.3847/1538-4357/834/1/28}

\bibitem[{{Nava} {et~al.}(2012){Nava}, {Salvaterra}, {Ghirlanda}, {Ghisellini},
  {Campana}, {Covino}, {Cusumano}, {D'Avanzo}, {D'Elia}, {Fugazza}, {Melandri},
  {Sbarufatti}, {Vergani}, \& {Tagliaferri}}]{Nava12}
{Nava}, L., {Salvaterra}, R., {Ghirlanda}, G., {et~al.} 2012, \mnras, 421,
  1256, \dodoi{10.1111/j.1365-2966.2011.20394.x}

\bibitem[{{Norris} {et~al.}(2000){Norris}, {Marani}, \& {Bonnell}}]{Norris00}
{Norris}, J.~P., {Marani}, G.~F., \& {Bonnell}, J.~T. 2000, \apj, 534, 248,
  \dodoi{10.1086/308725}

\bibitem[{{Norris} {et~al.}(1996){Norris}, {Nemiroff}, {Bonnell}, {Scargle},
  {Kouveliotou}, {Paciesas}, {Meegan}, \& {Fishman}}]{Norris96}
{Norris}, J.~P., {Nemiroff}, R.~J., {Bonnell}, J.~T., {et~al.} 1996, \apj, 459,
  393, \dodoi{10.1086/176902}

\bibitem[{{O'Connor} {et~al.}(2024){O'Connor}, {Beniamini}, \&
  {Gill}}]{Connor24}
{O'Connor}, B., {Beniamini}, P., \& {Gill}, R. 2024, \mnras, 533, 1629,
  \dodoi{10.1093/mnras/stae1941}

\bibitem[{{Paciesas} \& {Fermi GBM Collaboration}(2012)}]{Paciesas12}
{Paciesas}, W.~S., \& {Fermi GBM Collaboration}. 2012, in American Astronomical
  Society Meeting Abstracts, Vol. 219, American Astronomical Society Meeting
  Abstracts \#219, 149.12

\bibitem[{{Paczynski} \& {Xu}(1994)}]{Paczynski94}
{Paczynski}, B., \& {Xu}, G. 1994, \apj, 427, 708, \dodoi{10.1086/174178}

\bibitem[{{Pan} {et~al.}(2024){Pan}, {Zhao}, {Peng}, {Jin}, {Ling}, {Liu},
  {Ling}, {Zhang}, {Cheng}, {Chen}, {Cui}, {Fan}, {Hu}, {Hu}, {Huang}, {Li},
  {Liu}, {Liu}, {Lv}, {Lian}, {Mao}, {Pan}, {Sun}, {Wang}, {Wang}, {Wu}, {Xu},
  {Xu}, {Yang}, {Yuan}, {Zhang}, {Zhang}, {Zhang}, {Zhang}, {Chen}, {Jia},
  {Zhang}, {Kuulkers}, {Santovincenzo}, {O'Brien}, {Nandra}, {Rau}, {Cordier},
  \& {Einstein Probe Team}}]{Pan24}
{Pan}, X., {Zhao}, D.~H., {Peng}, J.~Q., {et~al.} 2024, GRB Coordinates
  Network, 36330, 1

\bibitem[{{Pe'er} {et~al.}(2006){Pe'er}, {M{\'e}sz{\'a}ros}, \&
  {Rees}}]{Pe'er06}
{Pe'er}, A., {M{\'e}sz{\'a}ros}, P., \& {Rees}, M.~J. 2006, \apj, 642, 995,
  \dodoi{10.1086/501424}

\bibitem[{{Peng} {et~al.}(2024){Peng}, {Shui}, {Dai}, {Wang}, {Jin}, {Ling},
  {Yuan}, {Liu}, {Zhang}, {Cheng}, {Cui}, {Fan}, {Hu}, {Hu}, {Huang}, {Li},
  {Liu}, {Liu}, {Lv}, {Lian}, {Mao}, {Pan}, {Pan}, {Sun}, {Wang}, {Wang},
  {Wen}, {Wu}, {Xu}, {Xu}, {Yang}, {Zhang}, {Zhang}, {Zhang}, {Zhang}, {Zhao},
  {Chen}, {Jia}, {Zhang}, {Kuulkers}, {Santovincenzo}, {O'Brien}, {Nandra},
  {Rau}, {Cordier}, \& {Einstein Probe Team}}]{PengJQ24}
{Peng}, J.~Q., {Shui}, Q.~C., {Dai}, C.~Y., {et~al.} 2024, GRB Coordinates
  Network, 36810, 1

\bibitem[{{Pescalli} {et~al.}(2015){Pescalli}, {Ghirlanda}, {Salafia},
  {Ghisellini}, {Nappo}, \& {Salvaterra}}]{Pescalli15}
{Pescalli}, A., {Ghirlanda}, G., {Salafia}, O.~S., {et~al.} 2015, \mnras, 447,
  1911, \dodoi{10.1093/mnras/stu2482}

\bibitem[{{Petrosian} {et~al.}(2015){Petrosian}, {Kitanidis}, \&
  {Kocevski}}]{Petrosian15}
{Petrosian}, V., {Kitanidis}, E., \& {Kocevski}, D. 2015, \apj, 806, 44,
  \dodoi{10.1088/0004-637X/806/1/44}

\bibitem[{{Piran}(2004)}]{Piran04}
{Piran}, T. 2004, Reviews of Modern Physics, 76, 1143,
  \dodoi{10.1103/RevModPhys.76.1143}

\bibitem[{{Preece} {et~al.}(1998){Preece}, {Briggs}, {Mallozzi}, {Pendleton},
  {Paciesas}, \& {Band}}]{Preece98}
{Preece}, R.~D., {Briggs}, M.~S., {Mallozzi}, R.~S., {et~al.} 1998, \apjl, 506,
  L23, \dodoi{10.1086/311644}

\bibitem[{{Preece} {et~al.}(2000){Preece}, {Briggs}, {Mallozzi}, {Pendleton},
  {Paciesas}, \& {Band}}]{Preece00}
---. 2000, \apjs, 126, 19, \dodoi{10.1086/313289}

\bibitem[{{Qin} \& {Chen}(2013)}]{Qin13}
{Qin}, Y.-P., \& {Chen}, Z.-F. 2013, \mnras, 430, 163,
  \dodoi{10.1093/mnras/sts547}

\bibitem[{{Ramirez-Ruiz} {et~al.}(2005){Ramirez-Ruiz}, {Granot}, {Kouveliotou},
  {Woosley}, {Patel}, \& {Mazzali}}]{Ramirez-Ruiz05}
{Ramirez-Ruiz}, E., {Granot}, J., {Kouveliotou}, C., {et~al.} 2005, \apjl, 625,
  L91, \dodoi{10.1086/431237}

\bibitem[{{Rees} \& {M{\'e}sz{\'a}ros}(1994)}]{Rees94}
{Rees}, M.~J., \& {M{\'e}sz{\'a}ros}, P. 1994, \apjl, 430, L93,
  \dodoi{10.1086/187446}

\bibitem[{{Rees} \& {M{\'e}sz{\'a}ros}(2005)}]{Rees05}
---. 2005, \apj, 628, 847, \dodoi{10.1086/430818}

\bibitem[{{Rhoads}(1997)}]{Rhoads97}
{Rhoads}, J.~E. 1997, \apjl, 487, L1, \dodoi{10.1086/310876}

\bibitem[{{Ridnaia} {et~al.}(2024){Ridnaia}, {Frederiks}, {Lysenko}, {Svinkin},
  {Tsvetkova}, {Ulanov}, {Cline}, \& {Konus-Wind Team}}]{Ridnaia24}
{Ridnaia}, A., {Frederiks}, D., {Lysenko}, A., {et~al.} 2024, GRB Coordinates
  Network, 37982, 1

\bibitem[{{Rossi} {et~al.}(2002){Rossi}, {Lazzati}, \& {Rees}}]{Rossi02}
{Rossi}, E., {Lazzati}, D., \& {Rees}, M.~J. 2002, \mnras, 332, 945,
  \dodoi{10.1046/j.1365-8711.2002.05363.x}

\bibitem[{{Ryan} {et~al.}(2020){Ryan}, {van Eerten}, {Piro}, \&
  {Troja}}]{Ryan20}
{Ryan}, G., {van Eerten}, H., {Piro}, L., \& {Troja}, E. 2020, \apj, 896, 166,
  \dodoi{10.3847/1538-4357/ab93cf}

\bibitem[{{Rybicki} \& {Lightman}(1979)}]{Rybicki79}
{Rybicki}, G.~B., \& {Lightman}, A.~P. 1979, {Radiative processes in
  astrophysics}

\bibitem[{{Ryde} {et~al.}(2011){Ryde}, {Pe'er}, {Nymark}, {Axelsson},
  {Moretti}, {Lundman}, {Battelino}, {Bissaldi}, {Chiang}, {Jackson},
  {Larsson}, {Longo}, {McGlynn}, \& {Omodei}}]{Ryde11}
{Ryde}, F., {Pe'er}, A., {Nymark}, T., {et~al.} 2011, \mnras, 415, 3693,
  \dodoi{10.1111/j.1365-2966.2011.18985.x}

\bibitem[{{Sakamoto} {et~al.}(2008){Sakamoto}, {Barthelmy}, {Barbier},
  {Cummings}, {Fenimore}, {Gehrels}, {Hullinger}, {Krimm}, {Markwardt},
  {Palmer}, {Parsons}, {Sato}, {Stamatikos}, {Tueller}, {Ukwatta}, \&
  {Zhang}}]{Sakamoto08b}
{Sakamoto}, T., {Barthelmy}, S.~D., {Barbier}, L., {et~al.} 2008, \apjs, 175,
  179, \dodoi{10.1086/523646}

\bibitem[{{Sakamoto} {et~al.}(2011){Sakamoto}, {Barthelmy}, {Baumgartner},
  {Cummings}, {Fenimore}, {Gehrels}, {Krimm}, {Markwardt}, {Palmer}, {Parsons},
  {Sato}, {Stamatikos}, {Tueller}, {Ukwatta}, \& {Zhang}}]{Sakamoto11}
{Sakamoto}, T., {Barthelmy}, S.~D., {Baumgartner}, W.~H., {et~al.} 2011, \apjs,
  195, 2, \dodoi{10.1088/0067-0049/195/1/2}

\bibitem[{{Salvaterra} \& {Chincarini}(2007)}]{Salvaterra07}
{Salvaterra}, R., \& {Chincarini}, G. 2007, \apjl, 656, L49,
  \dodoi{10.1086/512606}

\bibitem[{{Salvaterra} {et~al.}(2009){Salvaterra}, {Della Valle}, {Campana},
  {Chincarini}, {Covino}, {D'Avanzo}, {Fern{\'a}ndez-Soto}, {Guidorzi},
  {Mannucci}, {Margutti}, {Th{\"o}ne}, {Antonelli}, {Barthelmy}, {de Pasquale},
  {D'Elia}, {Fiore}, {Fugazza}, {Hunt}, {Maiorano}, {Marinoni}, {Marshall},
  {Molinari}, {Nousek}, {Pian}, {Racusin}, {Stella}, {Amati}, {Andreuzzi},
  {Cusumano}, {Fenimore}, {Ferrero}, {Giommi}, {Guetta}, {Holland}, {Hurley},
  {Israel}, {Mao}, {Markwardt}, {Masetti}, {Pagani}, {Palazzi}, {Palmer},
  {Piranomonte}, {Tagliaferri}, \& {Testa}}]{Salvaterra09b}
{Salvaterra}, R., {Della Valle}, M., {Campana}, S., {et~al.} 2009, \nat, 461,
  1258, \dodoi{10.1038/nature08445}

\bibitem[{{Salvaterra} {et~al.}(2012){Salvaterra}, {Campana}, {Vergani},
  {Covino}, {D'Avanzo}, {Fugazza}, {Ghirlanda}, {Ghisellini}, {Melandri},
  {Nava}, {Sbarufatti}, {Flores}, {Piranomonte}, \&
  {Tagliaferri}}]{Salvaterra12}
{Salvaterra}, R., {Campana}, S., {Vergani}, S.~D., {et~al.} 2012, \apj, 749,
  68, \dodoi{10.1088/0004-637X/749/1/68}

\bibitem[{{Spruit} {et~al.}(2001){Spruit}, {Daigne}, \& {Drenkhahn}}]{Spruit01}
{Spruit}, H.~C., {Daigne}, F., \& {Drenkhahn}, G. 2001, \aap, 369, 694,
  \dodoi{10.1051/0004-6361:20010131}

\bibitem[{{Sun} {et~al.}(2024{\natexlab{a}}){Sun}, {Li}, {Liu}, {Gao}, {Wang},
  {Yuan}, {Zhang}, {Filippenko}, {Xu}, {An}, {Ai}, {Brink}, {Liu}, {Liu},
  {Wang}, {Wu}, {Wu}, {Yang}, {Zhang}, {Zheng}, {Ahumada}, {Dai}, {Delaunay},
  {Elias-Rosa}, {Benetti}, {Fu}, {Howell}, {Huang}, {Kasliwal}, {Karambelkar},
  {Stein}, {Lei}, {Lian}, {Peng}, {Ridnaia}, {Svinkin}, {Wang}, {Wang}, {Wei},
  {An}, {Andrews}, {Bai}, {Dai}, {Ehgamberdiev}, {Fan}, {Farah}, {Feng},
  {Fynbo}, {Guo}, {Guo}, {Hu}, {Hu}, {Jiang}, {Jin}, {Li}, {Li}, {Li}, {Liang},
  {Ling}, {Liu}, {Mao}, {McCully}, {Mirzaqulov}, {Newsome}, {Padilla Gonzalez},
  {Pan}, {Terreran}, {Tinyanont}, {Wang}, {Wang}, {Wen}, {Xiang}, {Xue},
  {Yang}, {Zhu}, {Cai}, {Castro-Tirado}, {Chen}, {Chen}, {Chen}, {Chen},
  {Chen}, {Chen}, {Chen}, {Cheng}, {Cordier}, {Cui}, {Cui}, {Dai}, {Fan},
  {Feng}, {Guan}, {Han}, {Hou}, {Hu}, {Huang}, {Huo}, {Jia}, {Jia}, {Jiang},
  {Jin}, {Jin}, {Kuulkers}, {Li}, {Li}, {Li}, {Li}, {Li}, {Li}, {Li}, {Liu},
  {Liu}, {Liu}, {Liu}, {Lu}, {Luo}, {Ma}, {Mao}, {Nandra}, {O'Brien}, {Pan},
  {Rau}, {Rea}, {Sanders}, {Song}, {Sun}, {Sun}, {Tan}, {Tang}, {Tao}, {Wang},
  {Wang}, {Wang}, {Wang}, {Wang}, {Wang}, {Xiong}, {Xu}, {Xu}, {Xu}, {Xu},
  {Xu}, {Xue}, {Xue}, {Yan}, {Yang}, {Yang}, {Yang}, {Zhang}, {Zhang}, {Zhang},
  {Zhang}, {Zhang}, {Zhang}, {Zhang}, {Zhang}, {Zhang}, {Zhang}, {Zhao},
  {Zhao}, {Zhao}, {Zhao}, {Zhou}, {Zhu}, \& {Zhu}}]{SunH24}
{Sun}, H., {Li}, W.~X., {Liu}, L.~D., {et~al.} 2024{\natexlab{a}}, arXiv
  e-prints, arXiv:2410.02315, \dodoi{10.48550/arXiv.2410.02315}

\bibitem[{{Sun} {et~al.}(2024{\natexlab{b}}){Sun}, {Geng}, {Yan}, {Hu}, {Wu},
  {Castro-Tirado}, {Yang}, {Ping}, {Hu}, {Xu}, {Gao}, {Jiang}, {Zhu}, {Xue},
  {P{\'e}rez-Garc{\'\i}a}, {Wu}, {Fern{\'a}ndez-Garc{\'\i}a},
  {Caballero-Garc{\'\i}a}, {S{\'a}nchez-Ram{\'\i}rez}, {Guziy}, {Olivares},
  {P{\'e}rez del Pulgar}, {Castell{\'o}n}, {Castillo}, {Xiong}, {Pandey},
  {Hiriart}, {Garc{\'\i}a-Segura}, {Lee}, {Carrasco-Garc{\'\i}a}, {Park},
  {Meintjes}, {van Heerden}, {Mart{\'\i}n-Carrillo}, {Hanlon}, {Zhang},
  {Maury}, {Hern{\'a}ndez-Garc{\'\i}a}, {Gritsevich}, {Rossi}, {Maiorano},
  {Cusano}, {D'Avanzo}, {Ferro}, {Melandri}, {De Pasquale}, {Brivio}, {Fang},
  {Fan}, {Hu}, {Wan}, {Hu}, {Zuo}, {Tang}, {Zhang}, {Zheng}, {Li}, {Luo},
  {Liu}, {Wang}, {Zhang}, {Liu}, {Gao}, {Liang}, {Wang}, {Yao}, {Cheng},
  {Zhao}, \& {Dai}}]{Sun24}
{Sun}, T.-R., {Geng}, J.-J., {Yan}, J.-Z., {et~al.} 2024{\natexlab{b}}, ApJL in
  press, arXiv:2409.17983, \dodoi{10.48550/arXiv.2409.17983}

\bibitem[{{Svinkin} {et~al.}(2024){Svinkin}, {Frederiks}, {Lysenko}, {Ridnaia},
  {Tsvetkova}, {Ulanov}, {Cline}, \& {Konus-Wind Team}}]{Svinkin24}
{Svinkin}, D., {Frederiks}, D., {Lysenko}, A., {et~al.} 2024, GRB Coordinates
  Network, 37927, 1

\bibitem[{{Tavani}(1996)}]{Tavani96}
{Tavani}, M. 1996, \apj, 466, 768, \dodoi{10.1086/177551}

\bibitem[{{Tian} {et~al.}(2024){Tian}, {Peng}, {Mao}, {Jin}, {Liu}, {Liu},
  {Ling}, {Zhang}, {Cheng}, {Chen}, {Cui}, {Fan}, {Hu}, {W.}, {Huang}, {Li},
  {Liu}, {Lv}, {Lian}, {Pan}, {Pan}, {Sun}, {Wang}, {Wang}, {Xu}, {Xu}, {Yang},
  {Yuan}, {Zhang}, {Zhang}, {Zhang}, {Zhang}, {Zhao}, {Chen}, {Jia}, {Cui},
  {Han}, {Li}, {Song}, {Zhao}, {Zhang}, {Zhang}, {Liang}, {Kuulkers},
  {Santovincenzo}, {O'Brien}, {Nandra}, {Rau}, {Cordier}, \& {Einstein Probe
  Team}}]{Tian24}
{Tian}, X., {Peng}, H.~L., {Mao}, X., {et~al.} 2024, GRB Coordinates Network,
  37648, 1

\bibitem[{{Troja} {et~al.}(2017){Troja}, {Piro}, {van Eerten}, {Wollaeger},
  {Im}, {Fox}, {Butler}, {Cenko}, {Sakamoto}, {Fryer}, {Ricci}, {Lien}, {Ryan},
  {Korobkin}, {Lee}, {Burgess}, {Lee}, {Watson}, {Choi}, {Covino}, {D'Avanzo},
  {Fontes}, {Gonz{\'a}lez}, {Khandrika}, {Kim}, {Kim}, {Lee}, {Lee}, {Kutyrev},
  {Lim}, {S{\'a}nchez-Ram{\'\i}rez}, {Veilleux}, {Wieringa}, \&
  {Yoon}}]{Troja17}
{Troja}, E., {Piro}, L., {van Eerten}, H., {et~al.} 2017, \nat, 551, 71,
  \dodoi{10.1038/nature24290}

\bibitem[{{Troja} {et~al.}(2019){Troja}, {van Eerten}, {Ryan}, {Ricci},
  {Burgess}, {Wieringa}, {Piro}, {Cenko}, \& {Sakamoto}}]{Troja19}
{Troja}, E., {van Eerten}, H., {Ryan}, G., {et~al.} 2019, \mnras, 489, 1919,
  \dodoi{10.1093/mnras/stz2248}

\bibitem[{Tsvetkova {et~al.}(2017)Tsvetkova, Frederiks, Golenetskii, Lysenko,
  Oleynik, Pal'shin, Svinkin, Ulanov, Cline, Hurley, \& Aptekar}]{Tsvetkova17}
Tsvetkova, A., Frederiks, D., Golenetskii, S., {et~al.} 2017, \apj, 850, 161,
  \dodoi{10.3847/1538-4357/aa96af}

\bibitem[{{Tsvetkova} {et~al.}(2021){Tsvetkova}, {Frederiks}, {Svinkin},
  {Aptekar}, {Cline}, {Golenetskii}, {Hurley}, {Lysenko}, {Ridnaia}, \&
  {Ulanov}}]{Tsvetkova21}
{Tsvetkova}, A., {Frederiks}, D., {Svinkin}, D., {et~al.} 2021, \apj, 908, 83,
  \dodoi{10.3847/1538-4357/abd569}

\bibitem[{{Uhm} \& {Zhang}(2014)}]{Uhm14}
{Uhm}, Z.~L., \& {Zhang}, B. 2014, Nature Physics, 10, 351,
  \dodoi{10.1038/nphys2932}

\bibitem[{{Uhm} \& {Zhang}(2016)}]{Uhm16}
---. 2016, \apjl, 824, L16, \dodoi{10.3847/2041-8205/824/1/L16}

\bibitem[{{Uhm} {et~al.}(2018){Uhm}, {Zhang}, \& {Racusin}}]{Uhm18}
{Uhm}, Z.~L., {Zhang}, B., \& {Racusin}, J. 2018, \apj, 869, 100,
  \dodoi{10.3847/1538-4357/aaeb30}

\bibitem[{van Putten \& Levinson(2003)}]{Putten03}
van Putten, M. H. P.~M., \& Levinson, A. 2003, The Astrophysical Journal, 584,
  937, \dodoi{10.1086/345900}

\bibitem[{{Vlahakis} {et~al.}(2003){Vlahakis}, {Peng}, \&
  {K{\"o}nigl}}]{Vlahakis03}
{Vlahakis}, N., {Peng}, F., \& {K{\"o}nigl}, A. 2003, \apjl, 594, L23,
  \dodoi{10.1086/378580}

\bibitem[{{von Kienlin} {et~al.}(2020){von Kienlin}, {Meegan}, {Paciesas},
  {Bhat}, {Bissaldi}, {Briggs}, {Burns}, {Cleveland}, {Gibby}, {Giles},
  {Goldstein}, {Hamburg}, {Hui}, {Kocevski}, {Mailyan}, {Malacaria},
  {Poolakkil}, {Preece}, {Roberts}, {Veres}, \& {Wilson-Hodge}}]{Kienlin20}
{von Kienlin}, A., {Meegan}, C.~A., {Paciesas}, W.~S., {et~al.} 2020, \apj,
  893, 46, \dodoi{10.3847/1538-4357/ab7a18}

\bibitem[{{Wanderman} \& {Piran}(2010)}]{Wanderman10}
{Wanderman}, D., \& {Piran}, T. 2010, \mnras, 406, 1944,
  \dodoi{10.1111/j.1365-2966.2010.16787.x}

\bibitem[{Wang {et~al.}(2018)Wang, Zhang, Liang, Lu, Lin, Li, \& Li}]{Wang18}
Wang, X.-G., Zhang, B., Liang, E.-W., {et~al.} 2018, \apj, 859, 160,
  \dodoi{10.3847/1538-4357/aabc13}

\bibitem[{{Wang} {et~al.}(2024){Wang}, {Sun}, {Wang}, {Zhang}, {Liu}, {Pan},
  {Ling}, {Jin}, {Zhang}, {Cheng}, {Chen}, {Cui}, {Fan}, {Hu}, {Hu}, {Huang},
  {Li}, {Liu}, {Liu}, {Lv}, {Lian}, {Mao}, {Pan}, {Wang}, {Wang}, {Wu}, {Xu},
  {Xu}, {Yang}, {Yuan}, {Zhang}, {Zhang}, {Zhang}, {Zhao}, {Chen}, {Jia},
  {Cui}, {Han}, {Li}, {Song}, {Zhao}, {Zhang}, {Zhang}, {Kuulkers},
  {Santovincenzo}, {O'Brien}, {Nandra}, {Rau}, {Cordier}, \& {Einstein Probe
  Team}}]{WangY24}
{Wang}, Y., {Sun}, H., {Wang}, Y.~L., {et~al.} 2024, GRB Coordinates Network,
  37034, 1

\bibitem[{{Wichern} {et~al.}(2024){Wichern}, {Ravasio}, {Jonker},
  {Quirola-V{\'a}squez}, {Levan}, {Bauer}, \& {Kann}}]{Wichern24}
{Wichern}, H.~C.~I., {Ravasio}, M.~E., {Jonker}, P.~G., {et~al.} 2024, \aap,
  690, A101, \dodoi{10.1051/0004-6361/202450116}

\bibitem[{{Wu} \& {Fenimore}(2000)}]{Wu00}
{Wu}, B., \& {Fenimore}, E. 2000, \apjl, 535, L29, \dodoi{10.1086/312700}

\bibitem[{{Wu} {et~al.}(2024){Wu}, {Zhang}, {Wang}, {Sun}, {Wang}, {Zhang},
  {Liu}, {Pan}, {Ling}, {Jin}, {Zhang}, {Cheng}, {Chen}, {Cui}, {Fan}, {Hu},
  {Huang}, {Li}, {Liu}, {Liu}, {Lv}, {Lian}, {Mao}, {Pan}, {Wang}, {Wang},
  {Xu}, {Xu}, {Yang}, {Yuan}, {Zhang}, {Zhang}, {Zhang}, {Zhao}, {Chen}, {Jia},
  {Cui}, {Han}, {Li}, {Song}, {Zhao}, {Zhang}, {Zhang}, {Kuulkers},
  {Santovincenzo}, {O'Brien}, {Nandra}, {Rau}, {Cordier}, \& {Einstein Probe
  Team}}]{Wu24}
{Wu}, Q.~Y., {Zhang}, Y.~J., {Wang}, Y., {et~al.} 2024, GRB Coordinates
  Network, 37063, 1

\bibitem[{{Xiao} {et~al.}(2017){Xiao}, {Liu}, {Dai}, \& {Wu}}]{Xiao17}
{Xiao}, D., {Liu}, L.-D., {Dai}, Z.-G., \& {Wu}, X.-F. 2017, \apjl, 850, L41,
  \dodoi{10.3847/2041-8213/aa9b2b}

\bibitem[{{Xu} {et~al.}(2023){Xu}, {Huang}, {Geng}, {Wu}, {Li}, \&
  {Zhang}}]{Xu23}
{Xu}, F., {Huang}, Y.-F., {Geng}, J.-J., {et~al.} 2023, \aap, 673, A20,
  \dodoi{10.1051/0004-6361/202245414}

\bibitem[{{Yamazaki} {et~al.}(2004){Yamazaki}, {Ioka}, \&
  {Nakamura}}]{Yamazaki04}
{Yamazaki}, R., {Ioka}, K., \& {Nakamura}, T. 2004, \apjl, 606, L33,
  \dodoi{10.1086/421084}

\bibitem[{{Yang} {et~al.}(2024{\natexlab{a}}){Yang}, {Zhang}, {Wang}, {Yuan},
  {Ling}, {Jin}, {Liu}, {Zhang}, {Cheng}, {Chen}, {Cui}, {Fan}, {Hu}, {Hu},
  {Huang}, {Li}, {Liu}, {Liu}, {Lv}, {Lian}, {Mao}, {Pan}, {Pan}, {Sun},
  {Wang}, {Wu}, {Xu}, {Xu}, {Zhang}, {Zhang}, {Zhang}, {Zhao}, {Chen}, {Jia},
  {Cui}, {Han}, {Li}, {Song}, {Zhao}, {Zhang}, {Zhang}, {Kuulkers},
  {Santovincenzo}, {O'Brien}, {Nandra}, {Rau}, {Cordier}, \& {Einstein Probe
  Team}}]{YangHN24a}
{Yang}, H.~N., {Zhang}, W.~J., {Wang}, W.~X., {et~al.} 2024{\natexlab{a}}, GRB
  Coordinates Network, 37188, 1

\bibitem[{{Yang} {et~al.}(2024{\natexlab{b}}){Yang}, {Zhang}, {Wang}, {Yuan},
  {Ling}, {Jin}, {Liu}, {Zhang}, {Cheng}, {Chen}, {Cui}, {Fan}, {Hu}, {Hu},
  {Huang}, {Li}, {Liu}, {Liu}, {Lv}, {Lian}, {Mao}, {Pan}, {Pan}, {Sun},
  {Wang}, {Wu}, {Xu}, {Xu}, {Zhang}, {Zhang}, {Zhang}, {Zhao}, {Chen}, {Jia},
  {Cui}, {Han}, {Li}, {Song}, {Zhao}, {Zhang}, {Zhang}, {Kuulkers},
  {Santovincenzo}, {O'Brien}, {Nandra}, {Rau}, {Cordier}, \& {Einstein Probe
  Team}}]{YangHN24b}
---. 2024{\natexlab{b}}, GRB Coordinates Network, 37185, 1

\bibitem[{{Yin} {et~al.}(2024){Yin}, {Zhang}, {Yang}, {Sun}, {Zhang}, {Shao},
  {Hu}, {Zhu}, {Xu}, {An}, {Gao}, {Wu}, {Zhang}, {Castro-Tirado}, {Pandey},
  {Rau}, {Lei}, {Xie}, {Ghirlanda}, {Piro}, {O'Brien}, {Troja}, {Jonker}, {Yu},
  {An}, {Chen}, {Chen}, {Dong}, {Eyles-Ferris}, {Fan}, {Fu}, {Fynbo}, {Gao},
  {Huang}, {Jiang}, {Jiang}, {Julakanti}, {Kuulkers}, {Lao}, {Li}, {Ling},
  {Liu}, {Liu}, {Mou}, {Varun}, {Wei}, {Wu}, {Yadav}, {Yang}, {Yuan}, \&
  {Zhang}}]{Yin24}
{Yin}, Y.-H.~I., {Zhang}, B.-B., {Yang}, J., {et~al.} 2024, arXiv e-prints,
  arXiv:2407.10156, \dodoi{10.48550/arXiv.2407.10156}

\bibitem[{{Yonetoku} {et~al.}(2004){Yonetoku}, {Murakami}, {Nakamura},
  {Yamazaki}, {Inoue}, \& {Ioka}}]{Yonetoku04}
{Yonetoku}, D., {Murakami}, T., {Nakamura}, T., {et~al.} 2004, \apj, 609, 935,
  \dodoi{10.1086/421285}

\bibitem[{{Yu} {et~al.}(2015){Yu}, {Wang}, {Dai}, \& {Cheng}}]{Yu15}
{Yu}, H., {Wang}, F.~Y., {Dai}, Z.~G., \& {Cheng}, K.~S. 2015, \apjs, 218, 13,
  \dodoi{10.1088/0067-0049/218/1/13}

\bibitem[{Yuan {et~al.}(2022)Yuan, Zhang, Chen, \& Ling}]{Yuan22}
Yuan, W., Zhang, C., Chen, Y., \& Ling, Z. 2022, The Einstein Probe Mission,
  ed. C.~Bambi \& A.~Santangelo (Singapore: Springer Nature Singapore), 1--30,
  \dodoi{10.1007/978-981-16-4544-0_151-1}

\bibitem[{Yuan {et~al.}(2024)Yuan, Zhang, Ling, Zhang, Cai, Liu, Sun, Sun,
  Chen, Cui, Han, Jia, Liu, Sun, Kuulkers, Nandra, \& Cordier}]{Yuan24}
Yuan, W., Zhang, C., Ling, Z., {et~al.} 2024, in Space Telescopes and
  Instrumentation 2024: Ultraviolet to Gamma Ray, ed. J.-W.~A. den Herder,
  S.~Nikzad, \& K.~Nakazawa, Vol. 13093, International Society for Optics and
  Photonics (SPIE), 130931C, \dodoi{10.1117/12.3023595}

\bibitem[{{Zhang}(2018)}]{Zhang18}
{Zhang}, B. 2018, {The Physics of Gamma-Ray Bursts (Cambridge: Cambridge Univ.
  Press)}, \dodoi{10.1017/9781139226530}

\bibitem[{{Zhang} \& {M{\'e}sz{\'a}ros}(2002{\natexlab{a}})}]{Zhang02}
{Zhang}, B., \& {M{\'e}sz{\'a}ros}, P. 2002{\natexlab{a}}, \apj, 571, 876,
  \dodoi{10.1086/339981}

\bibitem[{{Zhang} \& {M{\'e}sz{\'a}ros}(2002{\natexlab{b}})}]{Zhang02b}
---. 2002{\natexlab{b}}, \apj, 581, 1236, \dodoi{10.1086/344338}

\bibitem[{{Zhang} \& {Yan}(2011{\natexlab{a}})}]{ZhangB11}
{Zhang}, B., \& {Yan}, H. 2011{\natexlab{a}}, \apj, 726, 90,
  \dodoi{10.1088/0004-637X/726/2/90}

\bibitem[{{Zhang} \& {Yan}(2011{\natexlab{b}})}]{ZhangYang11}
---. 2011{\natexlab{b}}, \apj, 726, 90, \dodoi{10.1088/0004-637X/726/2/90}

\bibitem[{{Zhang} {et~al.}(2012){Zhang}, {Shao}, {Yan}, \& {Wei}}]{Zhang12d}
{Zhang}, F.-W., {Shao}, L., {Yan}, J.-Z., \& {Wei}, D.-M. 2012, \apj, 750, 88,
  \dodoi{10.1088/0004-637X/750/2/88}

\bibitem[{{Zhang} {et~al.}(2003){Zhang}, {Woosley}, \& {MacFadyen}}]{Zhang03}
{Zhang}, W., {Woosley}, S.~E., \& {MacFadyen}, A.~I. 2003, \apj, 586, 356,
  \dodoi{10.1086/367609}

\bibitem[{{Zhang} {et~al.}(2024{\natexlab{a}}){Zhang}, {Shui}, {Wu}, {He},
  {Yang}, {Hua}, {Yuan}, \& {Einstein Probe Team}}]{ZhangWJ24}
{Zhang}, W.~J., {Shui}, Q.~C., {Wu}, H.~Z., {et~al.} 2024{\natexlab{a}}, GRB
  Coordinates Network, 38281, 1

\bibitem[{{Zhang} {et~al.}(2024{\natexlab{b}}){Zhang}, {Mao}, {Zhang}, {Liu},
  {Liu}, {Zhang}, {Ling}, {Jin}, {Cheng}, {Chen}, {Cui}, {Fan}, {Hu}, {Hu},
  {Huang}, {Li}, {Lian}, {Liu}, {Lv}, {Pan}, {Pan}, {Sun}, {Wang}, {Wang},
  {Wu}, {Xu}, {Xu}, {Yang}, {Zhang}, {Zhang}, {Zhang}, {Zhao}, {Kuulkers},
  {O'Brien}, {Yuan}, \& {Einstein Probe team}}]{Zhang24}
{Zhang}, W.~J., {Mao}, X., {Zhang}, W.~D., {et~al.} 2024{\natexlab{b}}, GRB
  Coordinates Network, 35931, 1

\bibitem[{{Zhang} {et~al.}(2019){Zhang}, {Geng}, \& {Huang}}]{Zhang19}
{Zhang}, Y., {Geng}, J.-J., \& {Huang}, Y.-F. 2019, \apj, 877, 89,
  \dodoi{10.3847/1538-4357/ab1b10}

\bibitem[{{Zhang} {et~al.}(2024{\natexlab{c}}){Zhang}, {Liu}, {Jiang}, {Pan},
  {Jin}, {Ling}, {Yuan}, {Liu}, {Zhang}, {Chen}, {Cheng}, {Cui}, {Fan}, {Hu},
  {Hu}, {Huang}, {Li}, {Liu}, {Lv}, {Lian}, {Mao}, {Pan}, {Sun}, {Wang},
  {Wang}, {Wen}, {Wu}, {Xu}, {Xu}, {Yang}, {Zhang}, {Zhang}, {Zhang}, {Zhang},
  {Zhao}, {Chen}, {Jia}, {Zhang}, {Kuulkers}, {Santovincenzo}, {O'Brien},
  {Nandra}, {Rau}, {Cordier}, \& {Einstein Probe Team}}]{ZhangYJ24}
{Zhang}, Y.~J., {Liu}, M.~J., {Jiang}, S.~Q., {et~al.} 2024{\natexlab{c}}, GRB
  Coordinates Network, 36818, 1

\bibitem[{{Zhao} {et~al.}(2014){Zhao}, {Li}, {Liu}, {Zhang}, {Bai}, \&
  {M{\'e}sz{\'a}ros}}]{ZhaoXH14}
{Zhao}, X., {Li}, Z., {Liu}, X., {et~al.} 2014, \apj, 780, 12,
  \dodoi{10.1088/0004-637X/780/1/12}

\bibitem[{{Zheng} {et~al.}(2024){Zheng}, {Brink}, {Filippenko}, {Davies},
  {Deller}, \& {KAIT GRB team}}]{ZhengWK24}
{Zheng}, W., {Brink}, T.~G., {Filippenko}, A.~V., {et~al.} 2024, GRB
  Coordinates Network, 37228, 1

\bibitem[{{Zhou} {et~al.}(2024{\natexlab{a}}){Zhou}, {Chen}, {Sun}, {Zhang},
  {Hu}, {Li}, {Ling}, {Liu}, {Zhang}, {Jin}, {Cheng}, {Cui}, {Fan}, {Hu},
  {Huang}, {Liu}, {Liu}, {Lv}, {Lian}, {Mao}, {Pan}, {Pan}, {Wang}, {Wang},
  {Wu}, {Xu}, {Xu}, {Yang}, {Yuan}, {Zhang}, {Zhang}, {Zhang}, {Zhang}, {Zhao},
  {Yang}, {Dai}, {Liang}, {Chen}, {Jia}, {Zhang}, {Kuulkers}, {Santovincenzo},
  {O'Brien}, {Nandra}, {Rau}, {Cordier}, \& {Einstein Probe Team}}]{Zhou24a}
{Zhou}, H., {Chen}, W., {Sun}, H., {et~al.} 2024{\natexlab{a}}, GRB Coordinates
  Network, 36691, 1

\bibitem[{{Zhou} {et~al.}(2024{\natexlab{b}}){Zhou}, {Wang}, {Hu}, {Ling},
  {Jin}, {Liu}, {Zhang}, {Cheng}, {Chen}, {Cui}, {Fan}, {Hu}, {Huang}, {Li},
  {Liu}, {Liu}, {Lv}, {Lian}, {Mao}, {Pan}, {Pan}, {Sun}, {Wang}, {Wu}, {Xu},
  {Xu}, {Yang}, {Yuan}, {Zhang}, {Zhang}, {Zhang}, {Zhang}, {Zhao}, {Chen},
  {Jia}, {Zhang}, {Kuulkers}, {Santovincenzo}, {O'Brien}, {Nandra}, {Rau},
  {Cordier}, \& {Einstein Probe Team}}]{Zhou24b}
{Zhou}, H., {Wang}, W.~X., {Hu}, J.~W., {et~al.} 2024{\natexlab{b}}, GRB
  Coordinates Network, 36997, 1

\end{thebibliography}

\end{document}